\documentclass[apj]{emulateapj}

\usepackage{graphicx}

\newcommand{\vect}[1]{{\mathbf{#1}}}
\newcommand{\mydotfill}{\leaders\hbox to 2pt{\hss.\hss}\hfill\phantom{.}}

\newcommand{\tff}{t_\mathrm{ff}}
\newcommand{\racc}{r_\mathrm{acc}}
\newcommand{\rsoft}{r_\mathrm{soft}}
\newcommand{\rhores}{\rho_\mathrm{res}}
\newcommand{\lJ}{\lambda_\mathrm{J}}
\newcommand{\MJ}{M_\mathrm{J}}
\newcommand{\cs}{c_\mathrm{s}}
\newcommand{\Msol}{\mbox{$M_{\sun}$}}
\newcommand{\AU}{\mbox{AU}}

\newcommand{\cm}{\mbox{cm}}

\newcommand{\km}{\mbox{km}}
\newcommand{\rad}{\mbox{rad}}
\newcommand{\g}{\mbox{g}}

\newcommand{\pc}{\mbox{pc}}

\newcommand{\K}{\mbox{K}}
\newcommand{\yr}{\mbox{yr}}

\newcommand{\Ma}{{\cal M}}
\newcommand{\di}{\mbox{d}}

\slugcomment{draft \today}

\shorttitle{Modeling Collapse and Accretion in Turbulent Gas Clouds}

\shortauthors{Federrath, Banerjee, Clark, \& Klessen}

\begin{document}

\title{Modeling Collapse and Accretion in Turbulent Gas Clouds: Implementation and Comparison of Sink Particles in \textsc{amr} and \textsc{sph}}

\author{Christoph Federrath\altaffilmark{1,2}, Robi Banerjee\altaffilmark{1}, Paul C.~Clark\altaffilmark{1}, and Ralf S.~Klessen\altaffilmark{1}}

\altaffiltext{1}{Zentrum f\"ur Astronomie der Universit\"at Heidelberg, \\Institut f\"ur Theoretische Astrophysik, Albert-Ueberle-Str.~2, 69120 Heidelberg, Germany}
\altaffiltext{2}{Max-Planck-Institute for Astronomy, K\"onigstuhl 17, 69117 Heidelberg, Germany}

\email{chfeder@ita.uni-heidelberg.de}
\email{banerjee@ita.uni-heidelberg.de}
\email{pcc@ita.uni-heidelberg.de}
\email{rklessen@ita.uni-heidelberg.de}

\begin{abstract}
Star formation is such a complex process that accurate numerical tools are needed to quantitatively examine the mass distribution and accretion of fragments in collapsing, turbulent, magnetized gas clouds. To enable a numerical treatment of this regime, we implemented sink particles in the adaptive mesh refinement (\textsc{amr}) hydrodynamics code \textsc{flash}. Sink particles are created in regions of local gravitational collapse, and their trajectories and accretion can be followed over many dynamical times. We perform a series of tests including the time integration of circular and elliptical orbits, the collapse of a Bonnor-Ebert sphere and a rotating, fragmenting cloud core. We compare the collapse of a highly unstable singular isothermal sphere to the theory by Shu (1977), and show that the sink particle accretion rate is in excellent agreement with the theoretical prediction.

To model eccentric orbits and close encounters of sink particles accurately, we show that a very small timestep is often required, for which we implemented subcycling of the $N$-body system. We emphasize that a sole density threshold for sink particle creation is insufficient in supersonic flows, if the density threshold is below the opacity limit. In that case, the density can exceed the threshold in strong shocks that do not necessarily lead to local collapse. Additional checks for bound state, gravitational potential minimum, Jeans instability and converging flows are absolutely necessary for a meaningful creation of sink particles.

We apply our new sink particle module for \textsc{flash} to the formation of a stellar cluster, and compare to a smoothed particle hydrodynamics (\textsc{sph}) code with sink particles. Our comparison shows encouraging agreement of gas properties, indicated by column density distributions and radial profiles, and of sink particle formation times and positions. We find excellent agreement in the number of sink particles formed, and in their accretion and mass distributions.
\end{abstract}


\section{Introduction}
Molecular clouds are turbulent, magnetized, self-gravitating objects. Supersonic turbulence, in particular, plays an important role in shaping the cloud structure, and in controlling star formation, because it creates the seeds for local gravitational collapse \citep{ElmegreenScalo2004,ScaloElmegreen2004,MacLowKlessen2004,McKeeOstriker2007}. Due to the filamentary, fractal structure of the interstellar medium, and due to the large density contrasts, star formation in turbulent molecular clouds proceeds in multiple regions at the same time in parallel, reflecting the hierarchical and fractal nature of the gas density probability distribution \citep[e.g.,][]{Scalo1990,Vazquez1994,ElmegreenFalgarone1996,KlessenHeitschMacLow2000,FederrathKlessenSchmidt2008,GoodmanPinedaSchnee2009,FederrathKlessenSchmidt2009}. Moreover, stars typically form in clusters \citep{LadaLada2003}, showing similar fractal patterns as the gas clouds from which they form \citep[][]{SanchezAlfaro2009}.

To model this complex interplay of turbulence and gravity that eventually leads to cloud fragmentation and to stellar birth with a well-defined initial mass function, it is necessary to follow the freefall collapse of each individual fragment, while keeping track of the global evolution of the entire cloud at the same time. The fundamental numerical difficulty with this approach is that the freefall timescale $\tff$ decreases with increasing density,
\begin{equation} \label{eq:tff}
\tff = \left(\frac{3\pi}{32 G \rho}\right)^{1/2}\,.
\end{equation}
Following the freefall collapse from typical molecular cloud densities up to stellar densities requires the numerical scheme to cover about ten orders of magnitude in timescales. Modeling each individual collapse over such an enormous dynamic range, and following the large-scale evolution over several global freefall times in a single hydrodynamical simulation is beyond the capabilities of modern numerical schemes and supercomputers. Thus, if one wants to model the large-scale evolution of the molecular cloud alongside the collapse of individual regions far beyond the collapse of the first object, the individual runaway collapse must be cut-off in a controlled way, and replaced by a subgrid model.

There are two different subgrid models to tackle this problem in numerical simulations. The first approach is to heat up the gas that exceeds a given density threshold, $\rhores$, which is necessarily related to the achievable numerical resolution. We call this procedure `Jeans heating'. This relative heating of the gas is often modeled by changing the effective equation of state above the density threshold (e.g., by setting the adiabatic exponent to $\gamma>4/3$ for $\rho>\rhores$). Heating up the gas above a density threshold increases the sound speed locally and thus increases the Jeans length,
\begin{equation} \label{eq:lJ}
\lJ = \left(\frac{\pi\cs^2}{G\rho}\right)^{1/2}\;,
\end{equation}
until the gas is stabilized against gravitational collapse. The problem with the Jeans heating approach is that any parcel of gas above a given density threshold is heated artificially, although the actual gas equation of state should still be close to isothermal so long as the density threshold is below the opacity limit. This is the case for density thresholds smaller than about $10^{-14}\,\g\,\cm^{-3}$ \citep[e.g.,][and references therein]{Larson1969,Penston1969,Larson2005,JappsenEtAl2005}. Moreover, gas can become denser than the threshold value in shocks that do not necessarily lead to the formation of a gravitationally bound structure. Thus, shocked gas not going into freefall collapse will be heated up artificially. An additional problem of the Jeans heating approach is that the increasing sound speed in the heated regions can reduce the Courant timestep (see~\S~\ref{sec:timesteps}) to prohibitively small values.

The second type of subgrid model is to use so called `sink particles', a method invented by \citet*{BateBonnellPrice1995} for Smoothed Particle Hydrodynamics (\textsc{sph}), and first adopted for Eulerian, Adaptive Mesh Refinement (\textsc{amr}) by \citet*{KrumholzMcKeeKlein2004}. If the gas has reached a given density, a sink particle is introduced, which can accrete the gas exceeding the threshold, without altering the thermal physics. However, sink particles are supposed to represent bound objects that are/or will be going into freefall collapse, and thus, a density threshold for their creation is insufficient. As for the `Jeans heating', shock compression can temporarily create local densities larger than the threshold \emph{without} triggering gravitational collapse in that region. Previous grid-based implementations of sink particles are mostly based on a density threshold criterion. If the density threshold for sink particle creation is smaller than the opacity limit (about $10^{-14}\,\g\,\cm^{-3}$), we show that spurious sink particles are created in shocks that did not create a gravitationally bound and collapsing structure. Here, we present an implementation of sink particles for the Eulerian, \textsc{amr} code \textsc{flash} that avoids this problem by using a series of checks for gravitational collapse similar to \citet{BateBonnellPrice1995} prior to sink particle creation, such that only gravitationally bound and collapsing gas is turned into sink particles.

We show that the star formation efficiency and the number of fragments is overestimated, if additional, physical checks (e.g., checks for a gravitationally bound and collapsing state) in addition to a density threshold are ignored prior to sink particle formation. We believe that it is also crucial to investigate the resulting mass distribution of the sink particles, because any successful numerical and analytical model of cloud collapse and star formation is expected to account for the observed mass distribution of cores and stars, i.e., the clump, core and stellar initial mass functions \citep[e.g.,][]{KlessenBurkertBate1998,Klessen2001,Kroupa2001,PadoanNordlund2002,Chabrier2003,MacLowKlessen2004,ElmegreenScalo2004,Larson2005,BonnellBate2006,McKeeOstriker2007,KrumholzBonnell2007,HennebelleChabrier2008,HennebelleChabrier2009}. The sink particle implementation presented here enables us to address the star formation efficiency and rate, as well as the mass distribution of fragments obtained in numerical experiments in a robust and quantitative way.

We test our sink particle implementation against a number of standard fragmentation and orbit integration tests. Since stars typically form in dense cluster, close encounters of stars are common, and thus we dedicate significant attention to testing our scheme for its ability to capture close orbits and close encounters. Moreover, in the densest existing star clusters, even merging of stars might be possible \citep{ZinneckerYorke2007}. Our scheme also supports sink particle merging for such extreme cases.

We compare our sink particles with the original sink particle implementation of \citet{BateBonnellPrice1995} in their \textsc{sph} code. It is a standard \textsc{sph} code, and most \textsc{sph} implementations of sink particles are based on their approach \citep[e.g.,][]{JappsenEtAl2005}. We find that \textsc{flash} and \textsc{sph} show encouraging agreement in the obtained mass distribution of sink particles. This code comparison strengthens our confidence in numerical calculations of collapse and fragmentation, using \textsc{amr} on the one hand and \textsc{sph} on the other hand. Here, we primarily introduce the new sink particle module for \textsc{flash} and present a series of initial tests for follow-up studies using sink particles. For instance, there are various \textsc{sph} studies with sink particles of purely hydrodynamic collapse. Our comparison with \textsc{sph} shows that these results are robust. However, one would also like to test the influence of magnetic fields and ambipolar diffusion on star formation as well. The first significant steps toward this were taken in \textsc{sph} simulations of magnetized turbulent clouds recently \citep{PriceBate2008}. However, the numerical representation of tangled and wound-up magnetic fields in \textsc{sph} turbulence simulations is (still) limited \citep{Brandenburg2010,Price2010}. Our grid-based approach of sink particles should allow a refined modeling of collapsing, turbulent, magnetized clouds in follow-up studies, including additional physical processes like ambipolar diffusion and self-consistent jet and outflow formation.

In~\S~\ref{sec:numerics} the new sink particle implementation for the \textsc{amr} code \textsc{flash} is explained in detail. A series of simple and more complex tests of the sink particle implementation is presented in~\S~\ref{sec:tests}. In~\S~\ref{sec:AMRvsSPH} we analyze the formation of a star cluster, and compare column density images, radial profiles, and mass distributions of sink particles obtained with the \textsc{flash} code and with a standard \textsc{sph} code. In~\S~\ref{sec:conclusions}, we summarize our results and discuss the importance of checks for gravitational instability in addition to a density threshold to avoid spurious creation of sink particles, and to avoid overestimating the star formation efficiency and rate.

\section{Numerical Implementation of Sink Particles in the \textsc{flash} Code} \label{sec:numerics}

\subsection{The Basic \textsc{flash} Code} \label{sec:FLASHcode}
The \textsc{flash} code\footnote{\texttt{http://flash.uchicago.edu/website/home/}} \citep{FryxellEtAl2000,DubeyEtAl2008} is an adaptive mesh refinement (\textsc{amr}) code \citep{BergerColella1989}. For purely hydrodynamic studies, it uses the piecewise parabolic method \citep{ColellaWoodward1984} by default to integrate the equations of hydrodynamics. For magnetohydrodynamical (MHD) studies, \textsc{flash} provides an 8-wave Roe solver. In addition, a new approximate Riemann solver for ideal-MHD \citep{BouchutKlingenbergWaagan2007,BouchutKlingenbergWaagan2009}, which preserves positive states in highly supersonic MHD turbulence was recently developed for \textsc{flash} by \citet{Waagan2009}. The corresponding scheme for preserving positive states in purely hydrodynamic studies has been tested successfully in \citet{KlingenbergSchmidtWaagan2007}. Moreover, \citet{DuffinPudritz2008} have recently developed a non-ideal MHD scheme to model ambipolar diffusion. This module is also implemented in the \textsc{flash} code and works (like all other physics modules) within the \textsc{amr} framework. \textsc{flash} is parallelized with \textsc{mpi}, and output files are written in the versatile \textsc{hdf}5 format. The self-gravity of the gas is treated with an iterative multigrid solver \citep[here we use the multigrid solver implemented in \textsc{flash} v2.5, recently refined for v3 by][]{Ricker2008}. Moreover, a tree-based gravity solver was developed for \textsc{flash} (Richard W\"unsch 2009, priv.~comm.), which is currently being modified to run on graphics processing units. Our sink particle implementation is compatible with existing \textsc{flash} modules, i.e., it can be used with the different hydrodynamical and magnetohydrodynamical schemes, including the ambipolar diffusion module, and with either the multigrid or the tree gravity solver. \textsc{flash} has been extensively tested against laboratory experiments \citep{CalderEtAl2002} and other codes \citep{DimonteEtAl2004,HeitmannEtAl2005,AgertzEtAl2007,TaskerEtAl2008,KitsionasEtAl2009}.

For the implementation of the new sink particle module, we made use of the $N$-body capabilities of \textsc{flash}. Once created, sink particles are free to move within the Cartesian computational domain, independent of the underlying grid, i.e., they move in the Lagrangian frame of reference, while the grid points are fixed in space (Eulerian frame of reference). The outer boundary conditions, i.e., {\it outflow}, {\it reflecting}, or {\it periodic} also apply to the sink particles in the simulation box.

\subsection{Sink Particle Creation} \label{sec:creationchecks}
Prior to sink particle creation, it is necessary to perform a number of tests, since we want to avoid creating spurious sink particles in regions that are not undergoing freefall collapse. The basic idea is to first check each computational cell for whether it exceeds a given density threshold $\rhores$. If this is the case, a roughly spherical region with a given radius $\racc$ centered on that cell is temporarily created from the gas. This radius is usually the same as the accretion radius of the sink particle, so we will also call it $\racc$ in the following (see~\S~\ref{sec:accretion} for the implementation of accretion). We denote the region surrounding the cell with $\rho>\rhores$ the control volume $V$. It is defined such that it covers all computational cells with integer indexes $(i,j,k)$, such that
\begin{equation} \label{eq:controlvolume}
V = \sum_{ijk} \Delta V(i,j,k)
\end{equation}
for all $(i\Delta x)^2+(j\Delta y)^2+(k\Delta z)^2 \leq \racc^2$. The central cell of each temporary control volume is at $(i,j,k)=(0,0,0)$, and $\Delta V(i,j,k)=\Delta x\Delta y\Delta z$ is the computational cell volume at spatial position $(i,j,k)$. It is then checked whether the gas in the control volume $V$
\begin{itemize}
\item is on the highest level of refinement,
\item is converging,
\item has a central gravitational potential minimum,
\item is Jeans-unstable,
\item is bound,
\item is not within $\racc$ of an existing sink particle.
\end{itemize}
These checks are similar to the checks introduced in \citet{BateBonnellPrice1995} prior to sink particle formation in \textsc{sph}. Only after these conditions are fulfilled altogether, a sink particle is formed from the gas within the control volume and placed in the center of mass of the gas from which the particle forms. Each of the criteria for sink particle creation is discussed at more detail in the following. The order in which these checks are performed does not matter, however, during code development and tests it became clear that it is useful to do the least computationally expensive checks first to make a preselection of cells prior to the more expensive checks.

\subsubsection{Density Threshold} \label{sec:Truelove}
We introduced sink particles in the \textsc{flash} code to follow collapse calculations for many dynamical times without violating the Truelove criterion for the gas density \citep{TrueloveEtAl1997}. The Truelove criterion states that in order to avoid spurious fragmentation in numerical collapse calculations in grid codes, the Jeans length (eq.~\ref{eq:lJ}) must be resolved with at least four grid cells, $\lJ/\Delta x \geq 4$ \citep{TrueloveEtAl1997}. For \textsc{mhd} calculations, \citet{HeitschMacLowKlessen2001} find that more than four cells are required. It should be emphasized that the resolution criteria by \citet{TrueloveEtAl1997} and \citet{HeitschMacLowKlessen2001} are only meaningful in regions that are undergoing self-gravitational collapse (further discussed in~\S~\ref{sec:comparisonwithotherimplementations}). Usually, adaptive mesh refinement (\textsc{amr}) is used to guarantee that the resolution is always sufficient to satisfy these criteria. However, for collapse calculations involving multiple collapsing regions the pure \textsc{amr} approach only works for the first object going into collapse. This is because freefall collapse is a runaway process in which the first gravitationally bound over-density collapses the fastest. Due to the Courant condition (see \S~\ref{sec:timesteps}), this leads to smaller and smaller timesteps, which stalls the evolution of the entire simulation. Introducing sink particles in regions that are going into freefall collapse on the highest level of refinement provides a way for cutting-off this runaway process in a controlled fashion. However, as discussed in the introduction, it is insufficient to form sink particles solely based on a density threshold in supersonically turbulent gas. Additional checks are necessary.

\subsubsection{Refinement Check} \label{sec:refinement_check}
The Jeans refinement criterion discussed in the previous subsection is also used to resolve the Jeans length of the gas up to the highest level of the \textsc{amr} grid hierarchy. Only when the Jeans refinement reaches the highest \textsc{amr} level, sink particles are allowed to form. Since the accretion radius $\racc$ of a sink particle is coupled to the grid resolution criterion (see~\S~\ref{sec:resolutionFLASH}), sink particles should always be located on the highest level of the \textsc{amr} hierarchy. This is taken care of by an additional refinement criterion for sink particles, which guarantees that any grid cell located inside the accretion radius of any existing sink particle will always be adaptively refined up to the highest \textsc{amr} level.

\subsubsection{Converging Flow Check}
In order to guarantee that the gas supposed to form a sink particle is in freefall collapse, we introduced a check for convergence of the flow toward the center of the control volume, $\nabla\cdot\vect{v} < 0$. Unlike \citet{KrumholzMcKeeKlein2004} we implemented this criterion such that not just the total divergence toward the central cell must be negative, but also that the flows along each of the principal axes must be directed toward the center. The converging flow check alone is insufficient, because $\nabla\cdot\vect{v} < 0$ can also be fulfilled by a localized collision of multiple shocks that do not necessarily produce a gravitationally bound structure. Thus, here we use the converging flow check primarily as a preselection of cells that are considered for sink particle formation, before other, more computationally expensive checks are performed below. Only the combination of the converging flow check with the gravitational potential minimum, bound state and Jeans-instability checks actually guarantees that the gas is in freefall collapse.

\subsubsection{Gravitational Potential Minimum Check}
We guarantee that sink particles can only be created if the central cell of the control volume $V$ defined in equation~(\ref{eq:controlvolume}) is the minimum of the local gravitational potential $\phi$ inside the control volume. The central cell $(i,j,k)=(0,0,0)$ must fulfill the constraint
\begin{equation}
\phi(0,0,0) = \min_{ijk}\left[\phi(i,j,k)\right]
\end{equation}
for sink particle creation.

\subsubsection{Jeans Instability Check}
If the converging flow check and the potential minimum criterion are fulfilled, the control volume is checked for Jeans-instability. The thermal energy $E_\mathrm{th}$ and the gravitational energy $E_\mathrm{grav}$ of the gas in the control volume are calculated as follows:
\begin{eqnarray} 
E_\mathrm{th} & = & \frac{1}{2} \sum_{ijk} M(i,j,k) \cs^2(i,j,k) \label{eq:eth} \\
E_\mathrm{grav} & = & \sum_{ijk} M(i,j,k) \phi(i,j,k) \label{eq:egrav} \,,
\end{eqnarray}
where $\cs(i,j,k)$ is the sound speed, $\phi(i,j,k)$ is the gravitational potential due to the gas mass inside the control volume and
\begin{equation}
M(i,j,k) = \rho(i,j,k)\,\Delta V(i,j,k)\,
\end{equation}
is the mass inside each cell $(i,j,k)$. The relation
\begin{equation} \label{eq:JeansInstability}
\left|E_\mathrm{grav}\right| > 2 E_\mathrm{th}
\end{equation}
must hold for sink particle creation, which means that the gas exceeds the Jeans mass within the control volume. In the case of magnetic fields this criterion is modified such that the magnetic pressure is taken into account as well. The magnetic energy,
\begin{equation} \label{eq:emag}
E_\mathrm{mag} = \frac{1}{8\pi}\,\sum_{ijk} \left|\vect{B}(i,j,k)\right|^2\,\Delta V(i,j,k)\,,
\end{equation}
is computed and added to the right hand side of equation~(\ref{eq:JeansInstability}) to get a modified version of the Jeans criterion, which takes into account the additional pressure provided by magnetic field fluctuations.

\subsubsection{Check for Bound State}
For successful sink particle creation, the total gas energy inside the control volume must be negative,
\begin{equation} \label{eq:bound}
E_\mathrm{grav} + E_\mathrm{th} + E_\mathrm{kin} + E_\mathrm{mag} < 0\,.
\end{equation}
The gravitational, thermal, and magnetic energies are computed from equations~(\ref{eq:egrav}), (\ref{eq:eth}) and~(\ref{eq:emag}). The kinetic energy
\begin{equation}
E_\mathrm{kin} = \frac{1}{2} \sum_{ijk} M(i,j,k) \left|\vect{v}(i,j,k)-\vect{v}_\mathrm{cm}\right|^2
\end{equation}
is determined from the velocity dispersion of the gas, where the center of mass motion, 
\begin{equation}
\vect{v}_\mathrm{cm} = \frac{\sum_{ijk} M(i,j,k)\,\vect{v}(i,j,k)}{\sum_{ijk} M(i,j,k)}
\end{equation}
is subtracted. Only if equation~(\ref{eq:bound}) holds, the gas within the control volume is a bound system.

\subsubsection{Proximity Check}
A new sink particle cannot be created within the accretion radius of an already existing sink particle. Gas that exceeds the density threshold within the accretion radius of an existing particle is accreted by that particle, if the gas passes the accretion checks as explained in the following section.

\subsubsection{Comparison with Other Sink Particle Implementations} \label{sec:comparisonwithotherimplementations}
Except for the refinement and potential minimum checks, all checks used here are analogous to the sink particle implementation by \citet{BateBonnellPrice1995} for \textsc{sph}. Test simulations showed that the potential minimum check prevents almost all spurious sink particles (particles not representing bound and collapsing gas) from forming. However, the Jeans instability and bound state checks are still required in regions of large velocity dispersions, occurring in supersonic, shock-dominated flows. Moreover, magnetic effects are not taken into account by the potential minimum check, but are covered with the Jeans instability and bound state checks \citep[as in][except for our inclusion of the magnetic energy]{BateBonnellPrice1995}.

The Eulerian, \textsc{amr} implementations of sink particles by \citet{KrumholzMcKeeKlein2004} and \citet{WangLiAbelNakamura2010}, and the Eulerian, uniform grid implementation by \citet{PadoanNordlund2009} use the density threshold check, the refinement check \citep[only][]{KrumholzMcKeeKlein2004,WangLiAbelNakamura2010} and the converging flow check (only the total divergence is required to be negative, which is implicitly covered by the density threshold check), but no check for potential minimum, Jeans instability, bound state and proximity of existing sink particles. Thus, a significantly larger number of sink particles is typically created, which requires subsequent merging in \citet{KrumholzMcKeeKlein2004} and \citet{WangLiAbelNakamura2010}. This merging of sink particles is used as an effective accretion in addition to their Bondy-Hoyle accretion model, while \citet{PadoanNordlund2009} use no merging of sink particles, but direct accretion from the grid based on the density threshold.

It is important to note here that our sink particle implementation \citep[as the implementation by][]{BateBonnellPrice1995} allows the density to exceed the density threshold set by the \citet{TrueloveEtAl1997} criterion in some cases. However, this does not mean that our implementation violates the Truelove criterion. By definition, the Truelove criterion only applies to gravitationally bound gas in freefall collapse, for which the density reached the threshold due to gravitational collapse, but not due to purely hydrodynamical compression alone, typically occurring in supersonic turbulent flows. In such supersonic flows, the gas can reach and exceed the density threshold in strong shocks (e.g., in isothermal gas the postshock density increases proportional to the square of the Mach number). Thus, supersonic turbulent density fluctuations can cause the gas to exceed the density threshold. Some of those shocks are dominated by the velocity dispersion and \emph{not} by self-gravity. In such cases it is important to avoid spurious creation of sink particles, because they would not represent gravitationally bound and collapsing objects. Our sink particle implementation avoids spurious creation of sink particles by explicitly testing the gas for local gravitational collapse. It should be noted that the additional checks to the density threshold are only necessary, if sink particles are created at densities below the opacity limit (about $10^{-14}\,\g\,\cm^{-3}$), when the gas is still in the roughly isothermal regime, as for instance in the large-scale simulations of colliding flows by \citet{BanerjeeEtAl2009} and in the turbulent box calculations by \citet{PadoanNordlund2009}, where sink particles are formed at much lower densities to represent clusters of stars rather than individual stars.

\subsection{Gas Accretion} \label{sec:accretion}
As soon as a sink particle was created, it can gain further mass by accreting gas from the grid. The combined gravitational attraction of the sink particle and the gas typically lead to an increase of the gas density over the density threshold $\rhores$ within the accretion radius $\racc$ of an existing particle. If a cell $(i,j,k)$ within $\racc$ exceeds $\rhores$, the mass increment
\begin{equation}
\Delta M = \left[\rho(i,j,k)-\rhores\right] \Delta V(i,j,k)
\end{equation}
is calculated. If $\Delta M$ is bound to the collective mass of the central sink particle and the remaining gas within $\racc$, $\Delta M$ is accreted by that particle. To verify that $\Delta M$ is bound to the particle, the kinetic energy of $\Delta M$ is calculated in the reference frame of the particle and compared to its gravitational binding energy. We furthermore check that $\Delta M$ is moving toward the sink particle, i.e., the radial velocity of $\Delta M$ must be negative. This additional check will also allow us to model mass-loaded protostellar jets from within the control volume in follow-up studies.

If the mass increment $\Delta M$ is located inside a region of overlapping multiple sink accretion radii (which can happen due to particle motion), the gravitational binding energy of $\Delta M$ for each of these particles is calculated, and $\Delta M$ is accreted by the particle to which it is most strongly bound.

Gas exceeding the density threshold $\rhores$ within a given inner accretion radius will always be accreted without further checks for a bound state and convergence. The inner accretion radius, however, is always set such that accretion without further checks only applies to the single central cell in which the particle is located. This is only to avoid numerical problems like division by zero and infinite gravitational energies during the default checks.

If the gas mass $\Delta M$ has successfully passed all the aforementioned tests for accretion, it is accreted by the central particle, such that mass, linear momentum, and angular momentum are conserved (see Appendix~\S~\ref{app:conservationlaws} for a discussion of angular momentum conservation). The accreting particle is moved to the center of mass of the particle--gas configuration before the accretion step.

\subsection{Gravitational Interactions} \label{sec:gravity}
We compute four different contributions to the gravitational interactions between the gas on the grid and the sink particles:
\begin{enumerate}
\item gas--gas (g--g)
\item gas--sinks (g--s)
\item sinks--gas (s--g)
\item sinks--sinks (s--s)
\end{enumerate}
The modeling of these interactions is described in detail in the following sections. In the code, these interactions are computed in the order given above, so we also present them in this order below. Physically, however, the order does not matter.

\subsubsection{Gas--Gas}
The self-gravity of the gas is computed using the standard Poisson solver in \textsc{flash}. It is an iterative multigrid solver that cycles over the \textsc{amr} hierarchy to solve Poisson's equation
\begin{equation} \label{eq:poisson}
\nabla^2\Phi_\mathrm{gas} = 4\pi\,G\,\rho_\mathrm{gas}
\end{equation}
for the gas distribution on the grid. A tree-based gravity solver was developed for \textsc{flash} recently (Richard W\"unsch 2009, priv.~comm.), which can also be used to calculate the gravitational potential instead of the multigrid solver.

The Poisson solver returns the gravitational potential $\Phi_\mathrm{gas}$ in each grid cell, which is due to the whole gas distribution only (without the sink particle contribution). The gravitational acceleration of the gas, $\vect{g}_\mathrm{g-g}$ is then computed for each grid cell as
\begin{equation} \label{eq:gasgas}
\vect{g}_\mathrm{g-g} = -\nabla\Phi_\mathrm{gas}\;.
\end{equation}

\subsubsection{Gas--Sinks} \label{sec:gassinks}
Equation~(\ref{eq:gasgas}) is used to calculate the gravitational acceleration for the sink particles due to the gas component only. The acceleration $\vect{g}_\mathrm{g-g}$ is interpolated from the grid onto the sink particles with a first-order cloud-in-cell method at each position of a sink particle $n$, which yields $\vect{g}_{\mathrm{g-s},\,n}$. A higher-order interpolation scheme like the triangular-shaped cloud method or the tri-cubic interpolation \citep[e.g.,][]{LekienMarsden2005} does not yield significantly different results, because the gravitational acceleration of the gas is relatively smooth, and a linear interpolation scheme is sufficient.

\subsubsection{Sinks--Gas}
Due to their mass the sink particles can exert an appreciable gravitational acceleration onto the gas as well. This acceleration is computed by a direct sum involving all computational cells and all sink particles. For each computational cell center $(i,j,k)$, the distance $\vect{r}_n$ to each particle $n$ is calculated and the acceleration that the particles with masses $M_n$ exert onto the gas in each cell is then computed as
\begin{equation} \label{eq:sinksgas}
\vect{g}_\mathrm{s-g}(i,j,k) = -\sum_n \frac{G M_n}{\left|\vect{r}_n(i,j,k)\right|^3}\,\vect{r}_n(i,j,k)\;.
\end{equation}
Computing the acceleration of the gas due to the sinks thus involves a nested sum over all grid cells and all particles. This can become computationally expensive, because this operation scales as the number of grid cells times the number of particles. However, even for sink particle numbers up to $10^3$, the nested sum hardly affected the overall computational cost, which always remained dominated by the computation of the self-gravity of the gas component (eq.~\ref{eq:poisson}).

For cell centers $(i,j,k)$ very close to sink particles, the acceleration computed via equation~(\ref{eq:sinksgas}) can become very large and goes to infinity when a sink particle is exactly located in the center of a cell. It is thus necessary to use gravitational softening of the acceleration within a softening radius of each sink particle, which is discussed in \S~\ref{sec:softening}.

The acceleration of the gas due to the sink particles, $\vect{g}_\mathrm{s-g}$ is added to the gas--gas acceleration $\vect{g}_\mathrm{g-g}$ to yield the total gravitational acceleration for the gas:
\begin{equation}
\vect{g}_\mathrm{gas}(i,j,k) = \vect{g}_\mathrm{g-g}(i,j,k) + \vect{g}_\mathrm{s-g}(i,j,k)
\end{equation}

\subsubsection{Sinks--Sinks}
The gravitational acceleration for sink particle $n$ due to all other sink particles in the domain is computed as a direct sum involving all other sink particles $m$ with masses $M_m$:
\begin{equation} \label{eq:sinkssinks}
\vect{g}_{\mathrm{s-s},\,n} = -\sum_{m\ne n} \frac{G M_m}{\left|\vect{r}_{nm}\right|^3}\,\vect{r}_{nm}\,,
\end{equation}
where $\vect{r}_{nm}=\vect{r}_m-\vect{r}_n$, is the relative distance vector between two sink particles $n$ and $m$.

The sink--sink acceleration is added to the grid-interpolated acceleration caused by the gas (cf.~\S~\ref{sec:gassinks}) to yield the total gravitational acceleration for sink particle $n$:
\begin{equation} \label{eq:gravSinks}
\vect{g}_{\mathrm{sinks},\,n} = \vect{g}_{\mathrm{g-s},\,n} + \vect{g}_{\mathrm{s-s},\,n}
\end{equation}

Equation~(\ref{eq:sinkssinks}) is also subject to gravitational softening as explained in the next section.

\subsubsection{Gravitational Softening} \label{sec:softening}
The basic problem in computing the gravitational acceleration following equations~(\ref{eq:sinksgas}) and~(\ref{eq:sinkssinks}) is that it can yield extremely large values, if the distance between sink particles and cell centers or between adjacent sink particles becomes small. If the distance goes to zero, the acceleration becomes infinite. Hence, the timestep given by equation~(\ref{eq:timestepGravSinks}) goes to zero and the simulation grinds to a halt. It is therefore necessary to apply gravitational softening for distances smaller than a given softening radius $\rsoft$, such that the acceleration smoothly approaches zero as the distance between particles goes to zero. There are different approaches for gravitational softening. One of the standard softening types is the Plummer softening, $\vect{g}(\vect{r})\!\propto\!(|\vect{r}|^2+\rsoft^2)^{-3/2}\,\vect{r}$. In the case of Plummer softening the acceleration approaches Newton's acceleration for a point mass $g\propto 1/r^2$ in the limit $r\gg\rsoft$. Instead of Plummer softening, we use a type of spline softening that is typically used in \textsc{sph} and $N$-body simulations to soften the gravitational forces \citep[e.g.,][]{PriceMonaghan2007}. The functional form of the cubic spline softening used here is provided in Appendix~\ref{app:softening}.

Figure~\ref{fig:softening} shows a comparison of Plummer softening and spline softening, equation~(\ref{eq:spline}). For Plummer softening, the gravitational acceleration is modified for all distances $r$. In contrast, the spline softening exactly follows the asymptotic solution $g \propto 1/r^2$ for $r\geq\rsoft$ and smoothly approaches zero for $r<\rsoft$.

\subsection{Particle Timestep and Subcycling} \label{sec:timesteps}
Sink particles are evolved using a variable-timestep leapfrog integration scheme. We consider four timestep constraints to guarantee a stable numerical solution:
\begin{eqnarray}
\label{eq:timestepCFL}
\Delta t_\textsc{\scriptsize cfl} &=& C_\textsc{\scriptsize cfl}\,\min_{i,j,k}\left(\frac{\Delta x}{\max(|\vect{v}(i,j,k)|, \cs)}\right) \\
\label{eq:timestepGravGrid}
\Delta t_\mathrm{gg} &=& C_\mathrm{gg}\,\min_{i,j,k}\left(\frac{\Delta x}{|\vect{g}_\mathrm{gas}(i,j,k)|}\right)^{1/2} \\
\label{eq:timestepVelSinks}
\Delta t_\mathrm{vs} &=& C_\mathrm{vs}\,\min_n\left(\frac{\Delta x}{2 |\vect{v}_n|}\right) \\
\label{eq:timestepGravSinks}
\Delta t_\mathrm{gs} &=& C_\mathrm{gs}\,\min_{n,m}\left(\frac{\min(|\vect{r}_{nm}|, \Delta x)}{|\vect{g}_{\mathrm{sinks},\,n}|}\right)^{1/2}\;,
\end{eqnarray}
with $C_\textsc{\scriptsize cfl},\,C_\mathrm{gg},\,C_\mathrm{vs},\,C_\mathrm{gs}<1$, and $\Delta x$ is the smallest linear cell size in the computational domain. The smallest of the first three timestep constraints applies to the hydrodynamic solver to guarantee a stable hydrodynamical evolution of the gas. Equation~(\ref{eq:timestepCFL}) is the Courant-Friedrichs-Lewy condition \citep*{CourantFriedrichsLewy1928}. A modified version of this is used for magnetohydrodynamic studies, which takes into account the fastest possible magnetosonic wave. By solving equation~(\ref{eq:timestepGravGrid}), we also consider the gravitational forces of the self-gravity of the gas, and the gravitational forces of the sink particles exerted onto the gas. Equation~(\ref{eq:timestepVelSinks}) ensures that sink particles cannot cross more than half a grid cell ($\Delta x/2$) within a single timestep when they move with a velocity $\vect{v}_n$. The latter two constraints guarantee that the gravitational forces exerted by the sink particles onto the gas are consistently taken into account for the evolution of the hydrodynamics.

Since the sink particles are always moved within a hydrodynamical timestep, the sink particle timestep can never become larger than the smallest of the hydrodynamical timesteps given by equations~(\ref{eq:timestepCFL}), (\ref{eq:timestepGravGrid}) and~(\ref{eq:timestepVelSinks}). However, these three constraints are insufficient to guarantee a stable and accurate time integration of sink particles, because the sink particle timestep can become significantly smaller than the hydrodynamical timestep for close encounters of sink particles. One additional timestep constraint that captures closely approaching sink particles is needed.

The last timestep constraint given by equation~(\ref{eq:timestepGravSinks}) takes into account the total gravitational acceleration, $|\vect{g}_{\mathrm{sinks},\,n}|$ onto each sink particle (eq.~\ref{eq:gravSinks}), and the minimum distance $|\vect{r}_{nm}|$ between all sink particles $n$ and $m$. In the course of code development and initial tests of the scheme, it turned out that fulfilling equation~(\ref{eq:timestepGravSinks}) is extremely important to avoid inaccurate time integration and artificial acceleration, especially in cases of closely approaching or orbiting sink particles. In test simulations and applications, $\Delta t_\mathrm{gs}$ became up to three orders of magnitude smaller than the timestep given by any of the three constraints of equations~(\ref{eq:timestepCFL}), (\ref{eq:timestepGravGrid}) and~(\ref{eq:timestepVelSinks}). To avoid prohibitive small timesteps for the hydrodynamical evolution, we use subcycling for the time integration of sink particles, while keeping the hydrodynamical gas distribution fixed \citep[similar to][]{KrumholzMcKeeKlein2004}. During subcycling, sink particles cannot change their positions by more than half a grid cell due to equation~(\ref{eq:timestepVelSinks}), and thus they remain almost fixed relative to the grid. However, subcycling is absolutely necessary to guarantee accurate integration of sink particle orbits as shown in a series of $N$-body tests in the next section.

\section{Tests} \label{sec:tests}

In the following sections we describe a series of test simulations to analyze the performance of our sink particle scheme. In particular, we test the accuracy of the time integration of pure particle systems and systems involving gas-sink interactions as well as dynamical sink particle creation and their accretion.

\subsection{$N$-body Tests}

As described in~\S~\ref{sec:gravity} we modified the gravitational interactions of gas and particles compared to the standard \textsc{flash} implementation. Therefore, we test our new approach with the following setups.

\subsubsection{Circular Orbits}
\begin{figure}[t]
\centerline{\includegraphics[width=0.4\linewidth]{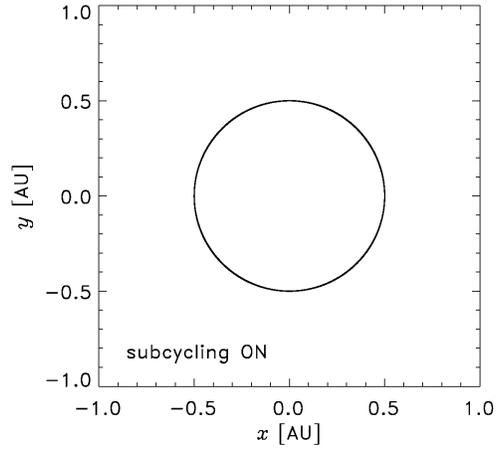}}
\caption{Trajectory of two point masses on circular orbits around their common center of mass after 1000 orbits of time integration. The leapfrog integrator is symplectic for constant timesteps, and thus the circular orbits remain numerically stable.}
\label{fig:circular_orbits}
\end{figure}

In the simplest tests we initialize two sink particles with equal masses of $1\,\Msol$ at a distance of $1\,\mathrm{AU}$, and let them orbit around their common center of mass. There are 8x8x8 grid cells in this test. However, the gas density was set small enough, so that this setup purely tests the $N$-body integration scheme. The results are shown in Figure~\ref{fig:circular_orbits}. The trajectory of the particles are plotted including all positions up to 1000 orbits around their common center of mass at $(x,y,z)=(0,0,0)$. The orbit is maintained circular for these 1000 orbits, and it is not expected to deviate significantly for longer integration times. This is because the leapfrog time-integration scheme is symplectic for constant timesteps. The scheme thus conserves energy and preserves time-reversibility in this test, such that the symmetry of the system is maintained to machine precision.

\subsubsection{Elliptical Orbits}
A more demanding test is the integration of an elliptical orbit. In this case the leapfrog integrator takes variable timesteps depending on the distance and gravitational acceleration according to the timestep criterion, equation~(\ref{eq:timestepGravSinks}). For this test we use the same initial conditions as for the circular orbit test, but rotate the initial velocity vectors by $45^\circ$ from the tangent of the circular orbit of the previous test. The system thus has the same total energy, but the sink particles move on elliptical orbits around their common center of mass. The results are shown in Figure~\ref{fig:elliptical_orbits}. The top panel shows the default time integration with subcycling, which means that equations~(\ref{eq:timestepCFL}), (\ref{eq:timestepGravGrid}), (\ref{eq:timestepVelSinks}) and~(\ref{eq:timestepGravSinks}) are all used in combination to determine the timestep. The bottom panel shows the same, but only taking into account the first three of these timestep constraints. It is important to note that equation~(\ref{eq:timestepVelSinks}) was the only additional timestep constraint for simulations with particles in the \textsc{flash} code before our modifications. Figure~\ref{fig:elliptical_orbits} (bottom panel) shows that using equations~(\ref{eq:timestepCFL}), (\ref{eq:timestepGravGrid}) and~(\ref{eq:timestepVelSinks}) is clearly insufficient, if close orbits are to be accurately reproduced. Even within a single orbit, a significant amount of artificial angular momentum is added to the particles if no subcycling is performed, and after ten orbits they have accumulated an artificial perihel shift of about $21^\circ$. In contrast, ten orbits are accurately followed without a noticeable shift with our present scheme. It should be noted however that even when using subcycling, the symplectic property of the leapfrog integrator is broken, and after roughly 100 elliptical orbits, an error leading to a perihel shift of about $1^\circ$ shows up.

\subsubsection{Gravitational Softening and Subcycling Test}
This setup tests the gravitational softening introduced in~\S~\ref{sec:softening}. We use the same computational domain as in the two previous tests with two $1\,\Msol$ sink particles separated by $1\,\mathrm{AU}$. One particle is initially located at $(x,y,z)=(-1.0,+0.5,\,0.0)\,\mathrm{AU}$, and the other one is at $(-1.0,-0.5,\,0.0)\,\mathrm{AU}$. Both particles have zero initial velocities in the $y$-direction, such that they accelerate toward each other by their mutual gravitational attraction. While approaching each other, the timestep drops according to the constraints given by equations~(\ref{eq:timestepVelSinks}) and~(\ref{eq:timestepGravSinks}). Due to the softening, however, the timestep does not go to zero as the particle distance goes to zero (cf.~Fig.~\ref{fig:softening}). This allows the two particles to pass through each other smoothly. By the time they pass through the point of zero distance, they have converted their initial potential energy completely into kinetic energy. The softening affects the specific value of this energy but nevertheless, the energy must be conserved. Thus, after one passage both particles have exactly exchanged their initial positions. This process then starts again and the system oscillates for an infinite time. In order to make these oscillations visible, we apply an initial constant velocity in $x$-direction to both particles. The result is shown in Figure~\ref{fig:oscis}. The top panel shows both particle trajectories after 1000 oscillations. Using subcycling, the system easily conserves energy during these 1000 oscillations. In contrast, if the timestep constraint given by equation~(\ref{eq:timestepGravSinks}) is ignored, energy conservation is broken already during the first oscillation as shown in Figure~\ref{fig:oscis} (bottom panel). Both particles gain energy artificially until they leave the domain. In the current test, this happens after about 68 oscillations. Scattering experiments have shown that violating the timestep condition given by equation~(\ref{eq:timestepGravSinks}), i.e., \emph{not} using subcycling can lead to enormous artificial accelerations already during a single close encounter of two particles.

\subsection{Gas--Sinks Gravity and Refinement Test}
The following setup tests the accuracy of the time integration of sink particles orbiting in the gravitational potential of a static gas distribution. We create a singular isothermal sphere with the following density profile on the \textsc{amr} grid:
\begin{equation} \label{eq:sis_density_profile}
\rho(r) = \rho(r_0)\,\left(\frac{r_0}{r}\right)^2\,,
\end{equation}
where $r_0=5\times10^{16}\,\cm$ and $\rho(r_0)=3.82\times10^{-18}\,\g\,\cm^{-3}$. The sphere thus has a mass of roughly $3\,\Msol$ for $r \leq r_0$. We place three sink particles at radii of $r=1,\,2,\,3\times10^{16}\,\cm$ with masses of $10^{-10}\,\Msol$. The particles thus represent test particles in a gravitational potential caused by the gas only, such that sink particle interaction is negligible. Since the singular isothermal gas distribution would immediately collapse and form a new sink particle in the center, we deliberately switch off the hydrodynamical evolution, and keep the density distribution and associated potential artificially static. However, we keep the \textsc{amr} scheme and the Poisson solver fully operational, and adaptively refine on Jeans length and sink particles as discussed in~\S~\ref{sec:refinement_check}. We use a base grid of $32^3$ grid cells plus two levels of refinement with isolated boundary conditions for the gravity. The Jeans length is resolved with 6 grid cells for this test, while de-refinement is triggered as soon as the Jeans length is resolved with more than 12 grid cells. The initial density and sink particle distributions are shown in Figure~\ref{fig:orbit_sis_amr} (left).

The three sink particles were given an initial velocity in the $y$-direction, such that they should remain on circular orbits around the center of the gas sphere. It is easy to determine this velocity analytically from the density distribution, equation~(\ref{eq:sis_density_profile}). The result is a Keplerian velocity of $v_\mathrm{K}=\sqrt{GM(r)/r}=0.895\,\mathrm{km\,s^{-1}}$ for each sink particle, independent of the radius. However, since the sphere is not singular in the center due to the finite resolution of the grid, the actual velocities necessary to keep the sink particles on circular orbits are slightly smaller than the analytical solution (by roughly 1-2\%).

Figure~\ref{fig:orbit_sis_amr} (right) shows the \textsc{amr} hierarchy and the sink particle positions after the innermost particle has finished roughly one orbit. At this time, the sink particle on the intermediate orbit with $r=2\times10^{16}\,\cm$ has finished almost half an orbit, and the outer sink particle has completed one third of an orbit. The grid is refined according to the position of the sink particles. The Jeans refinement criterion keeps the central part at the highest level of refinement and the outer parts at a resolution such that the Jeans length is refined with at least 6 grid cells. De-refinement to the base grid resolution was not triggered in the outer parts, because the Jeans length was resolved with 6-12 grid cells there.

We evolved this configuration for 100 orbits of the innermost sink particle. The trajectories of all three sink particles are shown in Figure~\ref{fig:orbit_sis} after 1, 10, and 100 orbits of the innermost sink particle (from left to right). After 100 orbits of the innermost sink particle, the intermediate particle has finished 50 orbits, while the outermost sink particle has finished about 33.3 orbits. The latter orbit as well as to a smaller degree also the intermediate orbit are slightly broadened, and they are not as accurately reproduced as the innermost orbit. This is because the Poisson solver introduces slight deviations from the spherical gravitational potential mainly in regions close to the boundaries of the Cartesian domain at the coarse grid resolution used in this test. For such a configuration and setup that incorporates the multigrid solver and the full \textsc{amr} scheme, it is hard to maintain perfect spherical symmetry for more than 10-100 sink particle orbits.

\subsection{Collapse of a Bonnor-Ebert Sphere}
The collapse of supercritical Bonnor-Ebert spheres~\citep[][]{Ebert1955,Bonnor1956}, is a well-studied problem~\citep[e.g.][]{Larson1969,FosterChevalier1993,BanerjeePudritzHolmes2004,AikawaEtAl2005}, and thus represents a good test case for our sink particle implementation. A Bonnor-Ebert sphere can be described by the following dimensionless parameters for the radius, time, mass, and mass accretion, respectively:
\begin{eqnarray}
\xi      & = & \frac{r}{\cs/\sqrt{4\pi\, G\, \rho_0}} \\
\tau     & = & \frac{t}{1/\sqrt{4\pi\, G\, \rho_0}} \\
m        & = & \frac{M}{\cs^3/\sqrt{4\pi\, G^3\, \rho_0}} \\
\dot{m}  & = & \frac{\dot{M}}{\cs^3/G}\;,
\end{eqnarray}
where $\cs$ is the sound speed and $\rho_0$ is the central density of the sphere. In the hydrostatic configuration of such a sphere, the dimensionless mass can be written as $m =\xi^2\, \phi^{\prime}$, where $\phi^{\prime} = -\di\ln{(\rho/\rho_0)}/\di \xi$ is the gravitational acceleration. The sphere becomes supercritical if its dimensionless radius exceeds the critical value $\xi_\mathrm{crit} = 6.451$. To trigger the collapse, we set the initial value to $\xi_\mathrm{max} = 7.0$, which gives a total mass, $m = 17.3$ in dimensionless units. We follow the collapse with two different density thresholds for sink particle creation and accretion: $\rhores=10^2\,\rho_0$ and $\rhores=10^4\,\rho_0$. These two density thresholds correspond to accretion radii of $\xi=0.1$ and $\xi=10^{-2}$, respectively.

\begin{figure}[t]
\centerline{\includegraphics[width=0.4\linewidth]{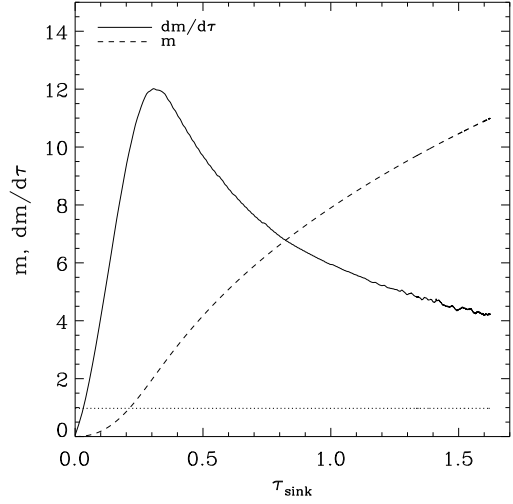}}
\caption{Time evolution of the mass accretion rate, $\di m /\di \tau$, and of the accreted mass, $m$, of the central sink particle in the case of a collapsing Bonnor-Ebert sphere. The quantities are given in dimensionless units (see text). The mass accretion rate increases rapidly until about 12\% of the gas mass is accreted onto the sink particle (the total mass of the Bonnor-Ebert sphere is $m=17.3$). Here, the density threshold for sink particle accretion was set to $10^2\,\rho_0$, which corresponds to an effective accretion radius of $\xi=0.1$. The time $\tau_\mathrm{sink} = 0$ corresponds to the time when the sink particle starts to accrete.}
\label{fig:BE_mdot_m_lowres}
\end{figure}

Figure~\ref{fig:BE_mdot_m_lowres} shows the mass accretion history of the sink particle in the case of the low density threshold, $\rhores=10^2\,\rho_0$. Initially, when the sink particle starts to accrete at $\tau_\mathrm{sink}=0$, the accretion rate increases quickly until it reaches a maximum. The accretion rate peaks at $\tau_\mathrm{sink}\approx0.3$ when about 12\% of the total mass of the sphere is accreted onto the sink particle. After that, the accretion rate decreases while the envelope runs out of gas. At $\tau_\mathrm{sink}=1.62$ about 64\% of the total gas mass has been accreted onto the sink particle.

\begin{figure}[t]
\centerline{\includegraphics[width=0.4\linewidth]{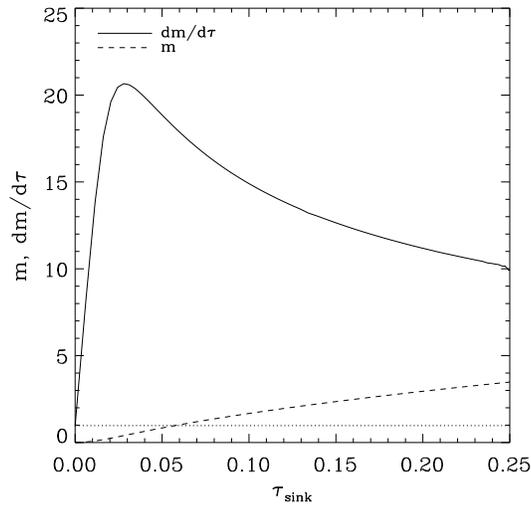}}
\caption{Same as Fig.~\ref{fig:BE_mdot_m_lowres} but in the case of $\rhores=10^4\,\rho_0$, corresponding to an effective accretion radius of $\xi=10^{-2}$, ten times smaller than in the previous test. The peak accretion rate is thus reached ten times faster than in Fig.~\ref{fig:BE_mdot_m_lowres}.}
\label{fig:BE_mdot_m_highres}
\end{figure}

The time evolution of the accretion rate is similar in the case of a higher density threshold ($\rhores=10^4\,\rho_0$) for the sink particle accretion. Figure~\ref{fig:BE_mdot_m_highres} shows the mass evolution for this case. Here, the effective accretion radius is smaller by a factor of ten compared to the $\rhores=10^2\,\rho_0$ case, and thus the peak accretion rate is reached ten times faster.

The time evolution of the mass accretion is in qualitative agreement with 1D simulations by \citet[e.g.,][]{FosterChevalier1993}, and with various analytical \citep[e.g.,][]{Hunter1977,WhitworthSummers1985} and numerical models of protostellar collapse, which predict a peak in the accretion followed by a smoothly declining accretion rate \citep[see also][and references therein]{SchmejaKlessen2004}.

\subsection{Collapse of a Singular Isothermal Sphere}
The collapse of a singular isothermal sphere with a $\rho(r)\propto r^{-2}$ density profile produces a constant flux of mass through spherical shells, i.e.~a constant accretion rate \citep[see e.g.][]{Shu1977}. We compute the collapse of a singular isothermal sphere to test whether our model of sink particle accretion reflects this collapse behavior. The sphere has a truncation radius of $R=5\times 10^{16}\,\cm$, a density at this radius of $\rho(R)=3.82\times 10^{-18}\,\g\,\cm^{-3}$, and therefore a total mass of $3.02\,\Msol$. The sphere is at a temperature of $10\,\K$ corresponding to a sound speed of $0.166\,\km\,\mathrm{s}^{-1}$. With these values the sphere has a large instability parameter of $A=29.3$, where $A=4\pi G\,\rho(R)\,R^2/\cs^2$ \citep[][]{Shu1977}.

\begin{figure}[t]
\centerline{\includegraphics[width=0.5\linewidth]{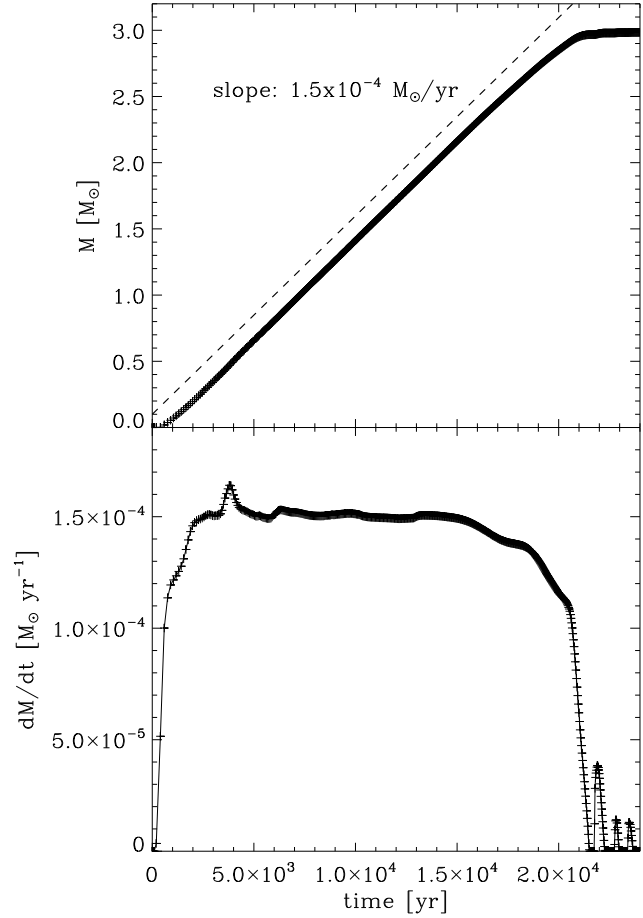}}
\caption{Mass accretion of the central sink particle during the collapse of a singular isothermal gas sphere. The mass accretion rate onto the sink particle is roughly constant in time with a value of $\dot{M}\approx 1.5\times 10^{-4}\Msol\,\yr^{-1}$, until the sphere runs out of gas at $t\approx 2\times 10^4\,\yr$. The collapse is highly supersonic due to the large instability parameter $A=4\pi G\,\rho(R)\,R^2/\cs^2=29.3$. Therefore, the mass accretion rate reaches values much larger than the expansion-wave collapse solution by \citet{Shu1977} that would give $m_0\cs^3/G = 1.06 \times 10^{-6}\,\Msol\,\yr^{-1}$ for the asymptotic solution: $A=2$ and $m_0=0.975$ \citep[see][Tab.~1]{Shu1977}. For our case, $A=29.3$ and $m_0\approx133$, \citet{Shu1977} predicts an accretion rate of $\dot{M}\approx 1.45\times 10^{-4}\,\Msol\,\yr^{-1}$, which is in excellent agreement with our numerical estimate of the central sink particle accretion rate.}
\label{fig:sis_mass_accretion}
\end{figure}

Figure~\ref{fig:sis_mass_accretion} shows the mass and the mass accretion rate onto the sink particle during the collapse of the singular isothermal sphere. As the gas of the sphere is initially at rest, the mass accretion reaches a close to constant value of about $1.5\times 10^{-4}\,\Msol\,\yr^{-1}$, until the entire gas of the sphere is accreted onto the central sink particle at $t\!\sim\!2\times10^4\,\yr$.

For comparison of the accretion history of the sink particle, we show the time evolution of radial profiles of the gas sphere in Figure~\ref{fig:sis_radius}. As expected, the density profile of the sphere evolves from $r^{-2}$ to $r^{-1.5}$ \citep[see e.g.,][]{Shu1977,OginoTomisakaNakamura1999}. The collapse proceeds highly supersonically ($\Ma > 15$) with increasing infall velocities toward the collapse center (Fig.~\ref{fig:sis_radius}, middle panel). The mass accretion, i.e. the mass flux through spherical shells of radius $r$ is constant in time as well as independent of the radial position with a value of $\dot{M}\approx 1.5\times 10^{-4}\,\Msol\,\yr^{-1}$. It is the exact same amount of gas that flows through these spherical shells into the control volume of the sink particle, where it is finally accreted. A comparison of the sink particle accretion rate in Figure~\ref{fig:sis_mass_accretion} with the mass flux through spherical shells in Figure~\ref{fig:sis_radius} (bottom panel) shows the high accuracy of the sink particle accretion mechanism.

For comparison, we calculated the mass accretion rate predicted by \citet{Shu1977}. The constant accretion rate often referred to in many studies citing \citet{Shu1977} is 
\begin{equation} \label{eq:shu}
\dot{M}=m_0\,\frac{\cs^3}{G}
\end{equation}
with $m_0=0.975$, which is the appropriate value for an instability parameter close to $A=2$ \citep[see][Tab.~1]{Shu1977}. The accretion rate depends on the coefficient $m_0$, which is controlled by $A$. Numerical integration of the dimensionless, one-dimensional equations of hydrodynamics gives the solutions of $m_0$ for given instability parameters $A$ as in \citet[see][Tab.~1]{Shu1977}. There, Shu only tabulated values up to $A=4$. Since our singular isothermal sphere has a much larger instability parameter ($A=29.3$), we repeated the numerical analysis by \citet{Shu1977} to find solutions for instability parameters up to $A=50$. The result for the coefficient $m_0$ of the Shu accretion rate in equation~(\ref{eq:shu}) is plotted as a function of the instability parameter $A$ in Figure~\ref{fig:shu1977}. The parameter $m_0$ increases from $m_0=0.975$ for $A=2$ to $m_0=279$ for $A=50$, more than two orders of magnitude higher! For the instability parameter $A=29.3$ of the singular isothermal sphere studied here, we obtain $m_0=133$. The accretion rate predicted by \citet{Shu1977} is thus $\dot{M}\approx 1.45\times 10^{-4}\,\Msol\,\yr^{-1}$, which is in excellent agreement with our numerical estimate of the sink particle accretion rate (cf.~Fig.~\ref{fig:sis_mass_accretion}).

\subsection{Rotating Cloud Core Fragmentation Test} \label{sec:bb}
We now analyze the collapse and fragmentation of a rotating cloud core, also known as the \citet{BossBodenheimer1979} test. It is a standard test for fragmentation in hydrodynamical codes \citep[e.g.,][]{BodenheimerBoss1981,BurkertBateBodenheimer1997,TrueloveEtAl1997,CommerconEtAl2008}. In particular we use a setup similar to \citet{BurkertBodenheimer1993} and \citet{BateBurkert1997}. The initial parameters for the rotating cloud core are: radius $R=5\times10^{16}\,\cm$, constant density $\rho_0=3.82\times10^{-18}\,\g\,\cm^{-3}$, mass $M=1\,\Msol$, angular velocity $\Omega=7.2\times10^{-13}\,\rad\,\mathrm{s}^{-1}$ (ratio of rotational to gravitational energy $\beta=0.16$), sound speed $\cs=0.166\,\km\,\mathrm{s}^{-1}$ (ratio of thermal to gravitational energy $\alpha=0.26$), global freefall time $t_\mathrm{ff}=1.075\times 10^{12}\,\mathrm{s}=3.41\times10^4\,\yr$, and a 10\% density perturbation with an $m=2$ mode: $\rho=\rho_0[1+0.1\,\cos(2\varphi)]$, where $\varphi$ is the azimuthal angle. The following polytropic equation of state,
\begin{equation}
P=\cs^2\rho^\Gamma\;,
\end{equation}
was used with the polytropic exponent
\begin{equation}
\def\arraystretch{1.5}
\Gamma = \left\{
   \begin{array}{ll}
      1   & \,\mbox{for}\;\, \rho/(10^{-15}\,\g\,\cm^{-3}) \leq 0.25\,, \\
      1.1 & \,\mbox{for}\;\, 0.25 < \rho/(10^{-15}\,\g\,\cm^{-3}) \leq 5.0\,, \\
      4/3 & \,\mbox{for}\;\, \rho/(10^{-15}\,\g\,\cm^{-3}) > 5.0\,.
   \end{array} \right.
\end{equation}
The initial rotation ($\beta=0.16$) forces the gas sphere to collapse to a disk with a central bar due to the $m=2$ initial density perturbation.
Sink particles are allowed to form at densities exceeding $10^{-14}\,\g\,\cm^{-3}$. The accretion radius was set to $\racc=39\,\AU$ corresponding to 2.5 grid cells at the highest level of refinement.

Figure~\ref{fig:coldens_bb} shows a face-on column density projection of the disk at different times along the collapse of the rotating cloud core. Two sink particles form at the location of the two fragments at $t=1.26\,\tff$. At $t=1.29\,\tff$ a bar forms that connects the two main fragments. A third fragment forms in the center of the cloud core. After that, the two main fragments move toward the central object due to the global collapse of the cloud core. At $t=1.33\,\tff$ two of the three fragments merge to a single particle close to the center of the rotating cloud core. Note that sink particle merging was used in this test. Sink particle merging is optional in our implementation in \textsc{flash}, and can be activated if desired. If activated, sink particles are only allowed to merge, if they are inside the accretion radii of one another, and if they are gravitationally bound and converging. The merged particle is moved to the center of mass of the merging particles, and their linear and angular momenta are assigned to the merged particle.

The conservation laws during accretion and merging of sink particles are discussed in Appendix~\ref{app:conservationlaws}. The initial total angular momentum of the rotating cloud core computed from the analytic solution is $L_0=1.44\times10^{54}\,\g\,\cm^2\,\mathrm{s}^{-1}$. The numerical solution conserves angular momentum to within 2\% for all times including the last snapshot shown in Figure~\ref{fig:coldens_bb} when 23\% of the mass has been accreted onto sink particles. This is comparable to the typical conservation of angular momentum achieved in numerical simulations of rotating self-gravitating cloud cores \citep[e.g.,][]{CommerconEtAl2008}.

\section{Star Cluster Formation: \textsc{amr} versus \textsc{sph}} \label{sec:AMRvsSPH}
To check our sink particle implementation for the \textsc{amr} code \textsc{flash} on a more complex physical problem, we apply it here to the formation of a star cluster. We furthermore want to compare our sink particles against an existing sink particle implementation in a typical \textsc{sph} code. We compare to the \textsc{sph} code developed by \citet{BateBonnellPrice1995}. \citet{BateBonnellPrice1995} were the first to use sink particles in an \textsc{sph} code, and thus, most \textsc{sph} implementations for sink particles are based on their approach \citep[e.g.,][]{JappsenEtAl2005}, as is the grid-based implementation presented here. A detailed description of the \textsc{flash} code and the our sink particle implementation was given in~\S~\ref{sec:FLASHcode}. In the following we briefly describe the \textsc{sph} code used for our sink particle comparison.

\subsection{The \textsc{sph} Code}
Smoothed particle hydrodynamics \citep[\textsc{sph},][]{Lucy1977,GingoldMonaghan1977} calculations presented in this study were performed using a code based on the version developed by Benz \citep{Benz1988,Benz1990,BenzEtAl1990}, which has since been modified by Bate to include individual particle timesteps \citep{NavarroWhite1993,HernquistKatz1989} and sink particles \citep{BateBonnellPrice1995}. To capture shocks, the code uses the standard artificial viscosity suggested by \citet{GingoldMonaghan1983} and \citet{MonaghanGingold1983}, with $\alpha = 1$ and $\beta = 2$. To provide adaptive resolution, the code allows the \textsc{sph} smoothing lengths ($h$) to vary in time and space, with the constraint that each particle maintains a roughly constant number of neighbors ($50\pm10$) within a distance $2h$. Gravitational forces are found using a binary tree \citep[][and references therein]{Press1987,Benz1988}, which is also used to obtain the neighbor lists required by the \textsc{sph} algorithm. The binary tree opening angle was set to 0.47 in this study, and so returns a more accurate calculation of the gravitational forces than the theoretical limit of 0.57 for 3D binary trees described by \citet{Press1987}. Gravitational forces are softened using the spline softening technique described in Appendix~\ref{app:softening}. In the \textsc{sph} code this softening is used between all particle types. Between the gas-gas interactions, the softening length $\rsoft$ is given by the mean smoothing length of the particle pair, and so the softening is zero for particles which are not neighbors, since the distance between them is greater than $2h$ by definition. This form of the softening assumes that the standard \textsc{sph} smoothing kernel \citep{MonaghanLattanzio1985} describes the radial density profile of the particles, and has been shown to reduce the effects of artificial fragmentation \citep{BateBurkert1997,Whitworth1998,HubberGoodwinWhitworth2006}. For the sink particles, one is free to chose the value of $\rsoft$ and in our current study this is set to the sink particle accretion radius, $\racc$, as for \textsc{flash}. The value of $\rsoft$ between gas-sink interactions is then set to $(\racc + h)/2$ for each pair-wise force. Finally, the \textsc{sph} equations and sink particle trajectories are integrated using a second-order Runge-Kutta-Fehlberg integrator. A full description of the timestepping constraints can be found in \citet{BateBonnellPrice1995}.

\subsection{Initial Conditions for the Code Comparison} \label{sec:initialconditions}
An isolated gas sphere of radius $R=5.0\times10^{17}\,\cm=0.16\,\pc$ with a uniform density of $\rho_0=3.85\times10^{-19}\,\g\,\cm^{-3}$ containing $M=100\,\Msol$ was initialized in both the \textsc{flash} and the \textsc{sph} code. The initial column density structure is shown in Figure~\ref{fig:coldens_init} (top panels) for the \textsc{flash} run (left panels) and the \textsc{sph} run (right panels). An initial random, divergence-free turbulent velocity field was generated on a grid with $128^3$ grid cells with a velocity power spectrum $P(k)\propto k^{-4}$, consistent with the observed velocity dispersion-size relation in molecular clouds \citep[e.g.,][]{Larson1981, HeyerBrunt2004}, and consistent with the velocity spectra usually obtained for driven supersonic turbulence \citep[e.g.,][]{FederrathKlessenSchmidt2009,SchmidtEtAl2009,FederrathDuvalKlessenSchmidtMacLow2009}. However, in the present study we are not driving the turbulence, i.e., we study decaying turbulence. The velocity dispersion was scaled to $0.89\,\km\,\mathrm{s}^{-1}$, consistent with the observed velocity dispersions in molecular clouds of the size studied here \citep[e.g.,][]{Larson1981,FalgaronePugetPerault1992}. With the isothermal sound speed of $\cs=0.19\,\km\,\mathrm{s}^{-1}$ (temperature $T=10\,\K$, mean molecular weight $\mu=2.3$), this velocity dispersion corresponds to an initial \textsc{rms} Mach number of about 4.5. Thus, the ratio of kinetic to gravitational energy of the gas sphere is $E_\mathrm{kin} / E_\mathrm{grav} \approx 0.25$. The equation of state is isothermal, and self-gravity is used throughout this code comparison. Sink particles are allowed to form in both codes at densities exceeding $\rhores=8.0\times10^{-17}\,\g\,\cm^{-3}$ with a sink particle accretion radius of $\racc=7.3\times10^{15}\,\cm=490\,\AU$. Both codes used the same spline softening (cf.~Appendix~\ref{app:softening}) for sink particle interactions, with the softening radius set equal to the accretion radius ($\rsoft=\racc$). The temporal evolution is followed in units of the initial freefall time $\tff=3.39\times10^{12}\,\mathrm{s}=1.07\times10^5\,\yr$.

For the \textsc{flash} run, we used a cubic computational domain with a boxsize of $0.4\,\pc$, slightly larger than the diameter of the initial gas sphere ($d=0.32\,\pc$). To keep the mass outside of the sphere small, the gas density was set two orders of magnitude smaller outside than the uniform density inside the sphere. The additional mass of $2.7\,\Msol$ outside the sphere is about 3\% of the total mass, and thus the total mass in the computational domain is 3\% larger in the \textsc{flash} run than in the \textsc{sph} run. Pressure equilibrium is provided by keeping the gas outside the sphere at a temperature two orders of magnitude higher than inside.

For the \textsc{sph} run the initial velocity field was interpolated from the grid to the \textsc{sph} particles using the standard \textsc{sph} smoothing kernel \citep{MonaghanLattanzio1985}. The initial density outside the isolated gas sphere is zero for the \textsc{sph} run since no particles were used there. However, a constant boundary pressure was applied in order to keep the sphere in pressure equilibrium with the surrounding. Both codes used isolated boundary conditions for the computation of the gravitational potential.

The initial conditions used in our comparison provide a more rigorous test of the performance of our codes than those commonly used in the study of cluster formation. Typically, initial conditions are set up with roughly equal kinetic and gravitational energies, $E_\mathrm{kin}\approx E_\mathrm{grav}$ \citep[e.g.,][]{BateBonnellBromm2003,BonnellBateVine2003}. These clouds tend to evolve to form several centers of star formation, each producing a small group of fragments that feed from the gas delivered by the large-scale flows in the cloud. Clouds with less kinetic energy, in contrast, have less support against global contraction, and so undergo global collapse. As a result, such clouds tend to produce a single, dense star-forming center that also has a higher rate of star formation \citep[e.g.,][]{ClarkBonnellKlessen2008}. As a consequence, the cloud modeled here enters a more violent and chaotic $N$-body phase, in which it is more difficult to obtain convergence for different numerical methods.

\subsubsection{Resolution Criterion in \textsc{flash}} \label{sec:resolutionFLASH}
In order to satisfy the resolution criterion in simulations including self-gravity, the gas density on the grid must not exceed a critical density $\rhores$ in regions of gravitational collapse \citep{TrueloveEtAl1997}. This density is related to the smallest resolvable Jeans length $\lJ$ on the highest level of the \textsc{amr} hierarchy. The density threshold $\rhores$ is obtained by solving equation~(\ref{eq:lJ}) for $\rho$, and using the Jeans length and sound speed of the gas resolved with \textsc{amr} on the highest level of refinement:
\begin{equation}
\rhores = \frac{\pi\cs^2}{G \lJ^2} = \frac{\pi\cs^2}{4\,G \racc^2}\;.
\end{equation}
Since the Jeans length should be resolved with at least 4 grid cells in \textsc{amr} simulations \citep{TrueloveEtAl1997}, the sink particle accretion radius $\racc$ must not be smaller than 2 grid cells. The sink particle accretion radius is thus determined by the smallest linear cell size $\Delta x$ that can be resolved with pure \textsc{amr}. We set $\racc\simeq2.5\Delta x$ to satisfy the Truelove criterion on the highest level of refinement. The Jeans length $\lJ=2\racc$ is thus resolved with 5 grid cells on the top of the \textsc{amr} hierarchy. However, on all \textsc{amr} levels smaller than the maximum level of refinement we resolve the Jeans length with at least 12 grid cells to follow features on their way to runaway collapse more accurately. We additionally apply the shock refinement criterion provided in \textsc{flash} \citep{FryxellEtAl2000}, and our additional refinement on sink particles (cf.~\S~\ref{sec:refinement_check}).

For the comparison test presented here, we used a fixed base grid with $128^3$ grid cells plus two levels of \textsc{amr} resulting in an effective resolution of $512^3$ grid cells.

\subsubsection{Resolution Criterion in \textsc{sph}} \label{sec:resolutionSPH}
\citet{BateBurkert1997} have shown that at least two times the number of \textsc{sph} neighbors, $2 N_\mathrm{neigh}$ particles are necessary for resolving the minimum Jeans mass to avoid artificial fragmentation in \textsc{sph} codes. The Jeans mass is defined as
\begin{equation} \label{eq:MJ}
\MJ(\rho) = \frac{4\pi}{3}\left(\frac{\lJ(\rho)}{2}\right)^3\rho=\frac{1}{6}\pi^{5/2}\left(\frac{k_\mathrm{B}T}{G\mu m_\mathrm{H}}\right)^{3/2}\rho^{-1/2}\,,
\end{equation}
where $k_\mathrm{B}$, $\mu$ and $m_\mathrm{H}$ are the Boltzmann constant, the mean molecular weight and the mass of a hydrogen atom, respectively. Assuming a number of \textsc{sph} neighbors of roughly $N_\mathrm{neigh}\simeq50$, we can estimate the mass resolution limit for an \textsc{sph} calculation as $M_\mathrm{res} \simeq \MJ(\rhores)/100$. The corresponding sink particle accretion radius is $\racc=\lJ(\rhores)/2$.

Since we resolve the Jeans length with 5 grid cells in the \textsc{flash} run, we chose to use a similar number of \textsc{sph} smoothing lengths to resolve the Jeans length in the \textsc{sph} run. Therefore, we used a total number of 2 million \textsc{sph} particles for our comparison, which roughly corresponds to resolving the Jeans length with 5 \textsc{sph} smoothing lengths. Furthermore, the initial velocity field was constructed on a grid with $128^3\sim2\times10^6$ grid cells. Thus, to accurately sample the initial velocity field with \textsc{sph} particles, it was reasonable to use a similar number of resolution elements in the \textsc{sph} code as grid cells for the initial velocity field.

\subsection{Results of the Sink Particle Code Comparison}

\subsubsection{Column Density Distributions}
Figure~\ref{fig:coldens_init} shows the column density distributions of the \textsc{flash} (left) and the \textsc{sph} run (right) at $t=0.0\,\tff$ (top) and $t=0.5\,\tff$ (bottom). The initial supersonic turbulent velocity field has generated a complex network of shocks at $t=0.5\,\tff$. Some of the shock-compressed gas becomes gravitationally unstable and goes into freefall collapse. This happens at roughly $t=0.8\,\tff$ as shown in the top panels of Figure~\ref{fig:coldens} (left and middle columns for \textsc{flash} and \textsc{sph}, respectively). Five sink particles have formed in both the \textsc{flash} and the \textsc{sph} run, containing a total mass of $0.9\,\Msol$ and $0.7\,\,\Msol$, respectively. They form at roughly the same locations in the gas distribution with slight differences in the exact positions. These differences are mainly due to the slightly different hydrodynamic evolution of the gas prior to sink particle formation in the \textsc{flash} and \textsc{sph} runs. There is striking agreement between the \textsc{flash} and \textsc{sph} runs at $t=0.9\,\tff$, indicated by the number of sink particles (15 versus 17) and their accreted mass ($4.9\,\Msol$).

At $t=0.9\,\tff$, the gas distribution has already started to collapse globally due to the decay of the initial turbulence. At $t=1.0\,\tff$, 32 sink particles containing a total of $14.5\,\Msol$ have formed in the \textsc{flash} run, while the \textsc{sph} run has produced 48 sink particles with $20.0\,\Msol$. Thus, the \textsc{flash} and \textsc{sph} runs start to diverge at $t=1.0\,\tff$. The star cluster produced in the \textsc{sph} run is slightly denser than in the \textsc{flash} run. This trend continues to the last snapshot at $t=1.1\,\tff$ when the \textsc{flash} run has produced 49 sink particles with $26.3\,\Msol$, while the \textsc{sph} run has produced 61 with $36.7\,\Msol$. The star cluster is clearly denser in the \textsc{sph} run at this last time. There are still new sink particles forming in the outskirts of the cluster. They from at very similar locations in both the \textsc{flash} and the \textsc{sph} runs. For example, at $t=1.1\,\tff$ the rightmost sink particle has formed at almost exactly the same location in both codes with respect to the position of the dense core forming there. However, this particular sink particle also forms slightly closer to the center of the star cluster, reflecting the faster global collapse in the \textsc{sph} run. This is discussed further in the following section.

\subsubsection{Radial Profiles} \label{sec:rad_profiles}
To quantify the differences in the \textsc{flash} and \textsc{sph} runs we show radial profiles of the mass, density and radial velocity in Figure~\ref{fig:rad_profiles} for times $t=0.0$, 0.5, 0.8, 0.9, 1.0 and $1.1\,\tff$ (from top to bottom). The initial conditions are compared in the three top panels. The density is constant inside the initial gas sphere with a radius of $R=5.0\times10^{17}\,\cm=3.3\times10^4\,\AU$. Thus, the integrated mass grows with the radius $r$ proportional to $r^3$ inside the sphere until it reaches $M=100\,\Msol$ at $r=R$. In the \textsc{flash} run, the mass grows slightly further for $r>R$ up to about $103\,\Msol$, because of the remaining low-density gas outside of the sphere as discussed in~\S~\ref{sec:initialconditions}. In contrast, the density is zero for $r>R$ in the \textsc{sph} run. Initially, the radial velocity is roughly zero, because the turbulent velocity field was constructed to be isotropic with zero mean.

At $t=0.5\,\tff$ the radial velocity indicates the onset of global infall with slightly supersonic speeds of $0.5\,\km\,\mathrm{s}^{-1}$ at $r\sim 1-2\times10^4\,\AU$. Both the \textsc{flash} and the \textsc{sph} run exhibit similar radial profiles at that time. When the first sink particles have formed ($t=0.8\,\tff$) the radial density profiles roughly follow $\rho(r)\propto r^{-3/2}$ \citep[see e.g.,][]{Shu1977} for $2\times10^3\lesssim r/\AU\lesssim 2\times10^4$. At $t=0.9\,\tff$ the \textsc{flash} and \textsc{sph} run have produced 15 and 17 sink particles, respectively, which is seen as density peaks in the radial density profiles. The infall velocities have reached Mach 5 at that time, and they are roughly consistent between the two codes. However, at $t=1.0\,\tff$ the \textsc{sph} run shows slightly larger infall speeds than the \textsc{flash} run for $2\times10^3\lesssim r/\AU\lesssim 2\times10^4$. This becomes more prominent at $t=1.1\,\tff$, reaching down to the cluster center, when the infall speeds in the \textsc{sph} run are almost twice as large as in the \textsc{flash} run. The star cluster produced in the \textsc{sph} run is denser, which is shown by the mass and density profiles at $t=1.1\,\tff$. This is consistent with the visual inspection of column density images discussed in the previous section. Thus, the global collapse of the gas cloud proceeds slightly faster in the \textsc{sph} run compared to the \textsc{flash} run.

There are three explanations for this behavior, probably all contributing to the faster collapse speeds in the \textsc{sph} run. First, the gas mass outside the sphere may contribute to slow down the collapse slightly in the \textsc{flash} run. \emph{Inflow} boundary conditions were used for the hydrodynamics in the \textsc{flash} run. Thus, low-density gas is streaming in from the boundaries during the global collapse, which further increases the filamentary excess mass outside of the cluster from $3\,\Msol$ to $5\,\Msol$. Secondly, both the \textsc{flash} and \textsc{sph} codes have different dissipation mechanisms. \citet{KitsionasEtAl2009} find that \textsc{sph} codes tend to dissipate small-scale, random, turbulent motions slightly faster than grid codes. Thus, our \textsc{sph} run may loose turbulent support faster than our \textsc{flash} run, which shifts the onset of global collapse to earlier times in the \textsc{sph} run. Finally, one must consider that the low-density gas is better resolved in the grid calculation \citep[e.g.,][]{PriceFederrath2010}. This becomes even stronger at later times when the number of \textsc{sph} particles decreases due to accretion, and the number of grid cells in the \textsc{flash} run increases due to the Jeans refinement and the formation of sink particles (cf.~\ref{sec:refinement_check}). Therefore, at late times, the numerical resolution of the \textsc{flash} run becomes increasingly superior to the \textsc{sph} run. For the star cluster simulation presented here, lower numerical resolution could lead to somewhat smaller collapse timescales, because small-scale turbulent pressure support is weaker at lower resolution. We conclude that all these three effects may contribute to the slightly faster global collapse in the \textsc{sph} run compared to the \textsc{flash} run.

\subsubsection{Mass Distributions of Fragments}
Figure~\ref{fig:smfs} shows a comparison of the mass distributions of sink particles obtained in the \textsc{flash} and \textsc{sph} runs at $t=0.8$, 0.9, 1.0 and $1.1\,\tff$ (from top to bottom). The left panels show the histograms of sink particle mass, while the right panels show the cumulative mass distributions from which the Kolmogorov-Smirnov probability (p-KS) was determined with a KS test. The KS test provides an estimate of the probability that the mass distributions obtained in the \textsc{flash} and in the \textsc{sph} run were drawn from the same basic distribution, and it thus provides an estimate of their similarity.

Figure~\ref{fig:smfs} shows that at $t=0.8\,\tff$ the KS probability is about 70\%, which means that the mass distributions obtained in the \textsc{flash} and \textsc{sph} run are very likely drawn from the same distribution. The similarity in the number of particles formed and in the mass accreted onto sink particles taken together with the similarity in the location of sink particles at that time is encouraging. This is an important result, because it shows that two independent implementations of sink particle formation in two fundamentally different hydrodynamic codes give almost identical results for the number, mass and location of formed fragments.

At later times, the KS probability drops to 55\%, 40\% and 8\% at $t=0.9$, 1.0 and $1.1\,\tff$, respectively. However, the general shape of the mass distributions obtained in the \textsc{flash} and \textsc{sph} runs is similar for all times (although the relatively small number of sink particles does not allow for a detailed comparison of the slope at the high-mass end of the distributions). The peaks of the mass distributions agree well for $t\leq1.0\,\tff$. However, at $t=1.1\,\tff$ the peak of the distribution is shifted to slightly higher masses in the \textsc{sph} run as a result of the faster global collapse in the \textsc{sph} run (cf.~\S~\ref{sec:rad_profiles}). The high-mass ends of the \textsc{flash} and \textsc{sph} mass distributions are roughly consistent with the \citet{Salpeter1955} power law, $N\propto M^{-1.35}$, as also found in previous \textsc{sph} studies. However, our statistical samples are not large enough to draw further quantitative conclusions on the slope of the high-mass end. It should also be noted that the low-mass end is probably affected by our assumption of an isothermal equation of state. Our models did not take into account the effects of radiation transfer, which is expected to suppress fragmentation \citep{KrumholzKleinMcKee2007,Bate2009}. However, the fact that our sink particle radius is about $500\,\AU$ in this simulation does not allow any fragmentation below this scale. Thus, we are more likely underestimating the amount of fragmentation, because we cannot follow possible fragmentation on scales smaller than about $500\,\AU$.

The fact that the KS probability becomes smaller at later times is a result of the faster global collapse in the \textsc{sph} run. To quantify this further, we compare the mass distributions at the times when the total mass of sink particles is roughly the same in the \textsc{flash} and \textsc{sph} runs. Figure~\ref{fig:smfs_sfe} shows the mass distribution of \textsc{flash} at $t=1.1\,\tff$ together with the mass distribution of \textsc{sph} at $t=1.04\,\tff$ (about 5\% earlier) when the sink particle formation efficiency is about 26\% in both runs. The mass distributions agree very well for similar sink formation efficiencies, which is indicated by a KS probability of 55\%. Moreover, the number of fragments is almost identical in the \textsc{flash} and \textsc{sph} run (49 versus 50). There is striking agreement between the sink particle properties obtained in our \textsc{flash} implementation and in the \citet{BateBonnellPrice1995} implementation for \textsc{sph} albeit the highly complex and chaotic nature of the problem modeled in this code comparison. The fact that two fundamentally different numerical schemes with complementary strengths and weaknesses give the same overall result is very encouraging. It strengthens our confidence in simulations of the turbulent collapse of clouds with \textsc{amr} on the one hand and \textsc{sph} on the other hand.

\subsubsection{Computational Efficiency of \textsc{flash} and \textsc{sph}}
We briefly mention the computational efficiency of our \textsc{flash} and \textsc{sph} runs. Both codes were run in a mode of parallel computation on the same supercomputer (HLRB-II: SGI Altix 4700) at the Leibniz-Rechenzentrum Garching\footnote{\texttt{http://www.lrz-muenchen.de}}. The \textsc{flash} run consumed about 10,300 CPU hours and was run on 128 CPUs. The \textsc{sph} run consumed roughly 2,400 CPU hours and was run on 16 CPUs, which is 4-5 times faster than the \textsc{flash} run. The \textsc{sph} code automatically increases resolution in high-density regions, without the necessity to increase the number of resolution elements there, because the \textsc{sph} particles move in the Lagrangian frame. It should be noted however that due to accretion of \textsc{sph} particles onto sink particles, the total number of \textsc{sph} particles decreases from the initial 2 million to about 1.3 million at $t=1.1\,\tff$. In contrast, the Eulerian code \textsc{flash} refined the grid adaptively in high-density regions. The number of resolution elements (grid cells) thus increases from about 32 million at $t=0$ to 56 million at $t=1.1\tff$, which is a factor of about 30 more resolution elements on average in the \textsc{flash} run compared to the \textsc{sph} run. The computational cost per resolution element was thus a factor of 6-8 smaller for \textsc{flash} than for \textsc{sph}. This is roughly consistent with the difference in the average computational efficiency per resolution element measured for typical grid-based and particle-based hydrodynamic codes in the case of non-selfgravitating, supersonic turbulence \citep[e.g.,][]{KitsionasEtAl2009,PriceFederrath2010}.

\section{Conclusions} \label{sec:conclusions}
We introduced an implementation of accreting sink particles for the adaptive mesh refinement code \textsc{flash}. Sink particles are used as a subgrid model to cut-off local runaway collapse in a controlled way to avoid artificial fragmentation and to avoid prohibitive small timesteps in numerical simulations involving multiple collapse regions. Sink particles interact gravitationally with the gas and with one another, and can accrete bound gas. The gas must pass a series of checks prior to sink particle formation as explained in detail in \S~\ref{sec:creationchecks}. In particular, in our grid-based implementation the gas considered for sink particle creation is not only required to exceed a given density threshold, but this gas must be gravitationally bound and collapsing at the same time, similar to the original sink particle implementation of \citet{BateBonnellPrice1995} for \textsc{sph}. Figure~\ref{fig:coldens} (right panels) shows that spurious sink particles would be created in the star cluster formation run discussed in~\S~\ref{sec:AMRvsSPH}, if a \emph{sole} density threshold was used to determine sink particle creation. The root cause of spurious sink particle creation is that the gas density can exceed the given density threshold in shocks that do not necessarily trigger gravitational collapse. Figure~\ref{fig:coldens} (right versus left panels) shows that the number of fragments would be overestimated by more than one order of magnitude, and that the star formation efficiency (total gas mass accreted by sink particles) would also be overestimated by 133\%, 51\%, 19\% and 14\% at $t/\tff=0.8,\,0.9,\,1.0$ and 1.1, respectively, if the additional checks described in~\S~\ref{sec:creationchecks} are switched off. Some sort of sink particle merging could be used to reduce the overestimated number of sink particles, but the mass accreted onto sink particles would still be overestimated if a sole density threshold was used for sink particle creation at densities below the opacity limit (about $10^{-14}\,\g\,\cm^{-3}$).

We performed a series of tests of our sink particle implementation in~\S~\ref{sec:tests}, including the time integration of circular and elliptical orbits, the collapse of a Bonnor-Ebert sphere, and a rotating cloud core fragmentation test. The sink particle accretion rate in the collapse of a singular isothermal sphere showed excellent agreement with the theoretical predictions of the \citet{Shu1977} model. Those tests showed that the dynamical creation, gravitational interaction, motion, timestepping and accretion of sink particles works in a way that enables us to analyze the trajectories, accretion rates and mass distributions of collapsing fragments quantitatively and reliably in follow-up studies that will also include \textsc{mhd} effects.

A comparison calculation of star cluster formation between the \textsc{sph} code by \citet{BateBonnellPrice1995} and our grid-based \textsc{flash} implementation showed encouraging agreement of gas and sink particle properties obtained in the \textsc{sph} and \textsc{flash} runs (cf.~\S~\ref{sec:AMRvsSPH}). Using column density images, we demonstrated that sink particles are created at roughly the same locations and times in both codes. Radial velocity profiles revealed that the global collapse in the \textsc{sph} run was about 5\% faster than in the \textsc{flash} run, probably due to the slightly higher numerical dissipation in \textsc{sph} codes \citep{KitsionasEtAl2009,PriceFederrath2010}. After correction for this time lag, we showed that the number of sink particles and their mass distributions in the \textsc{flash} and \textsc{sph} runs are in very good agreement. The agreement of our \textsc{flash} and \textsc{sph} runs is an encouraging result that strengthens our confidence in numerical simulations of gravoturbulent cloud fragmentation and collapse with \textsc{amr} and \textsc{sph}.

\acknowledgements
CF thanks Daniel Price for useful discussions on the gravitational softening and time integration schemes for sink particles during his visit at Monash University under the \textsc{g}o8/\textsc{daad} Joint Research Cooperation Scheme. We thank the anonymous referee for an unbiased and detailed report, which helped to improve the paper. CF is grateful for financial support from the International Max Planck Research School for Astronomy and Cosmic Physics (\textsc{imprs-a}) and the Heidelberg Graduate School of Fundamental Physics (\textsc{hgsfp}), which is funded by the Excellence Initiative of the Deutsche Forschungsgemeinschaft (\textsc{dfg}) \textsc{gsc} 129/1. RB is grateful to the \textsc{dfg}, which funds his Emmy Noether grant \textsc{ba} 3607/1-1. RB is also thankful for subsidies from the \textsc{frontier} initiative of the University of Heidelberg. PCC acknowledges support from the \textsc{dfg} under the Emmy Noether grant \textsc{kl} 1358/1 and the Priority Program \textsc{sfb} 439 `Galaxies in the Early Universe'. RSK thanks for financial resources provided by the \textsc{dfg} under grants no.~\textsc{kl} 1358/1, \textsc{kl} 1358/4, \textsc{kl} 1359/5. RSK furthermore thanks for subsidies from the \textsc{frontier} grant of Heidelberg University sponsored by the German Excellence Initiative and for support from the Landesstiftung Baden-W\"{u}rttemberg via their program International Collaboration II under grant \textsc{p-ls-spii}/18. All authors acknowledge financial support from the German Bundesministerium f\"{u}r Bildung und Forschung via the \textsc{astronet} project \textsc{star format} (grant 05A09\textsc{vha}). The \textsc{flash} and \textsc{sph} simulations used computational resources from the \textsc{hlrb ii} project grant pr32lo at the Leibniz Rechenzentrum Garching and from the \textsc{caspur} project grant. The test simulations were partly run on the \textsc{gpu} cluster \textsc{kolob} at the University of Heidelberg. Figure~\ref{fig:orbit_sis_amr} was produced with the visualization tool \textsc{visit}. The \textsc{flash} code was developed in part by the \textsc{doe}-supported Alliances Center for Astrophysical Thermonuclear Flashes (\textsc{asc}) at the University of Chicago.


\begin{appendix}

\section{Gravitational Spline Softening} \label{app:softening}
Cubic spline softening \citep[see][]{MonaghanLattanzio1985,PriceMonaghan2007} is used to soften the gravitational accelerations $\vect{g}(\vect{r})$ between sink particles and gas and among sink particles:
\begin{equation} \label{eq:spline}
\def\arraystretch{3.0}
\vect{g}(\vect{r}) \propto \left\{
   \begin{array}{ll}
      \frac{4}{\rsoft^2}\left[\frac{8}{3}\left(\frac{r}{\rsoft}\right)-\frac{48}{5}\left(\frac{r}{\rsoft}\right)^3+8\left(\frac{r}{\rsoft}\right)^4\right]\,\frac{\vect{r}}{r} & \quad\mbox{for}\quad 0\leq\frac{r}{\rsoft}<\frac{1}{2} \\
      \frac{4}{\rsoft^2}\left[\frac{16}{3}\left(\frac{r}{\rsoft}\right)-12\left(\frac{r}{\rsoft}\right)^2+\frac{48}{5}\left(\frac{r}{\rsoft}\right)^3-\frac{8}{3}\left(\frac{r}{\rsoft}\right)^4-\frac{1}{60}\left(\frac{\rsoft}{r}\right)^2\right]\,\frac{\vect{r}}{r} & \quad\mbox{for}\quad \frac{1}{2}\leq\frac{r}{\rsoft}<1 \\
      \frac{\vect{r}}{r^3} & \quad\mbox{for}\quad r \geq \rsoft\;, \\
   \end{array} \right.
\end{equation}
where $r=\left|\vect{r}\right|$ and $\rsoft$ is the softening radius. See Figure~\ref{fig:softening} for a graphical representation of this type of softening compared to Plummer softening and compared to Newton's acceleration of a point mass.

\section{Conservation Laws During Accretion of Gas} \label{app:conservationlaws}
We briefly discuss the conservation laws during accretion of gas onto sink particles. The index $i$ can be used to denote both the multiple gas portions to be accreted and the sink particle onto which these gas portions are being accreted within a single accretion step. Thus, mass and linear momentum conservation during accretion or merging of sink particles are given by the following equations:
\begin{equation}
M = \sum_i\,m_i
\end{equation}
\begin{equation}
M\,\vect{v}_\mathrm{cm} = \sum_i\,m_i\,\vect{v}_i\;.
\end{equation}
The last equation determines the center of mass velocity $\vect{v}_\mathrm{cm}$ to conserve linear momentum during an accretion or merging step. However, if these two conservation laws apply, then angular momentum conservation is generally broken in the process of accretion. To see that, consider the angular momentum before the accretion process
\begin{equation}
\vect{L} = \sum_i\,m_i\,\vect{r}_i\times\vect{v}_i\,.
\end{equation}
The angular momentum after accretion is
\begin{equation}
\vect{L}' = M\,\vect{R}\times\vect{v}_\mathrm{cm}\,.
\end{equation}
To fulfill angular momentum conservation $\vect{L}=\vect{L}'$, the following equation needs to be solved for the position $\vect{R}$ of the accreted gas plus sink particle:
\begin{equation} \label{eq:ang_mom_conserve}
\vect{R}\times\vect{v}_\mathrm{cm}=\frac{1}{M}\,\vect{L}\,.
\end{equation}
However, this equality \emph{cannot} be fulfilled in the general case, because $\vect{L}$ and $\vect{v}_\mathrm{cm}$ are generally \emph{not} orthogonal. Thus, it is impossible to construct a position vector $\vect{R}$, such that angular momentum conservation holds. Furthermore, if $\vect{R}$ is given by the center of mass
\begin{equation}
\vect{R}_\mathrm{cm} = \frac{1}{M}\,\sum_i\,m_i\,\vect{r}_i\,,
\end{equation}
angular momentum conservation is also typically broken, because eq.~(\ref{eq:ang_mom_conserve}) does not generally hold true. This is the case for any implementation of accreting sink particles. The only way to restore angular momentum conservation is to introduce an intrinsic angular momentum (spin) for each sink particle \citep[see][]{BateBonnellPrice1995,KrumholzMcKeeKlein2004,JappsenEtAl2005}. This spin compensates the deviation from global angular momentum conservation caused by accretion or merging of sink particles. It can also be used to determine the axis along which a bipolar outflow or jet would be launched in a subgrid model of stellar feedback.

\end{appendix}

\clearpage

\begin{figure}[t]
\centerline{\includegraphics[width=0.5\linewidth]{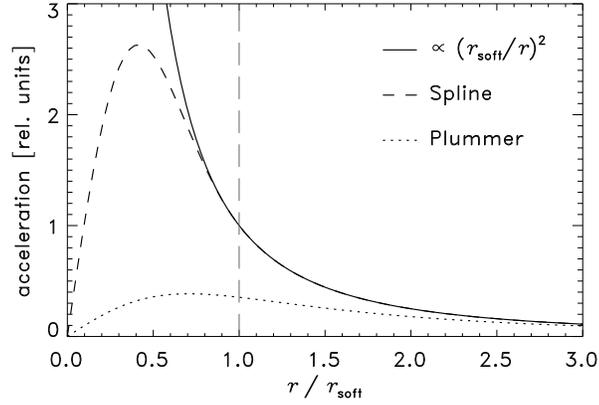}}
\caption{Plummer softening versus spline softening: The sink particle implementation presented here uses spline softening (Appendix~\ref{app:softening}) for the gravitational accelerations between sink particles and grid cell centers, and between adjacent sink particles.}
\label{fig:softening}
\end{figure}

\begin{figure}[t]
\begin{center}
\def\arraystretch{0.2}
\begin{tabular}{c}
\includegraphics[width=0.4\linewidth]{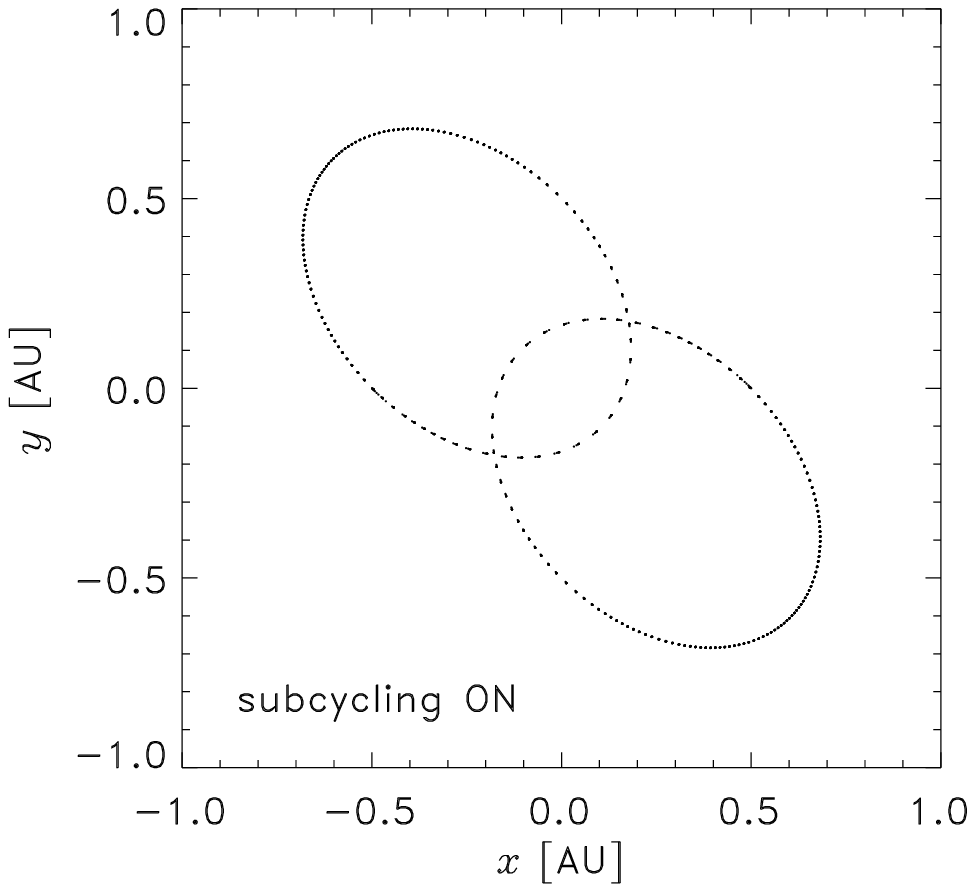} \\
\includegraphics[width=0.4\linewidth]{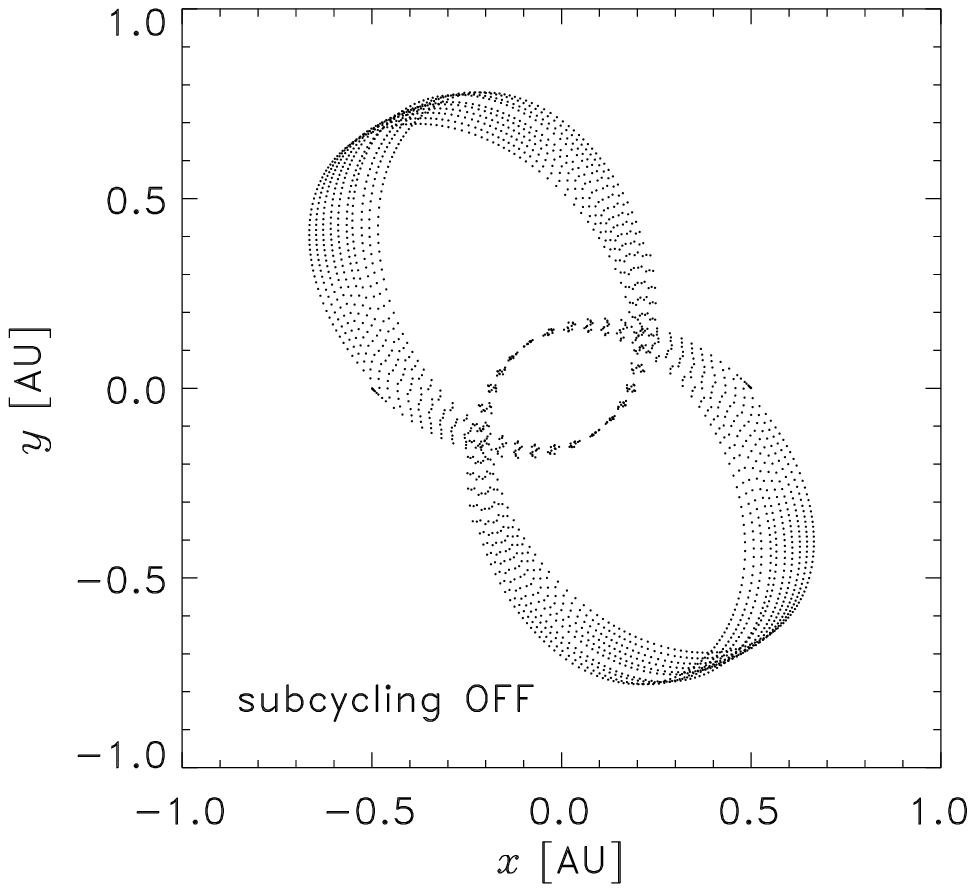}
\end{tabular}
\end{center}
\caption{Trajectories of two point masses on elliptical orbits around their common center of mass after 10 orbits. The symplectic property of the leapfrog integrator is broken due to variable timesteps in this type of problem. However, 10 orbits are reproduced quite accurately with subcycling turned ON, while for subcycling turned OFF (bottom), large errors occur already for single orbits, and a significant amount of artificial angular momentum is introduced.}
\label{fig:elliptical_orbits}
\end{figure}

\begin{figure}[t]
\begin{center}
\def\arraystretch{0.2}
\begin{tabular}{c}
\includegraphics[width=0.4\linewidth]{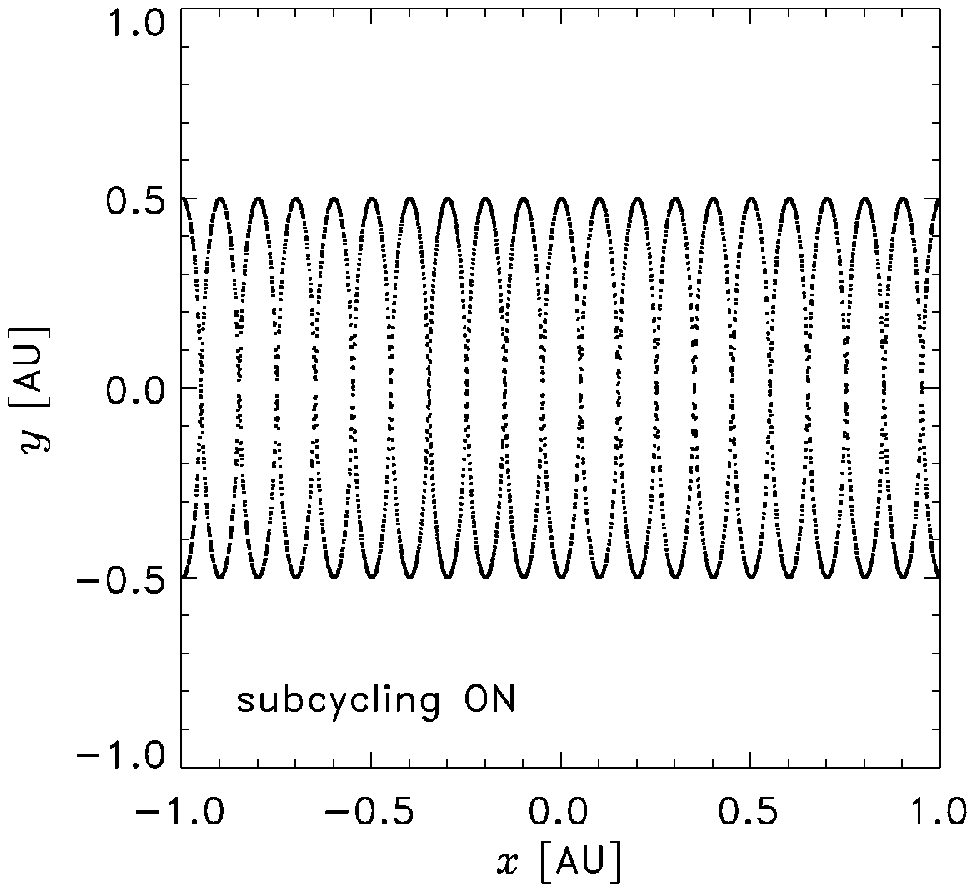} \\
\includegraphics[width=0.4\linewidth]{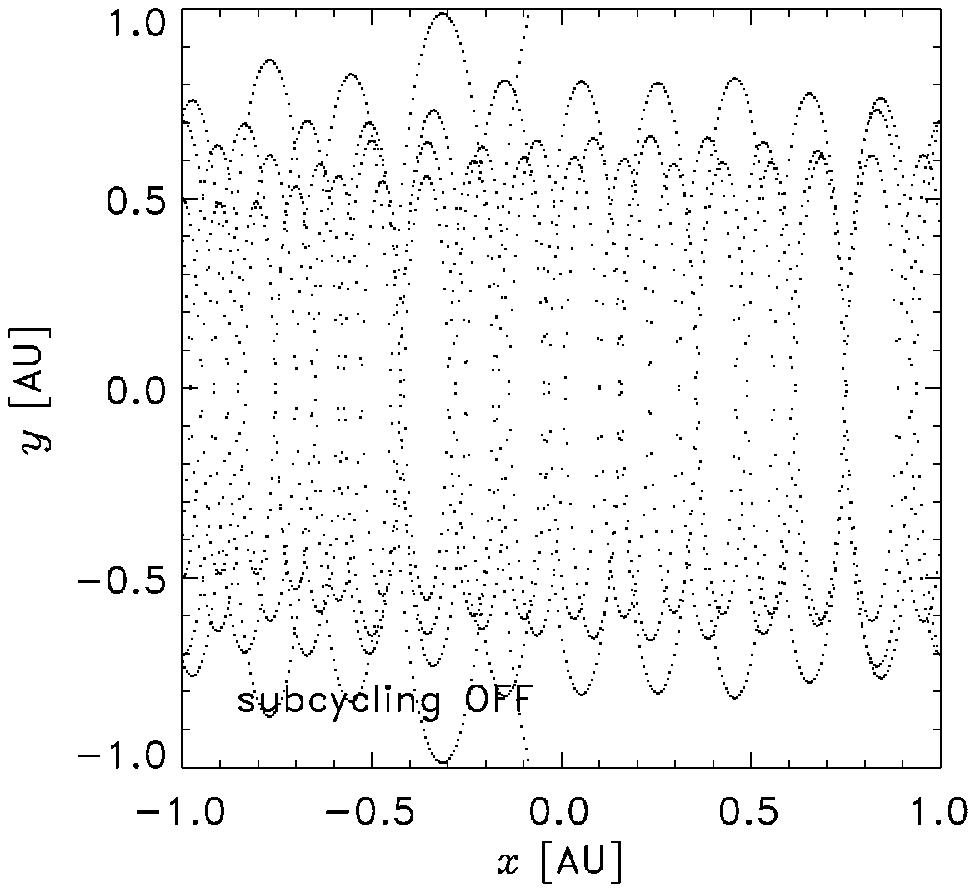}
\end{tabular}
\end{center}
\caption{Gravitational softening and subcycling test: The trajectories of two point masses starting with zero velocity along the $y$-axis, but with an initial velocity along the $x$-axis are shown. The two masses oscillate around their common center of mass, and draw a sine function. The boundary conditions are set to \emph{outflow} for the $y$-direction and to \emph{periodic} for the $x$-direction. The frequency of the oscillation in $y$ and the initial velocity in $x$ were chosen such that the two particles leaving on one side of the domain will exactly connect to their previous trajectories. Each sweep through $x$ contains 20 oscillations and a total number of 1000 oscillations is shown (top). When -- for the same setup -- subcycling is turned OFF (bottom) the sink particles get artificial accelerations during their close encounters, and energy conservation is broken. Already after roughly 68 oscillations, they have artificially gained enough energy to leave the domain at $x\approx-0.1\,\mathrm{AU}$.}
\label{fig:oscis}
\end{figure}

\begin{figure*}[t]
\begin{center}
\def\arraystretch{0.2}
\begin{tabular}{cc}
\includegraphics[width=0.5\linewidth]{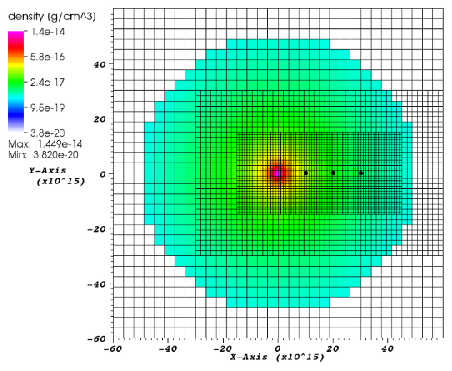} &
\includegraphics[width=0.5\linewidth]{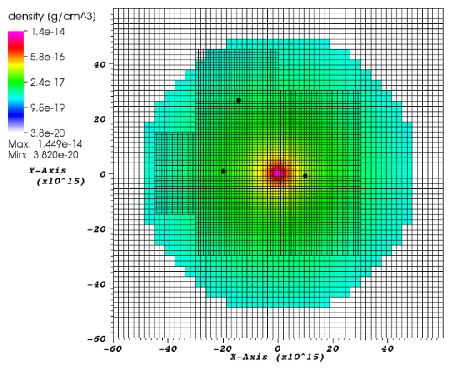}
\end{tabular}
\end{center}
\caption{Three sink particles orbiting a static spherical gas density profile with $\rho\propto r^{-2}$. Note that the hydrodynamical evolution was switched-off deliberately for this test, while the gravity solver and the \textsc{amr} scheme remained fully operational. The left image shows the initial conditions and the right image shows the sink particle positions after the innermost particle has completed roughly one full orbit. Adaptive mesh refinement is used to keep the sink particles at the highest level of refinement allowed in this setup. The particle on the middle orbit has completed half an orbit, while the outermost particle has completed 1/3 of an orbit at the time shown in the right-hand image. Figure~\ref{fig:orbit_sis} shows the full particle trajectories after one, ten and 100 orbits of the innermost sink particle.}
\label{fig:orbit_sis_amr}
\end{figure*}

\begin{figure*}[t]
\begin{center}
\def\arraystretch{0.2}
\begin{tabular}{ccc}
\includegraphics[width=0.32\linewidth]{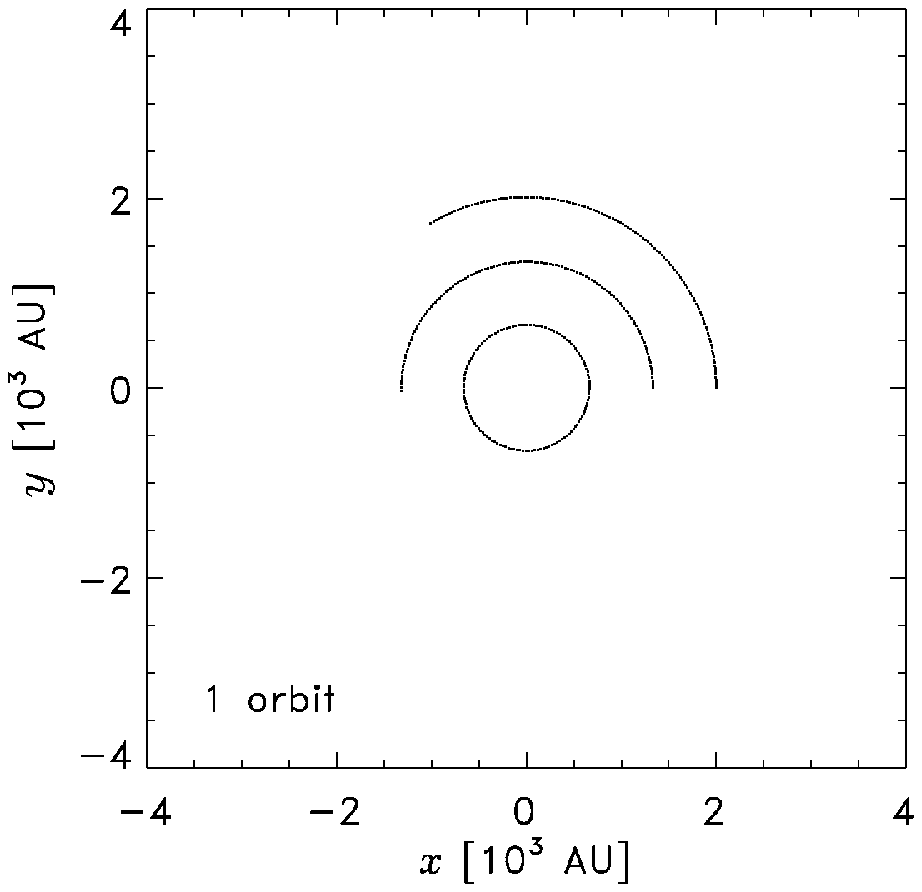} &
\includegraphics[width=0.32\linewidth]{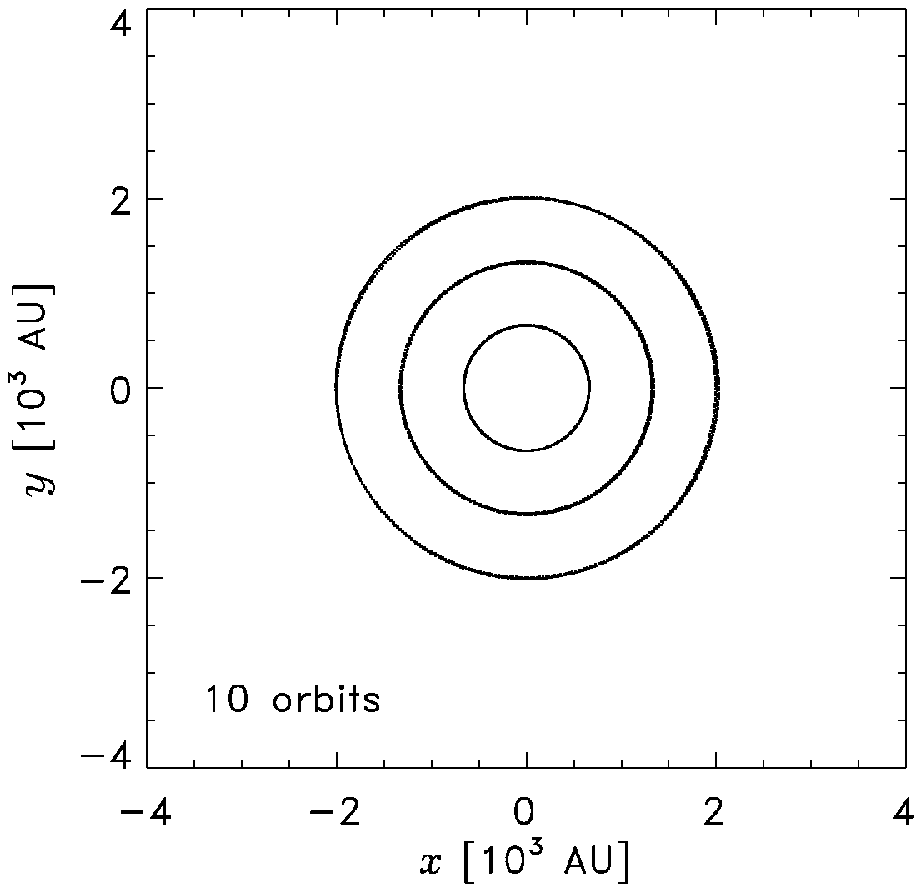} &
\includegraphics[width=0.32\linewidth]{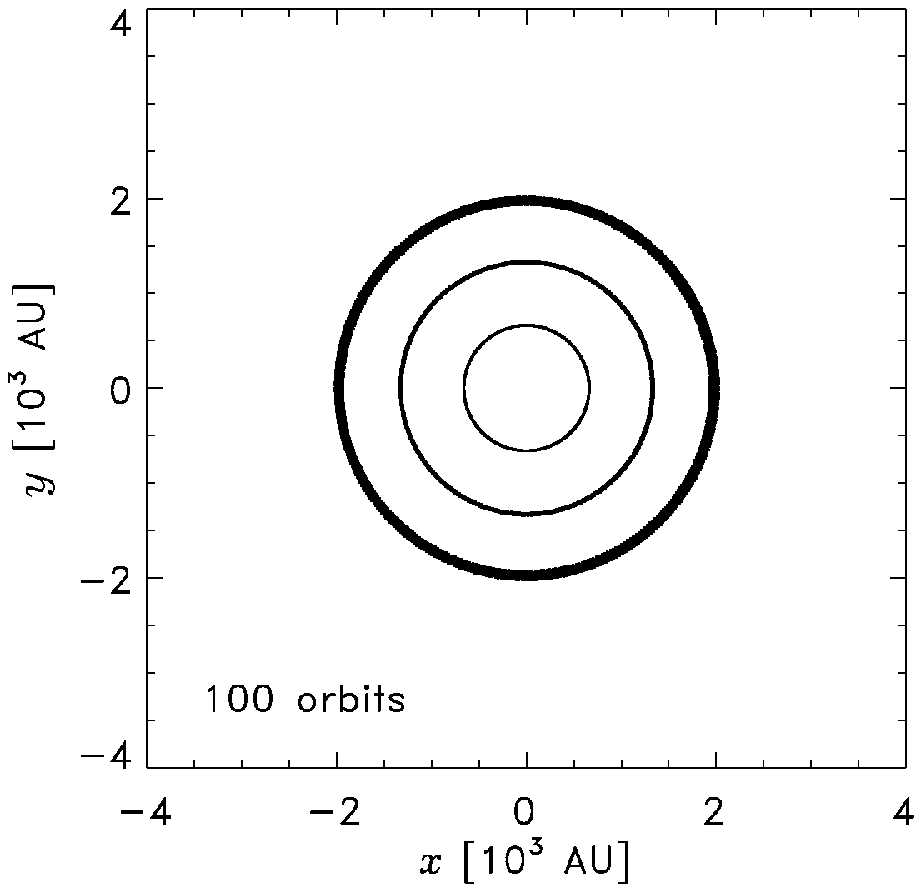}
\end{tabular}
\end{center}
\caption{Sink particle trajectories of the setup shown in Figure~\ref{fig:orbit_sis_amr} after one (left), ten (middle) and 100 (right) completed orbits of the innermost particle. The time in the left plot is equivalent to the image shown in Figure~\ref{fig:orbit_sis_amr} (right). Ten orbits (middle) are quite accurately followed. After 100 orbits of the innermost sink particle (right), the intermediate particle has finished 50 orbits, while the outermost sink particle has finished about 33 orbits. The outermost sink particle shows some deviation from a perfect spherical orbit, seen as a slight broadening of its circular trajectory, which is due to the accumulation of small time-integration errors, and due to deviations from spherical symmetry toward the edges of the Cartesian domain.}
\label{fig:orbit_sis}
\end{figure*}

\begin{figure}[t]
\begin{center}
\def\arraystretch{0.2}
\begin{tabular}{c}
\includegraphics[width=0.45\linewidth]{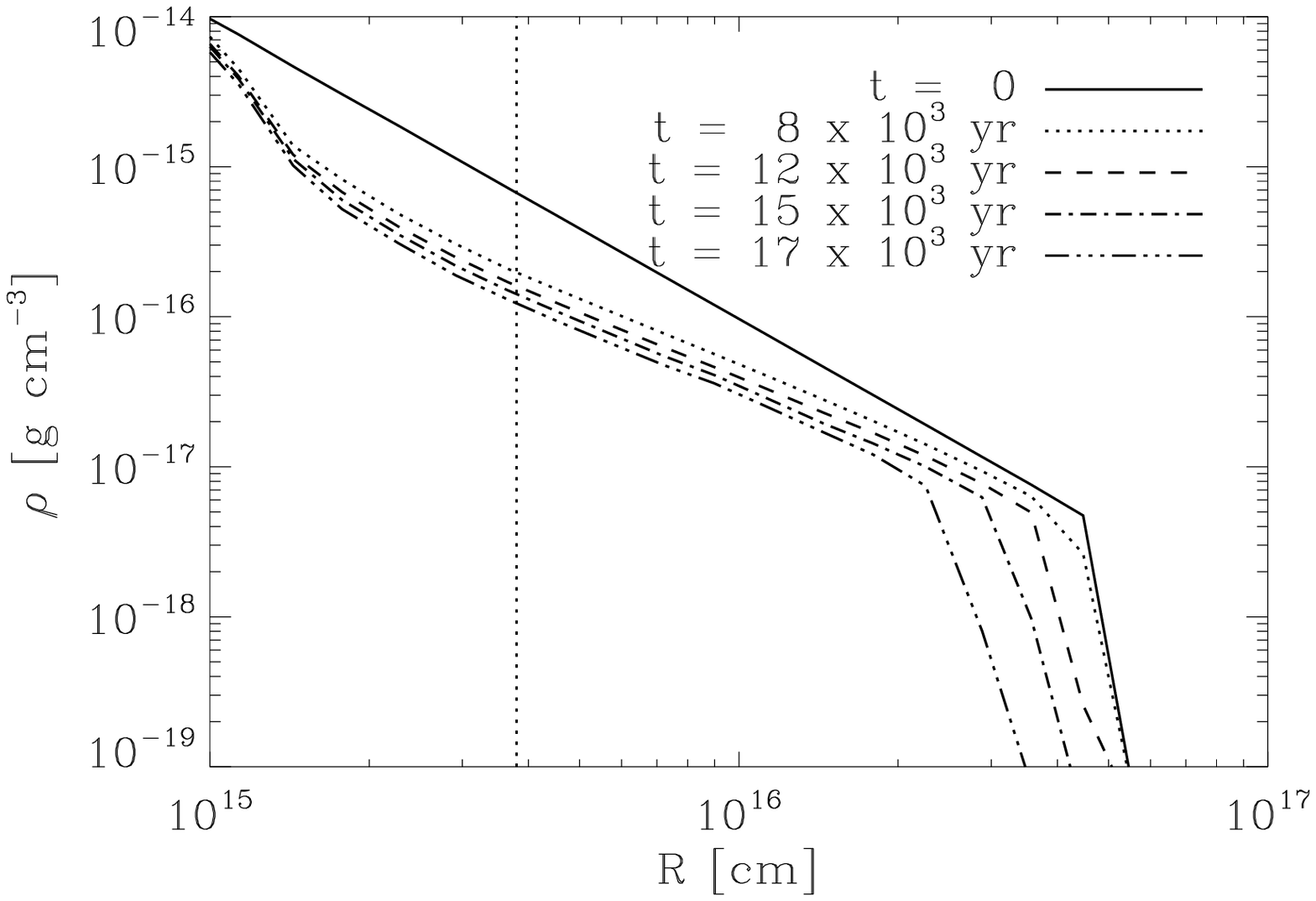} \\
\includegraphics[width=0.45\linewidth]{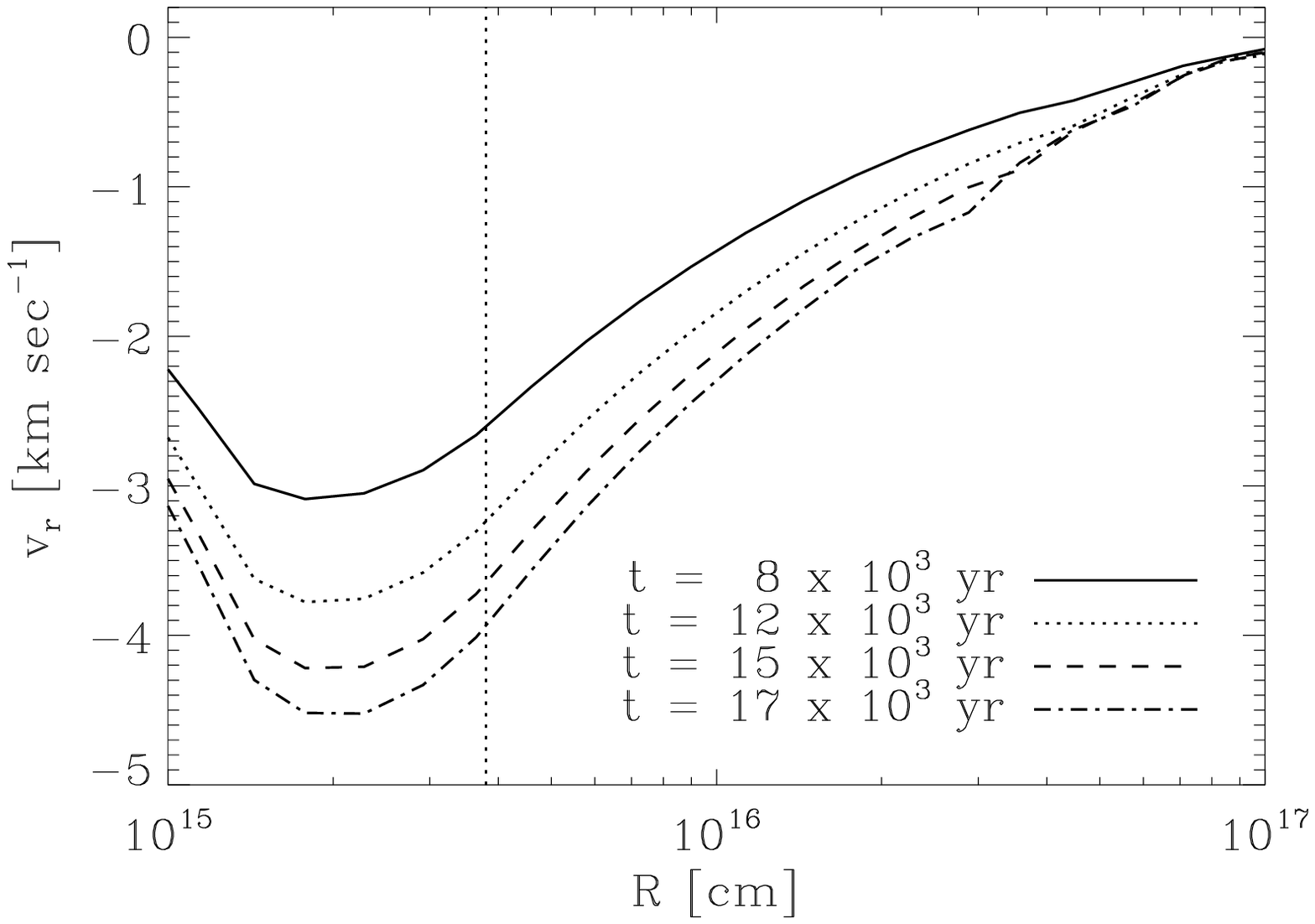} \\
\includegraphics[width=0.45\linewidth]{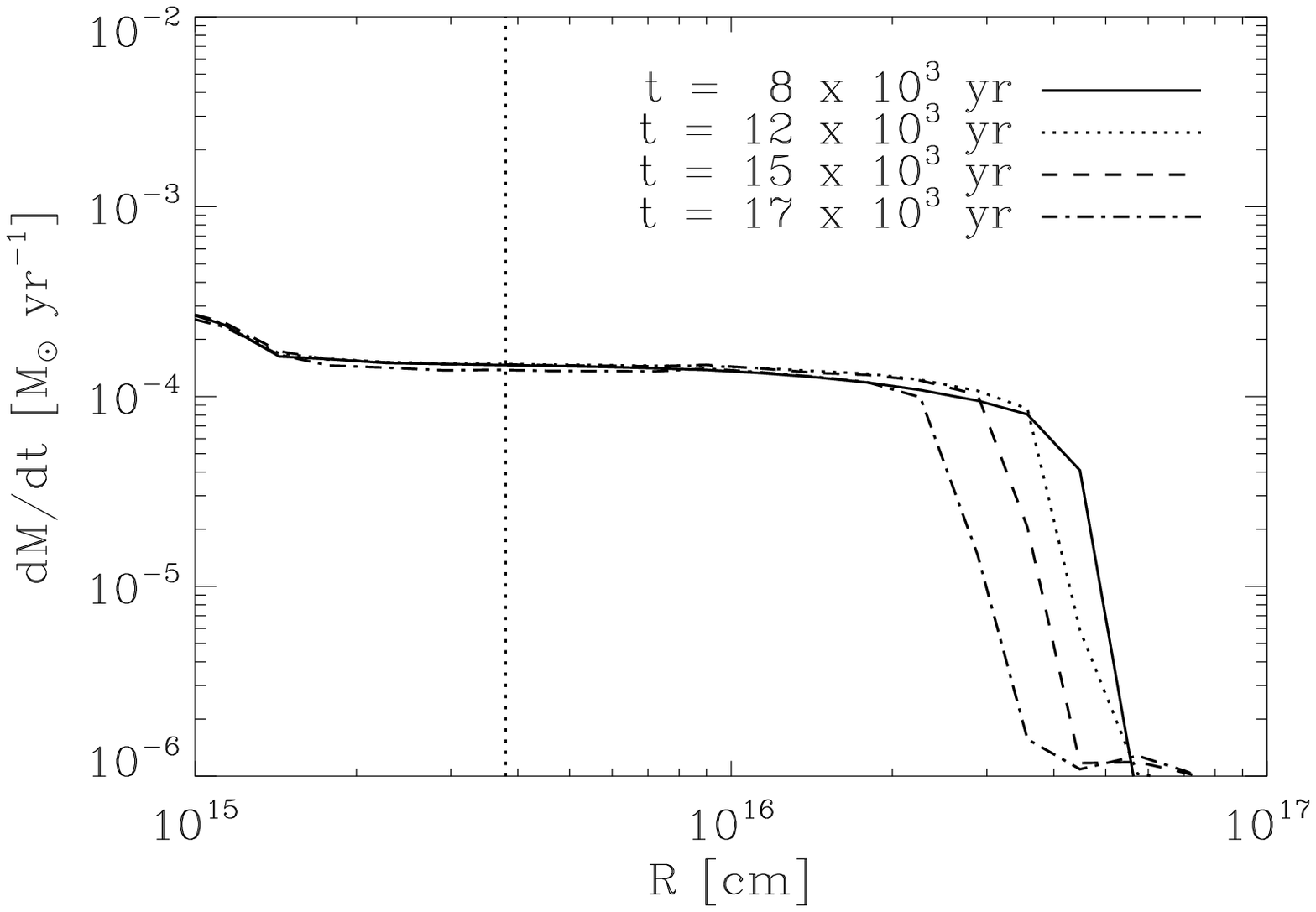}
\end{tabular}
\end{center}
\caption{Time evolution of the radial density profile (top panel), infall velocity (middle panel) and mass flux through radial shells (bottom panel) for the collapse of a singular isothermal sphere. The vertical dotted line marks the accretion radius of the sink particle. The density profile evolves from the initial $r^{-2}$ profile to $r^{-1.5}$ when the mass of the sink particle starts to dominate the gravitational potential~\citep[see e.g.][]{Shu1977,OginoTomisakaNakamura1999}. As expected for a highly unstable sphere the collapse proceeds with highly supersonic speeds (note that $\cs=0.166\,\km\,\mathrm{s}^{-1}$). The mass flux of gas through spherical shells of radius $r$ is constant in time. It is exactly the mass accretion rate of the central sink particle, $\dot{M}\approx 1.5\times 10^{-4}\Msol\,\yr^{-1}$ (cf.~Fig.~\ref{fig:sis_mass_accretion}).}
\label{fig:sis_radius}
\end{figure}

\begin{figure}[t]
\begin{center}
\includegraphics[width=0.4\linewidth]{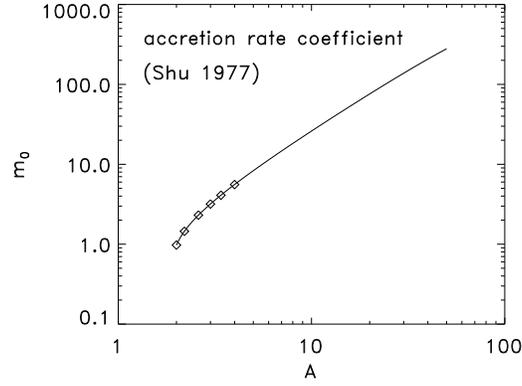}
\end{center}
\caption{Coefficient $m_0$ for the accretion rate in equation~(\ref{eq:shu}) as a function of the instability parameter $A=4\pi G\,\rho(R)\,R^2/\cs^2$ for the collapse of a singular isothermal sphere. The diamonds show tabulated values by \citet[][Tab.~1]{Shu1977}. For $A=29.3$ as in the collapse test shown in Figures~\ref{fig:sis_mass_accretion} and~\ref{fig:sis_radius}, the accretion rate is more than two orders of magnitude higher than $\cs^3/G$, in excellent agreement with the predicted accretion rate by \citet{Shu1977}.}
\label{fig:shu1977}
\end{figure}

\begin{figure*}[t]
\begin{center}
\def\arraystretch{0.6}
\begin{tabular}{cc}
\includegraphics[width=0.35\linewidth]{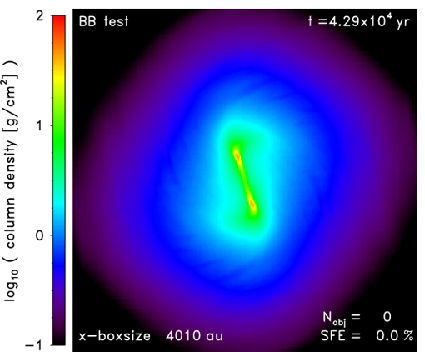} &
\includegraphics[width=0.35\linewidth]{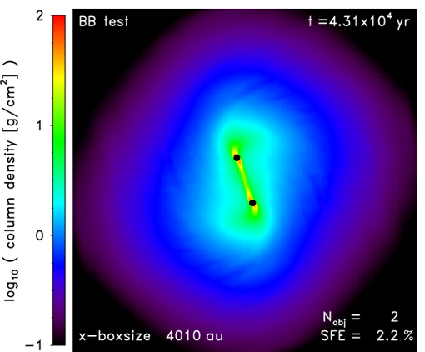} \\
\includegraphics[width=0.35\linewidth]{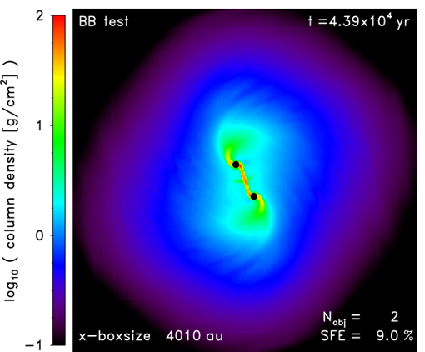} &
\includegraphics[width=0.35\linewidth]{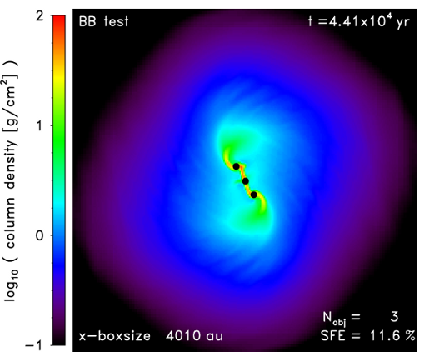} \\
\includegraphics[width=0.35\linewidth]{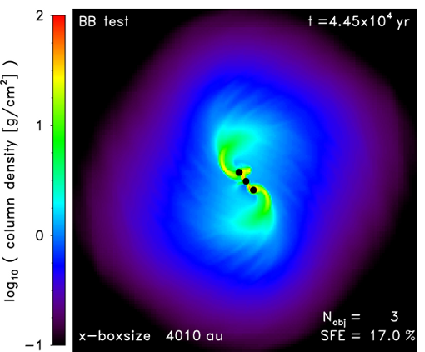} &
\includegraphics[width=0.35\linewidth]{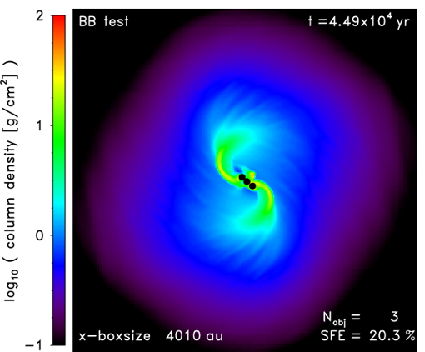} \\
\includegraphics[width=0.35\linewidth]{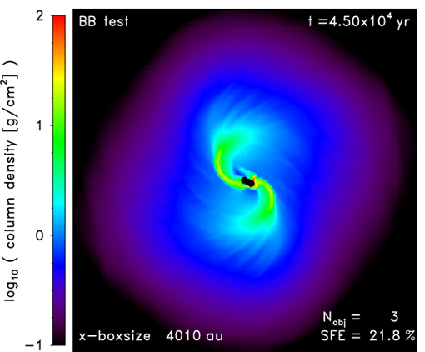} &
\includegraphics[width=0.35\linewidth]{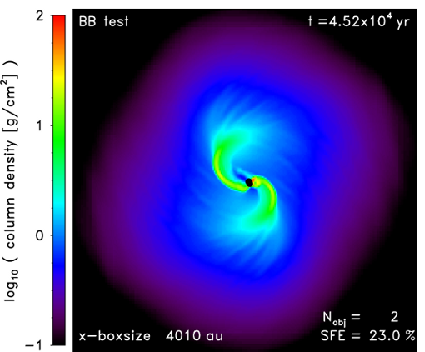}
\end{tabular}
\end{center}
\caption{Column density time sequence of the inner $4000\,\AU$ of a rotating cloud core (see initial conditions in~\S~\ref{sec:bb}), the so called `BB test' \citep{BossBodenheimer1979,BateBurkert1997}. Two sink particles form at the location of the two cloud fragments at $t=4.31\times10^4\,\mathrm{yr}$. At $t=4.39\times10^4\,\mathrm{yr}$ a bar forms that connects the two main fragments, in which a third particle is dynamically created at $t=4.41\times10^4\,\mathrm{yr}$ right in the center. The two main fragments move toward the central object, and at $t=4.52\times10^4\,\mathrm{yr}$ two of the three sink particles merged to a single particle close to the center of the rotating cloud core, because the optional sink particle merging was used in this test. The freefall time of the initial gas distribution is $\tff=3.41\times10^4\,\yr$. Sink particles are drawn as black circles with a radius of $\racc=39\,\AU$ in a true-to-scale representation.}
\label{fig:coldens_bb}
\end{figure*}

\begin{figure*}[t]
\begin{center}
\def\arraystretch{0.6}
\begin{tabular}{cc}
\includegraphics[width=0.4\linewidth]{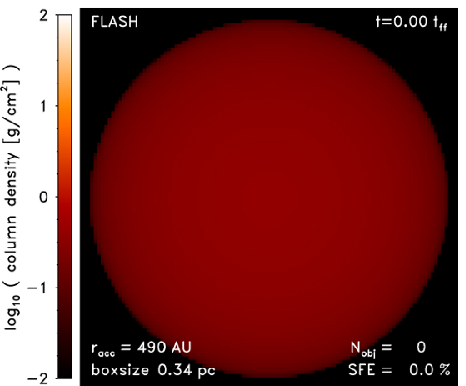} &
\includegraphics[width=0.4\linewidth]{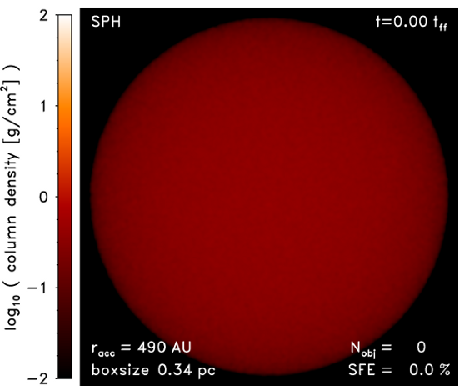} \\
\includegraphics[width=0.4\linewidth]{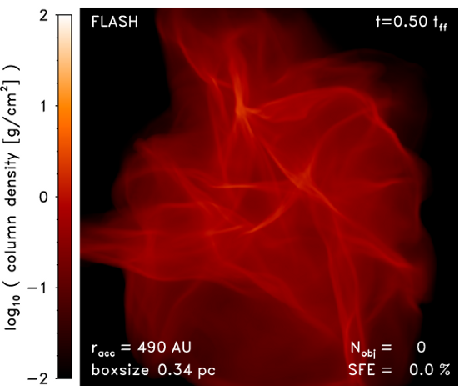} &
\includegraphics[width=0.4\linewidth]{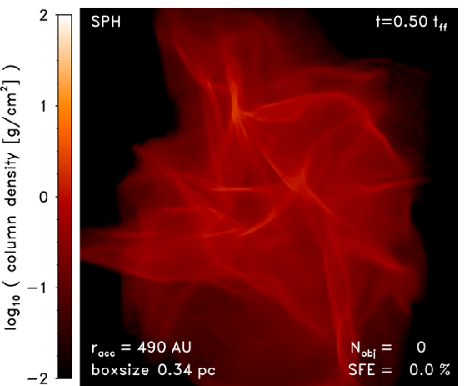}
\end{tabular}
\end{center}
\caption{Comparison of the column density distributions obtained with the \textsc{amr} code \textsc{flash} (left) and with the \textsc{sph} code by \citet{BateBonnellPrice1995} (right) at $t=0.0\,\tff$ (top) and $t=0.5\,\tff$ (bottom) prior to the formation of a stellar cluster. The initial supersonic, turbulent velocity field creates a complex network of filaments. Some of these shocked regions become gravitationally unstable and collapse, i.e., sink particles are created, while other dense regions are transient features, not leading to local collapse (see Fig.~\ref{fig:coldens}).}
\label{fig:coldens_init}
\end{figure*}

\begin{figure*}[t]
\begin{center}
\def\arraystretch{0.6}
\begin{tabular}{ccc}
\includegraphics[width=0.315\linewidth]{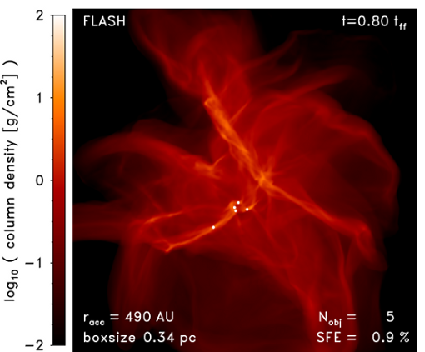} &
\includegraphics[width=0.315\linewidth]{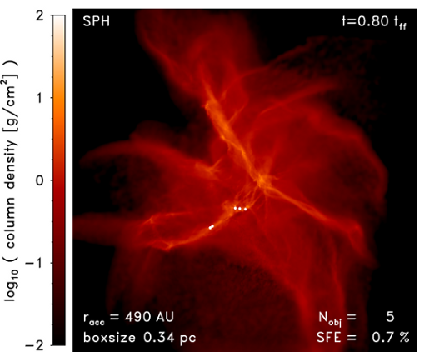} &
\includegraphics[width=0.315\linewidth]{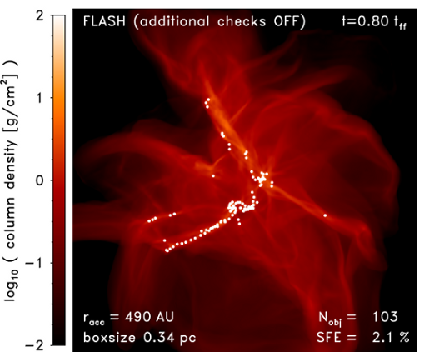} \\
\includegraphics[width=0.315\linewidth]{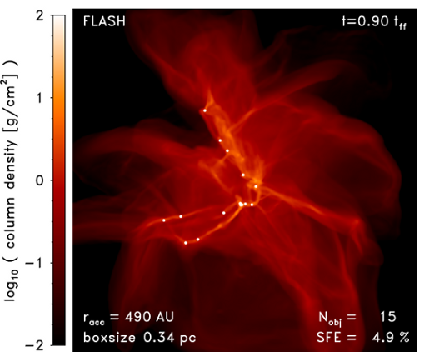} &
\includegraphics[width=0.315\linewidth]{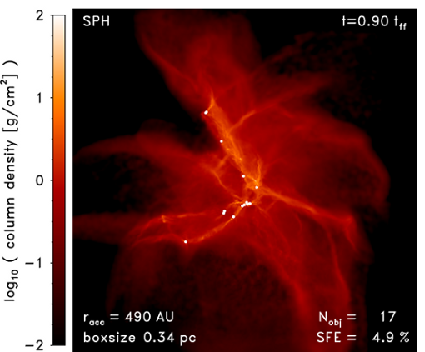} &
\includegraphics[width=0.315\linewidth]{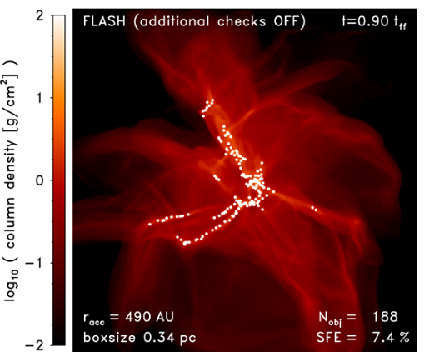} \\
\includegraphics[width=0.315\linewidth]{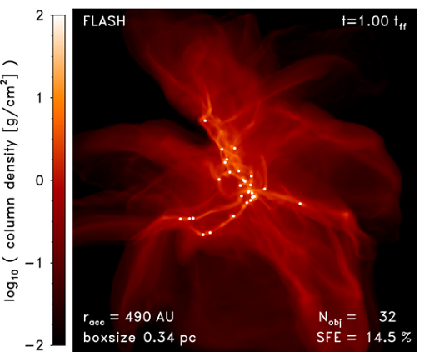} &
\includegraphics[width=0.315\linewidth]{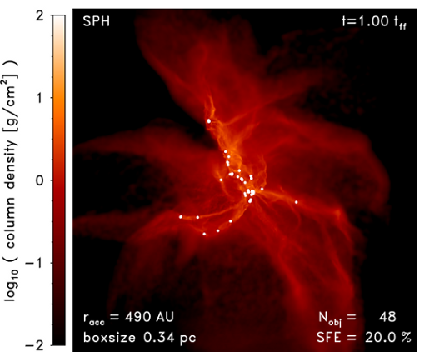} &
\includegraphics[width=0.315\linewidth]{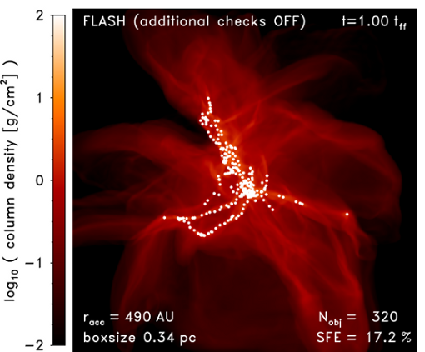} \\
\includegraphics[width=0.315\linewidth]{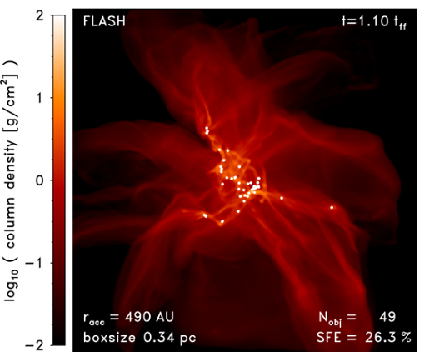} &
\includegraphics[width=0.315\linewidth]{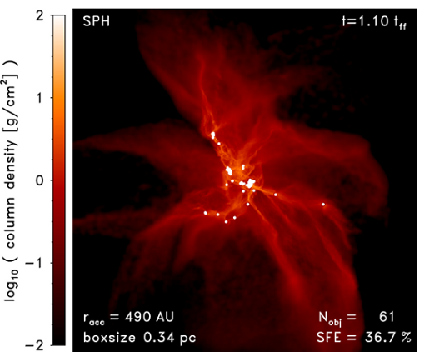} &
\includegraphics[width=0.315\linewidth]{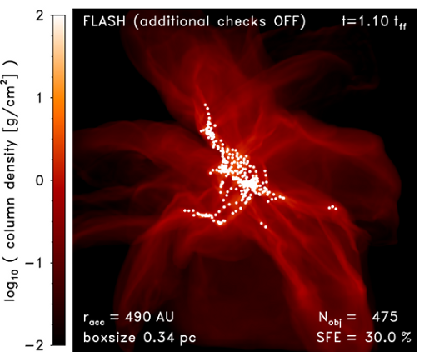}
\end{tabular}
\end{center}
\caption{Figure~\ref{fig:coldens_init} \emph{continued}. White filled circles show the projected position of sink particles obtained with \textsc{flash} (left panels) and \textsc{sph} (middle panels) as a function of the global freefall time ($t/\tff=0.8,\;0.9,\;1.0,\;1.1$, from top to bottom). Sink particles form at roughly the same positions in the \textsc{flash} and \textsc{sph} runs. The number and mass of sink particles is in excellent agreement for $t<1\,\tff$. At later times ($t>1\,\tff$), the stellar cluster produced in the \textsc{sph} run collapses on slightly smaller timescales than the \textsc{flash} cluster, and is more centrally condensed at $t=1.1\,\tff$. The right panels show the effect of \emph{only} using a density threshold for sink particle creation with \textsc{flash}, compared to the default \textsc{flash} run (left panels), for which \emph{all} sink particle creation checks (see Section~\ref{sec:creationchecks}) were used. Without additional checks for converging flows, gravitational potential minimum, bound state, and Jeans instability, a significant number of spurious sink particles are created, which do not represent collapsing objects. Most of the sink particles in the right hand panels were formed in transient density peaks caused by shocks that did not accumulate enough mass to create gravitationally bound and collapsing objects. The much larger number of particles created in that case may be reduced by some sort of merging algorithm, but the accreted mass in sink particles would still be overestimated, in particular at early times (by 133\%, 51\%, 19\% and 14\% compared to the default \textsc{flash} run at $t/\tff=0.8,\,0.9,\,1.0$ and 1.1, respectively), if a sole density threshold was used for sink particle creation at densities below the opacity limit. In contrast, the sink particles in the left hand plots (default \textsc{flash} runs) represent gravitationally bound and collapsing objects, as in the \textsc{sph} run (middle panels).}
\label{fig:coldens}
\end{figure*}

\begin{figure*}[t]
\begin{center}
\def\arraystretch{0.2}
\begin{tabular}{ccc}
\includegraphics[width=0.32\linewidth]{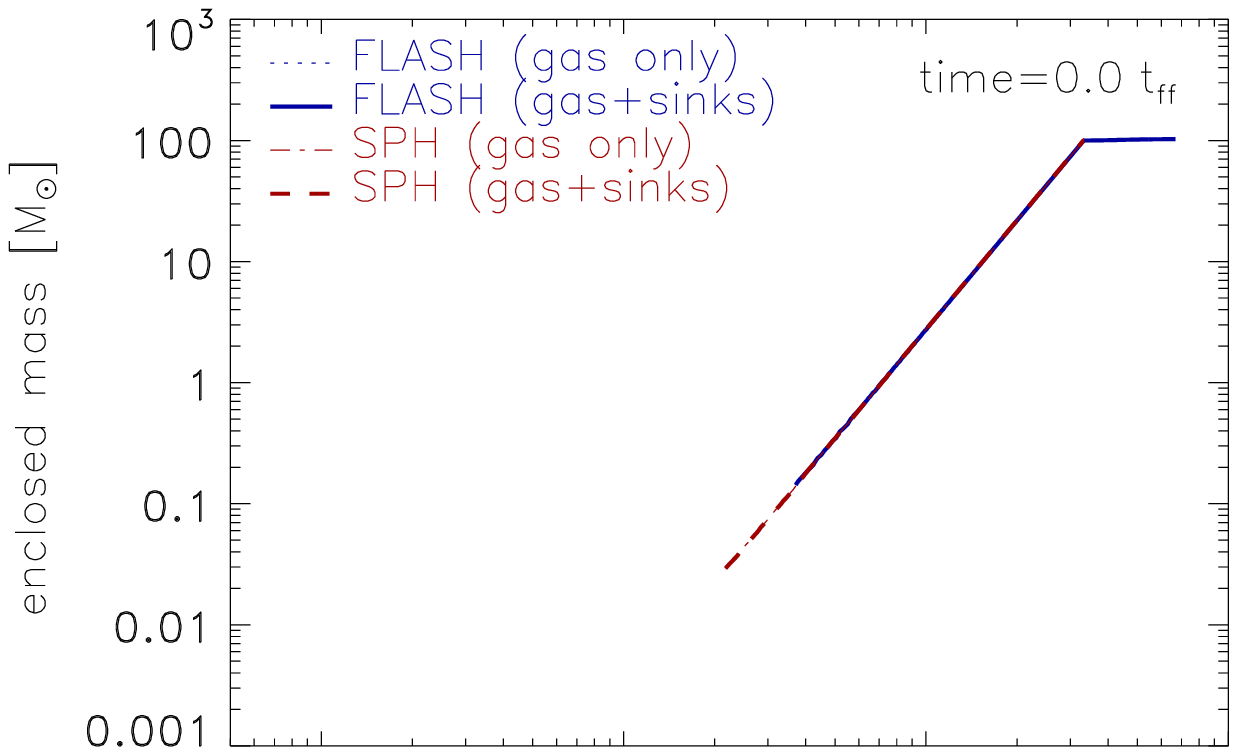} &
\includegraphics[width=0.32\linewidth]{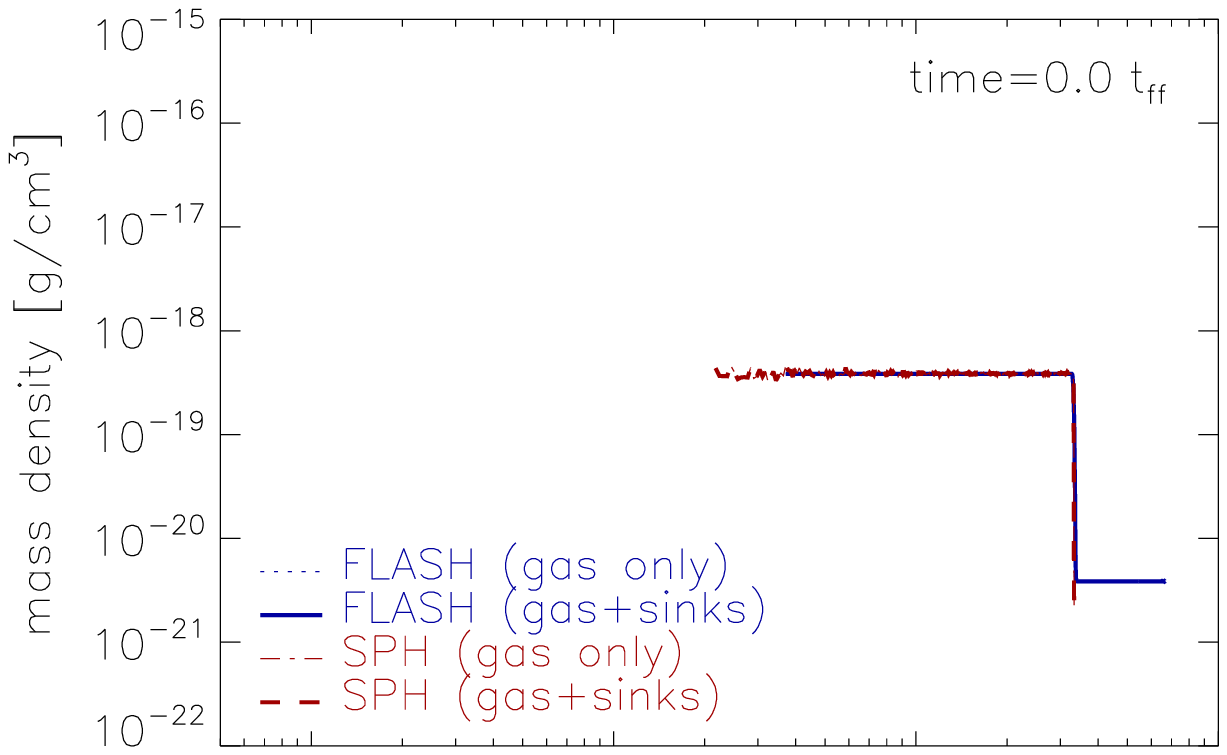} &
\includegraphics[width=0.32\linewidth]{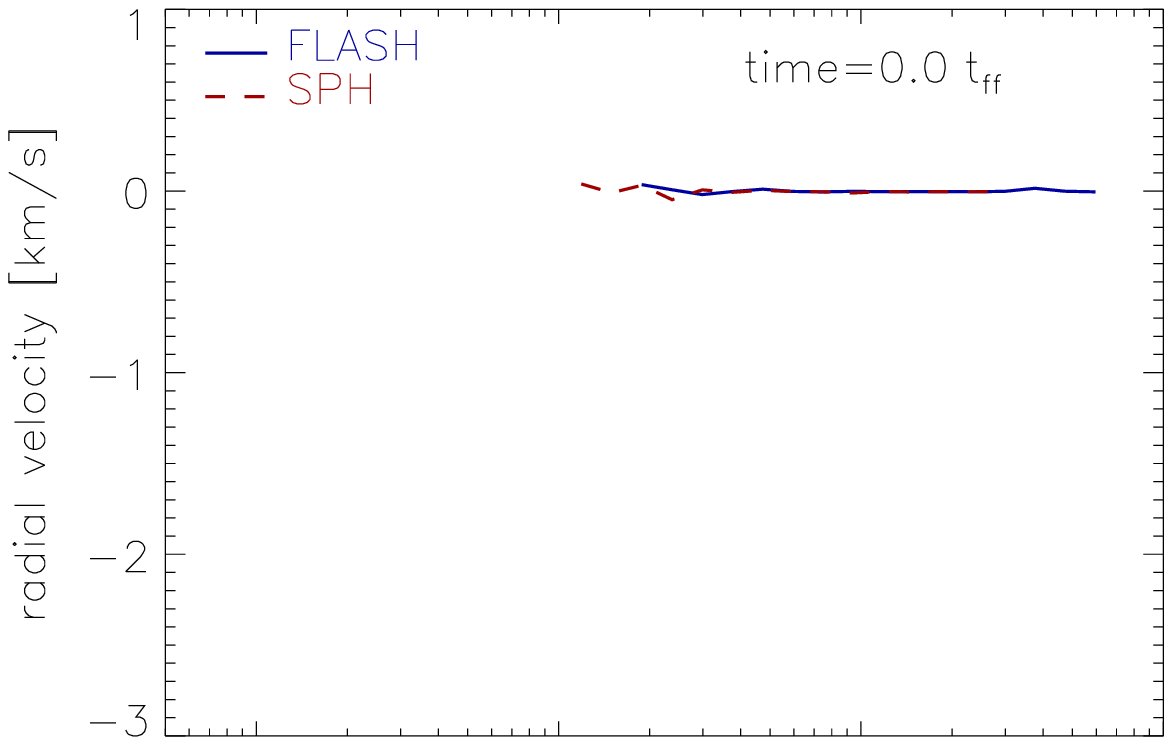} \\
\includegraphics[width=0.32\linewidth]{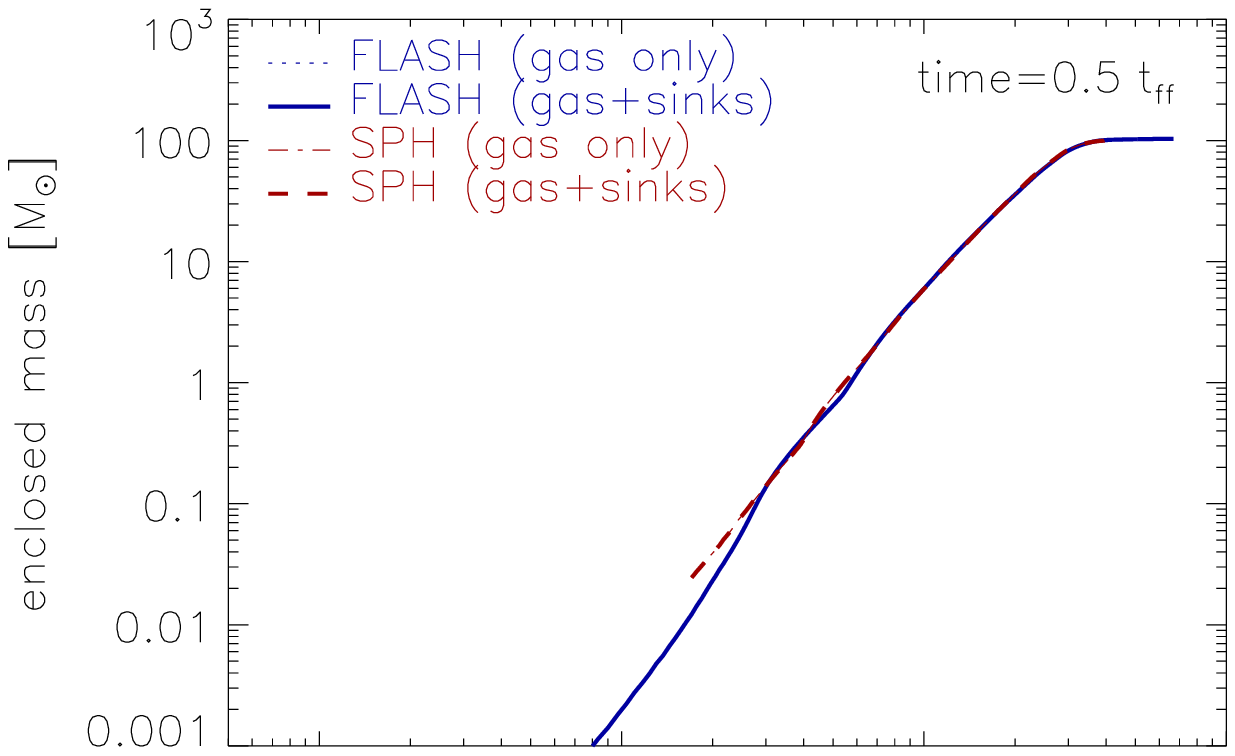} &
\includegraphics[width=0.32\linewidth]{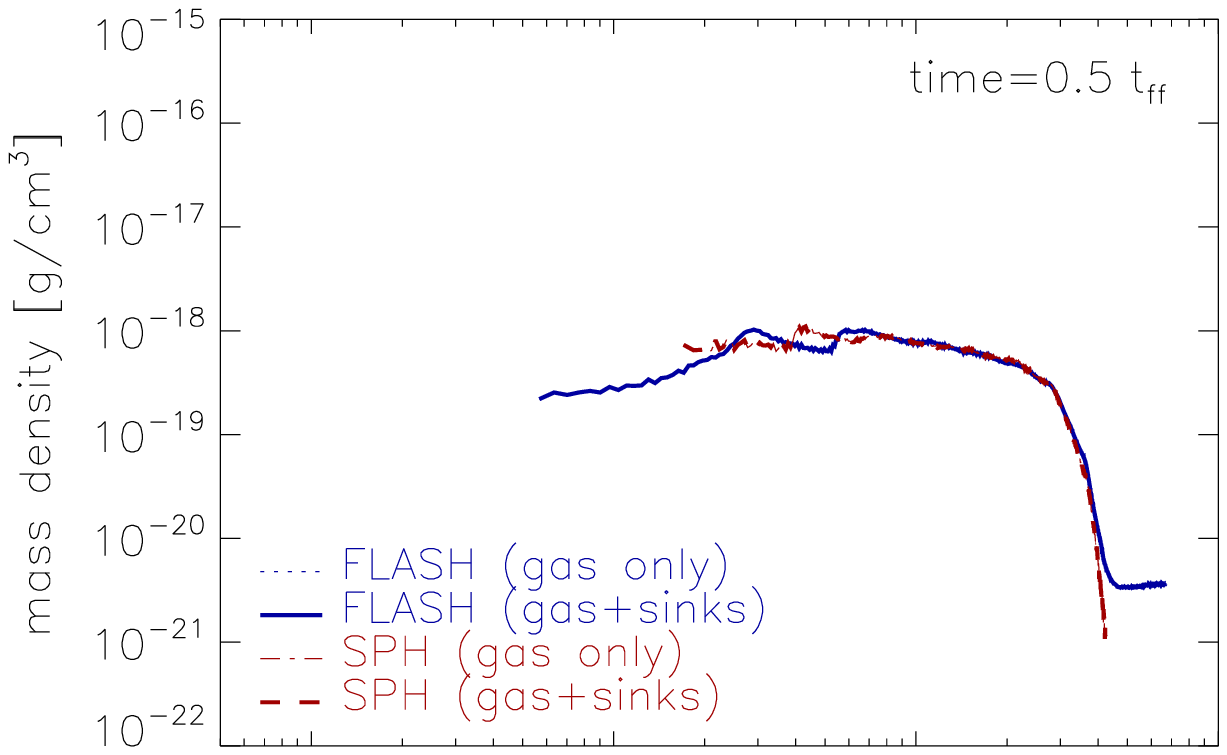} &
\includegraphics[width=0.32\linewidth]{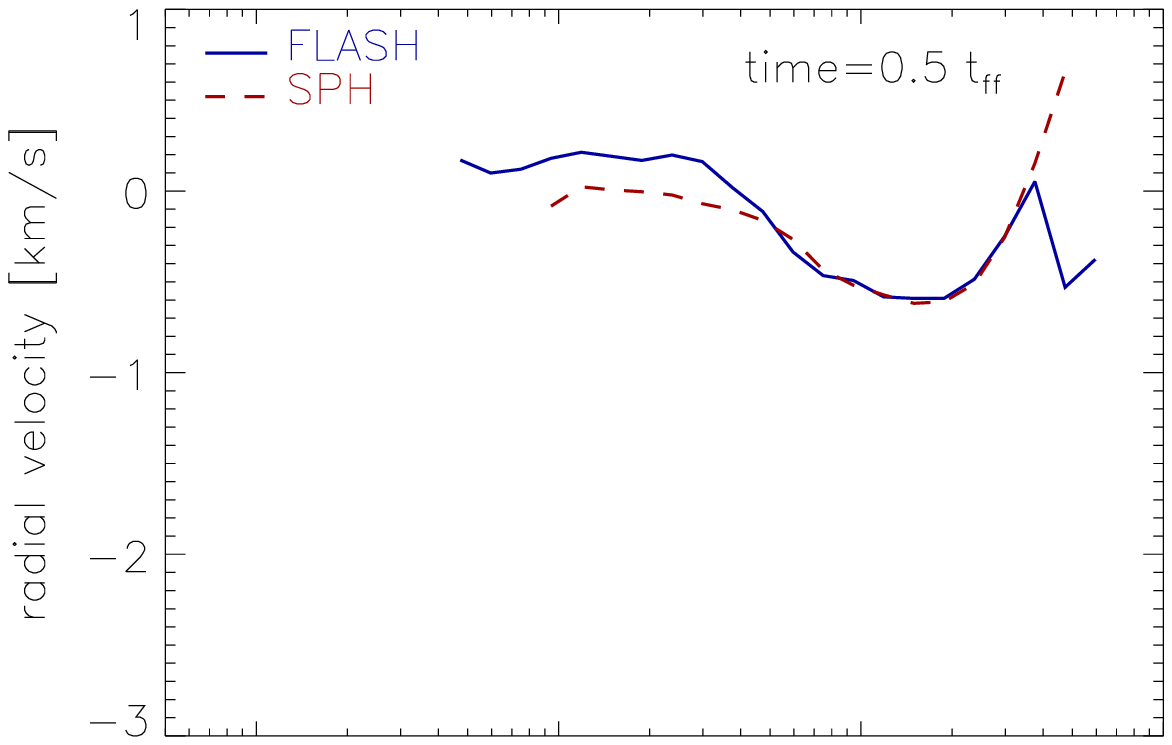} \\
\includegraphics[width=0.32\linewidth]{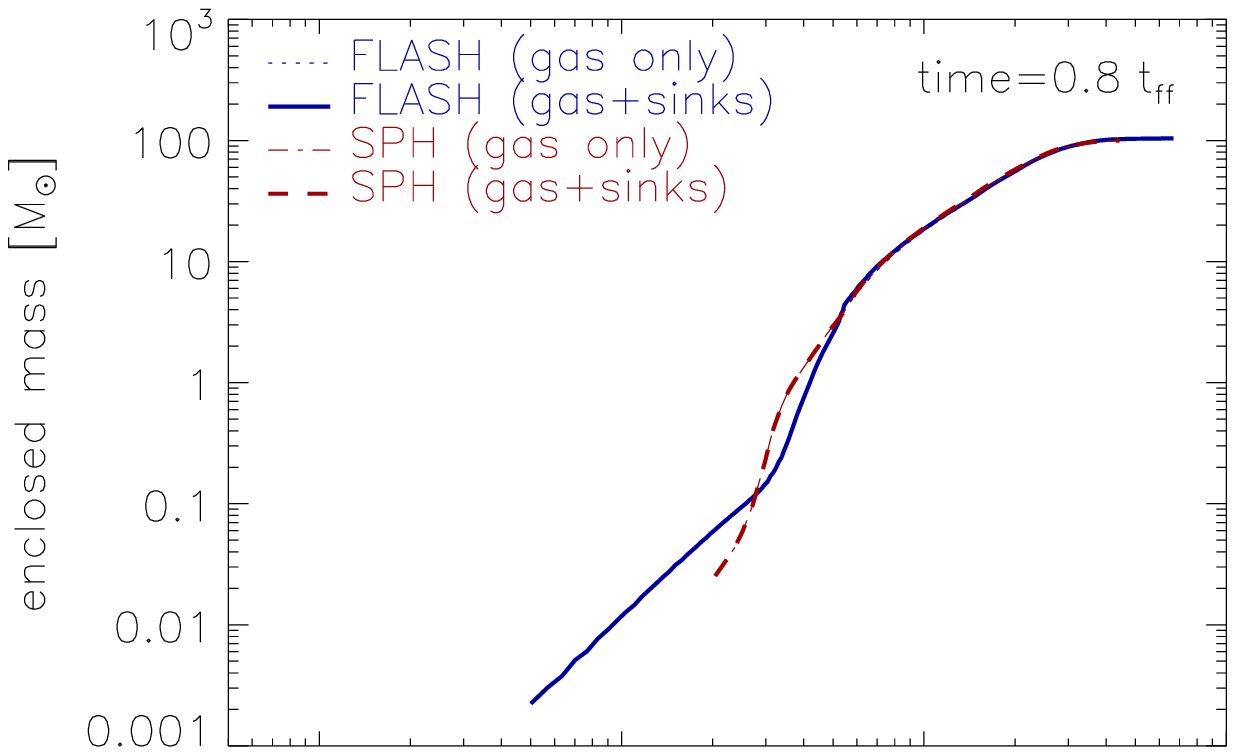} &
\includegraphics[width=0.32\linewidth]{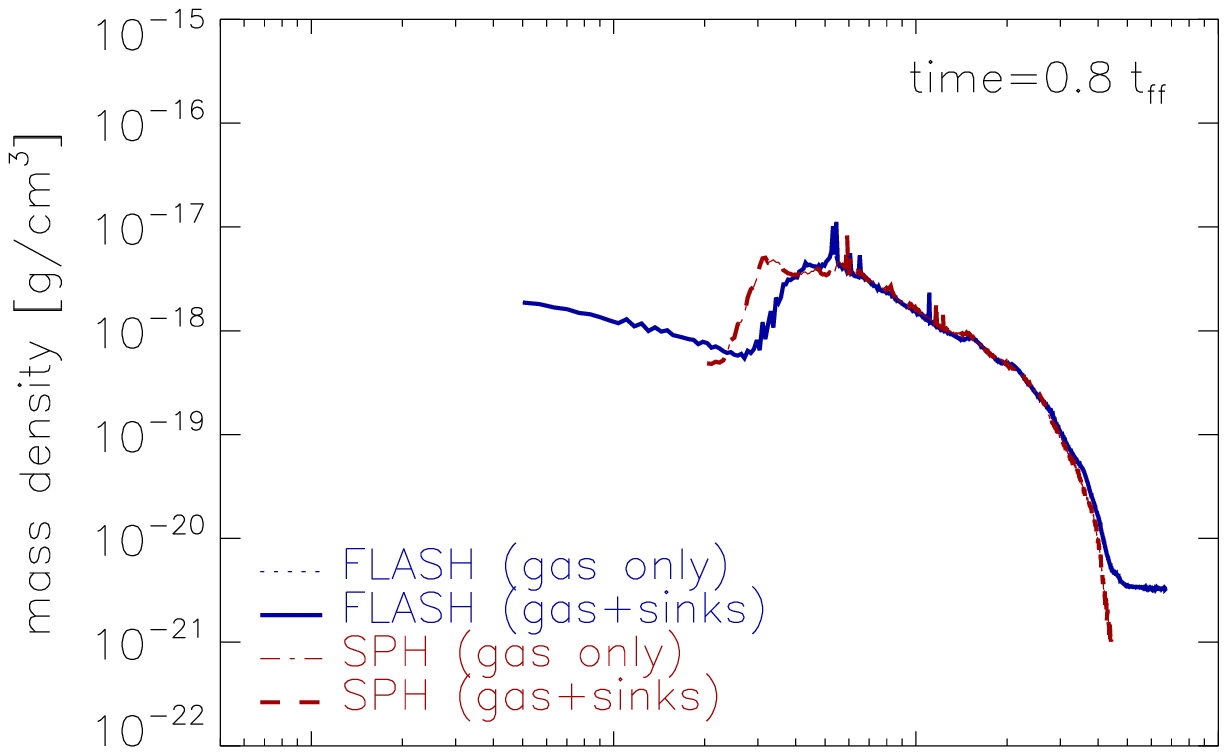} &
\includegraphics[width=0.32\linewidth]{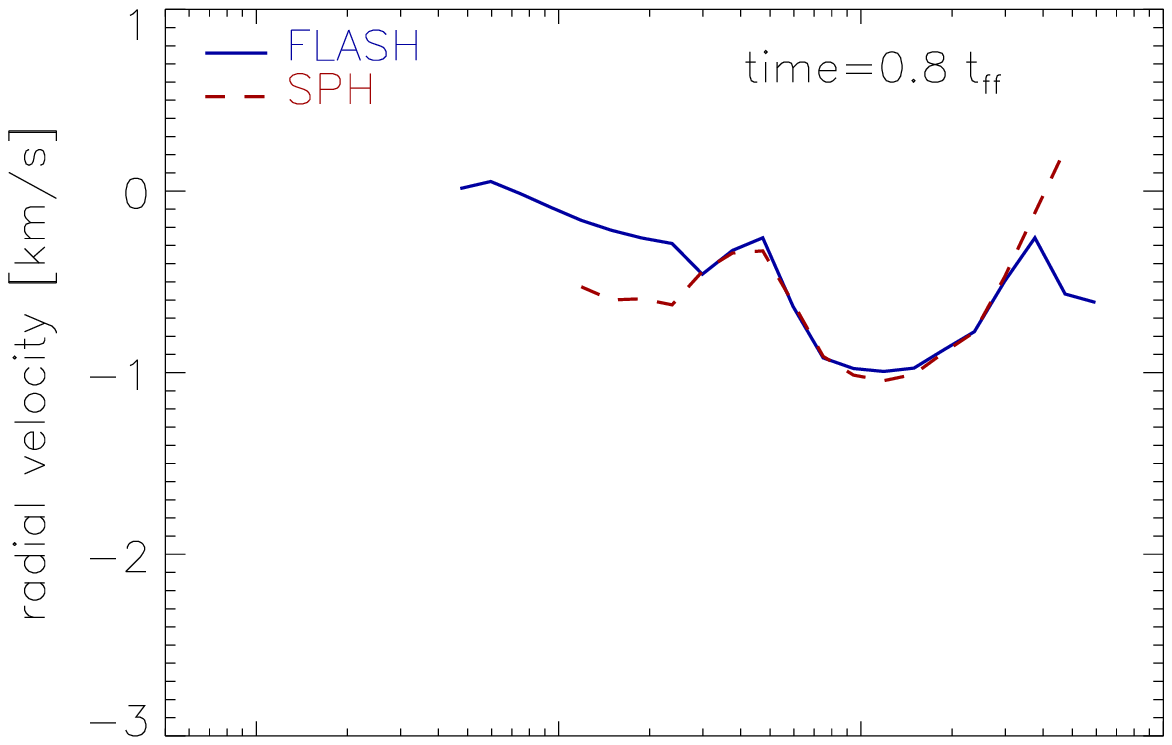} \\
\includegraphics[width=0.32\linewidth]{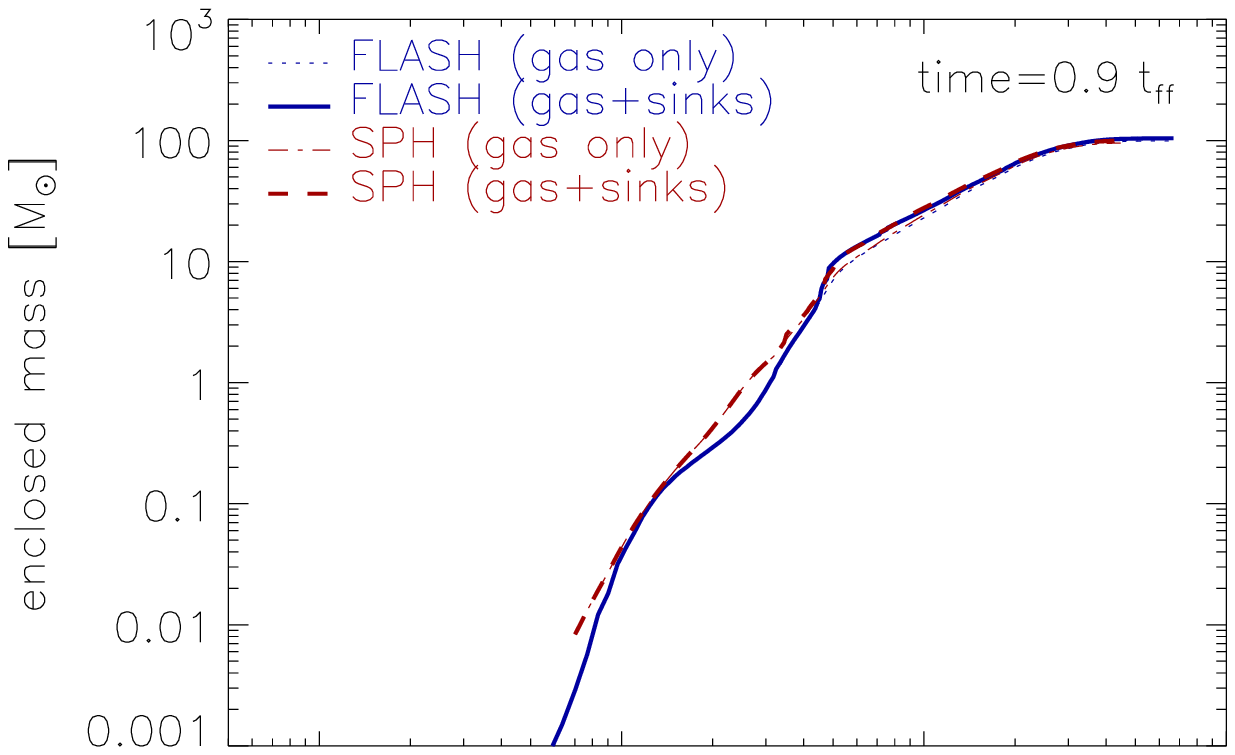} &
\includegraphics[width=0.32\linewidth]{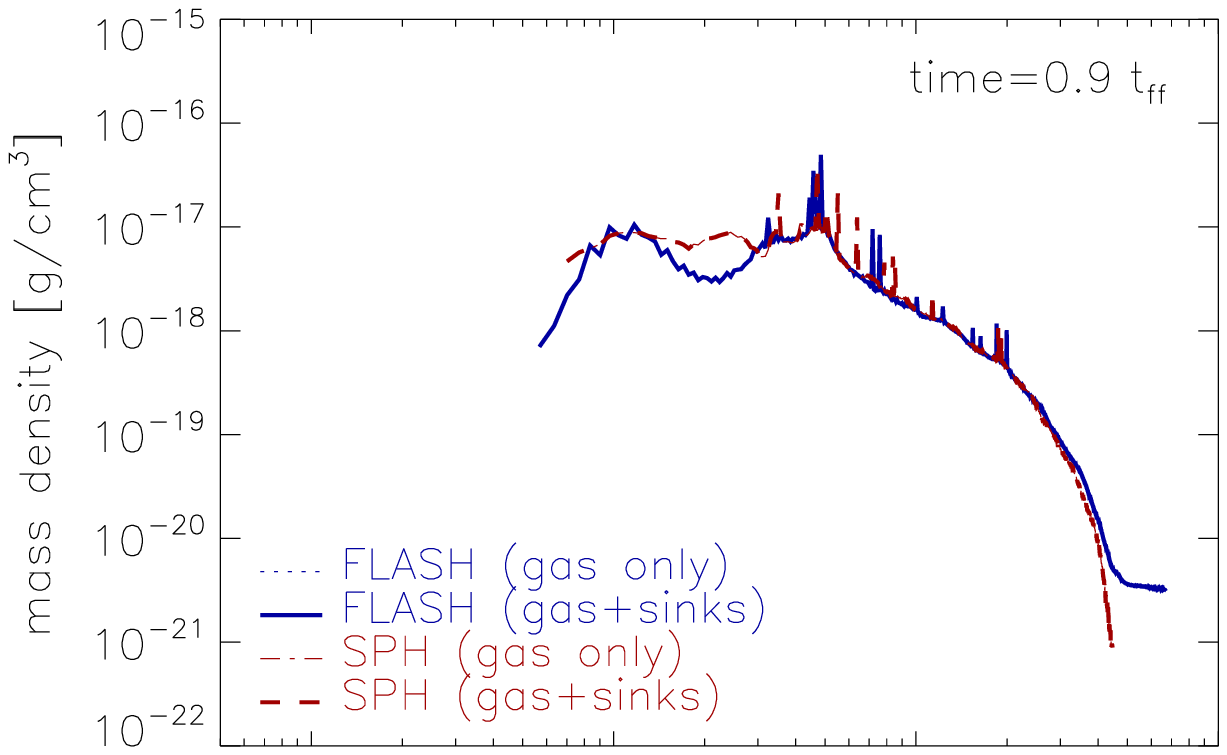} &
\includegraphics[width=0.32\linewidth]{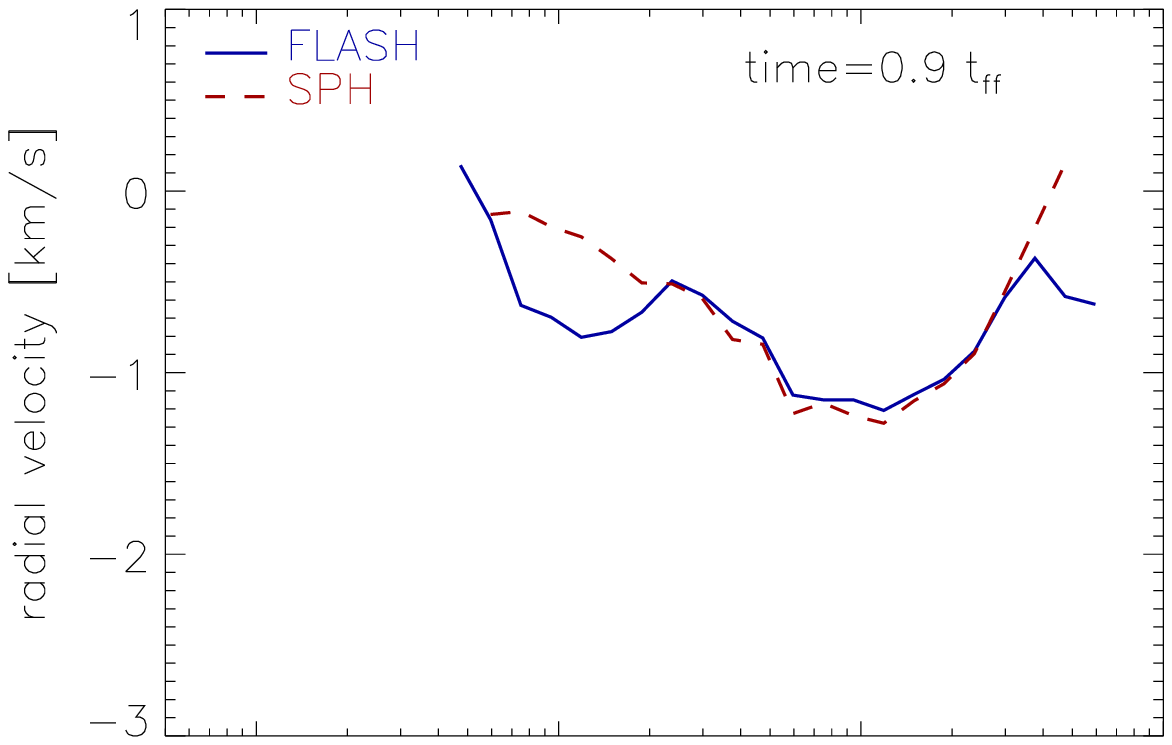} \\
\includegraphics[width=0.32\linewidth]{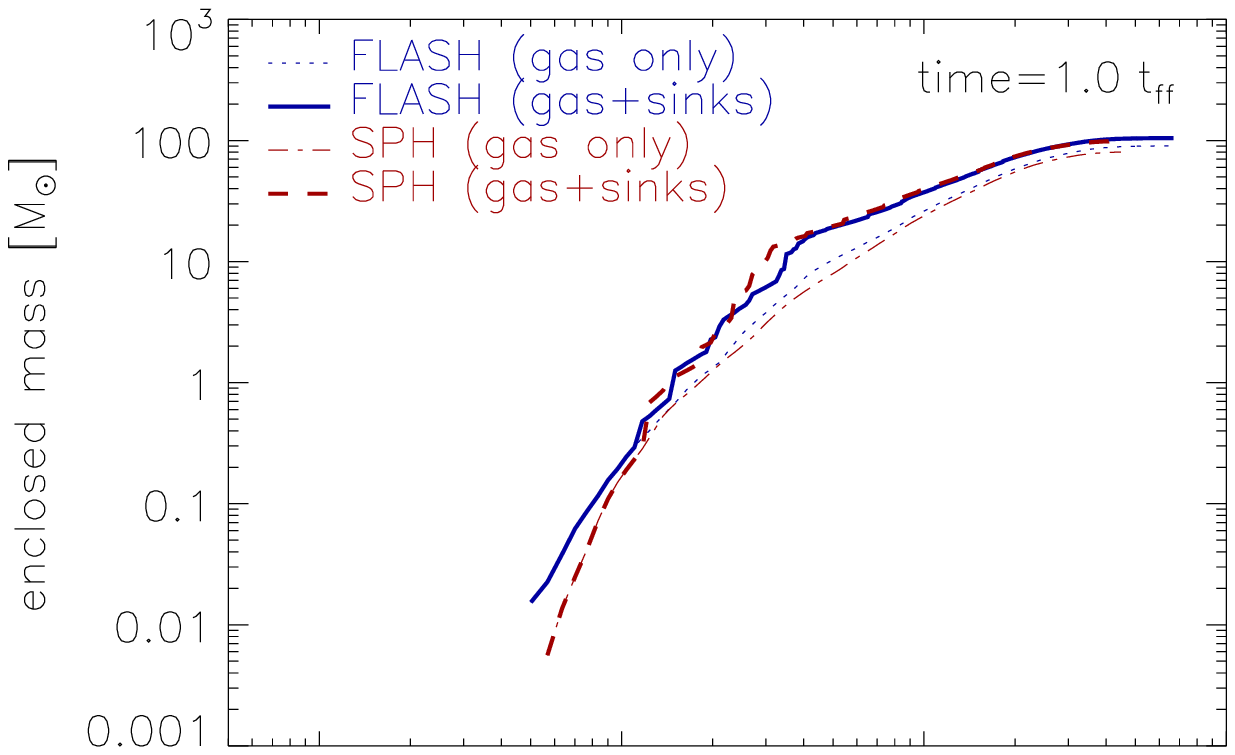} &
\includegraphics[width=0.32\linewidth]{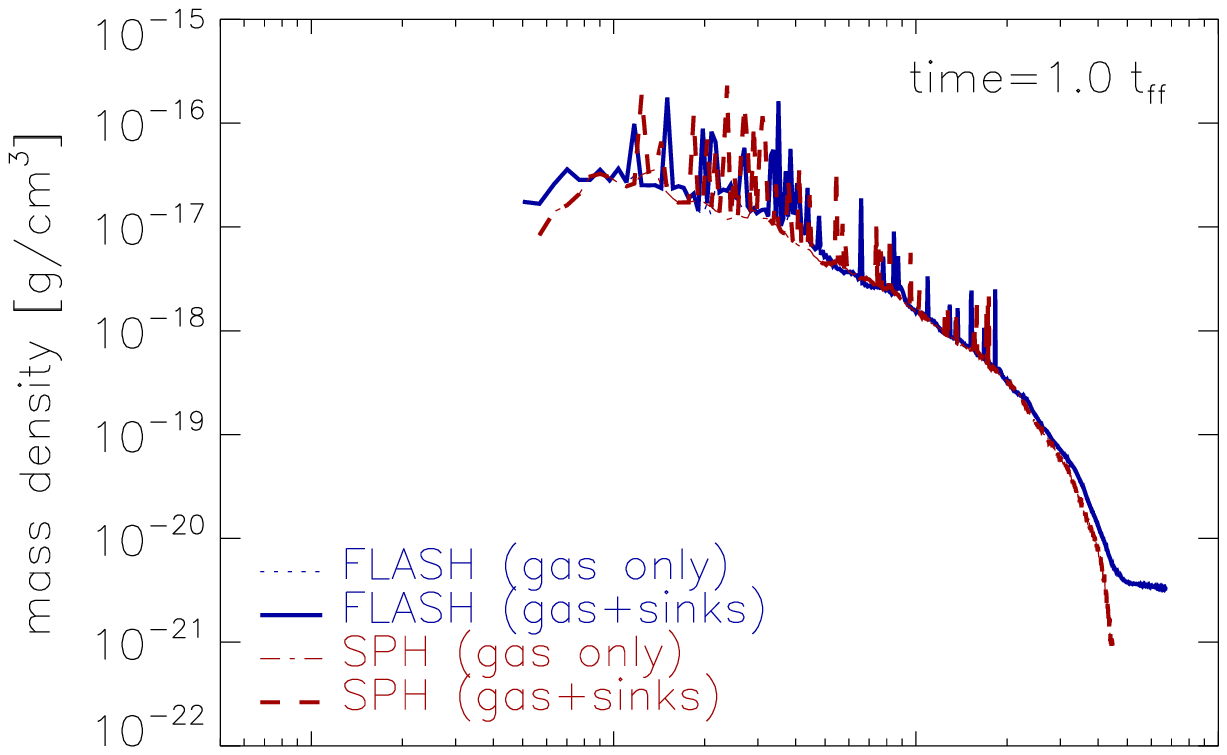} &
\includegraphics[width=0.32\linewidth]{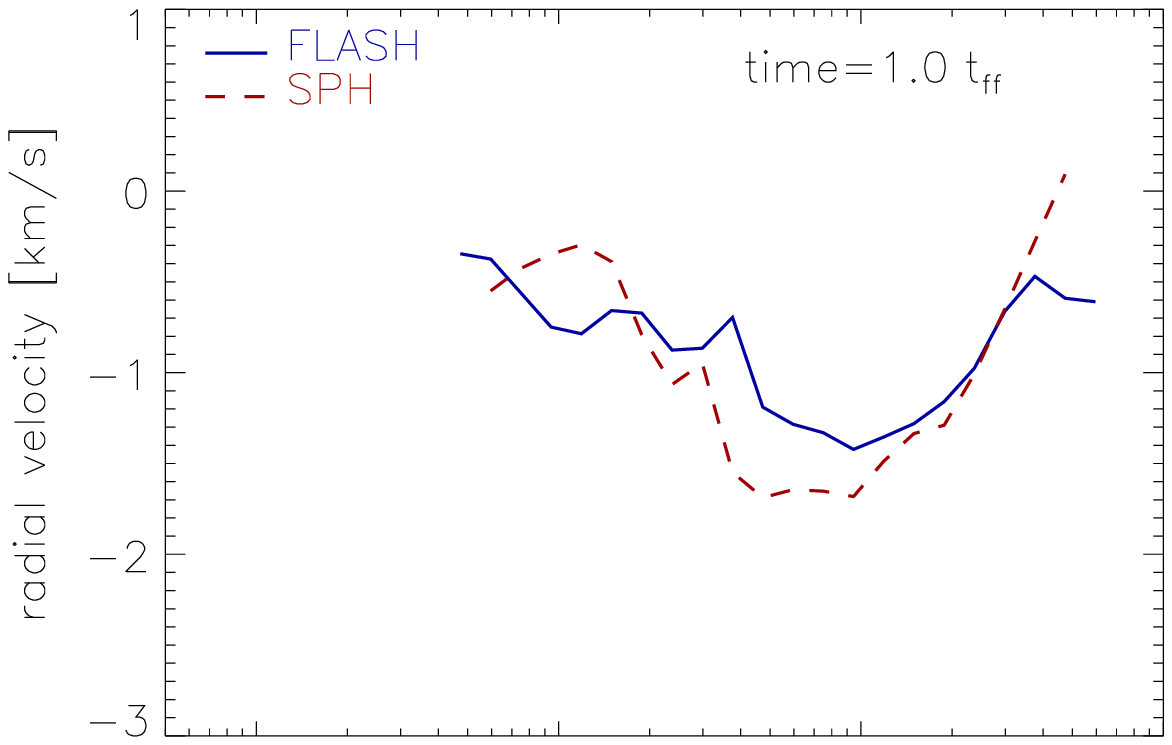} \\
\includegraphics[width=0.32\linewidth]{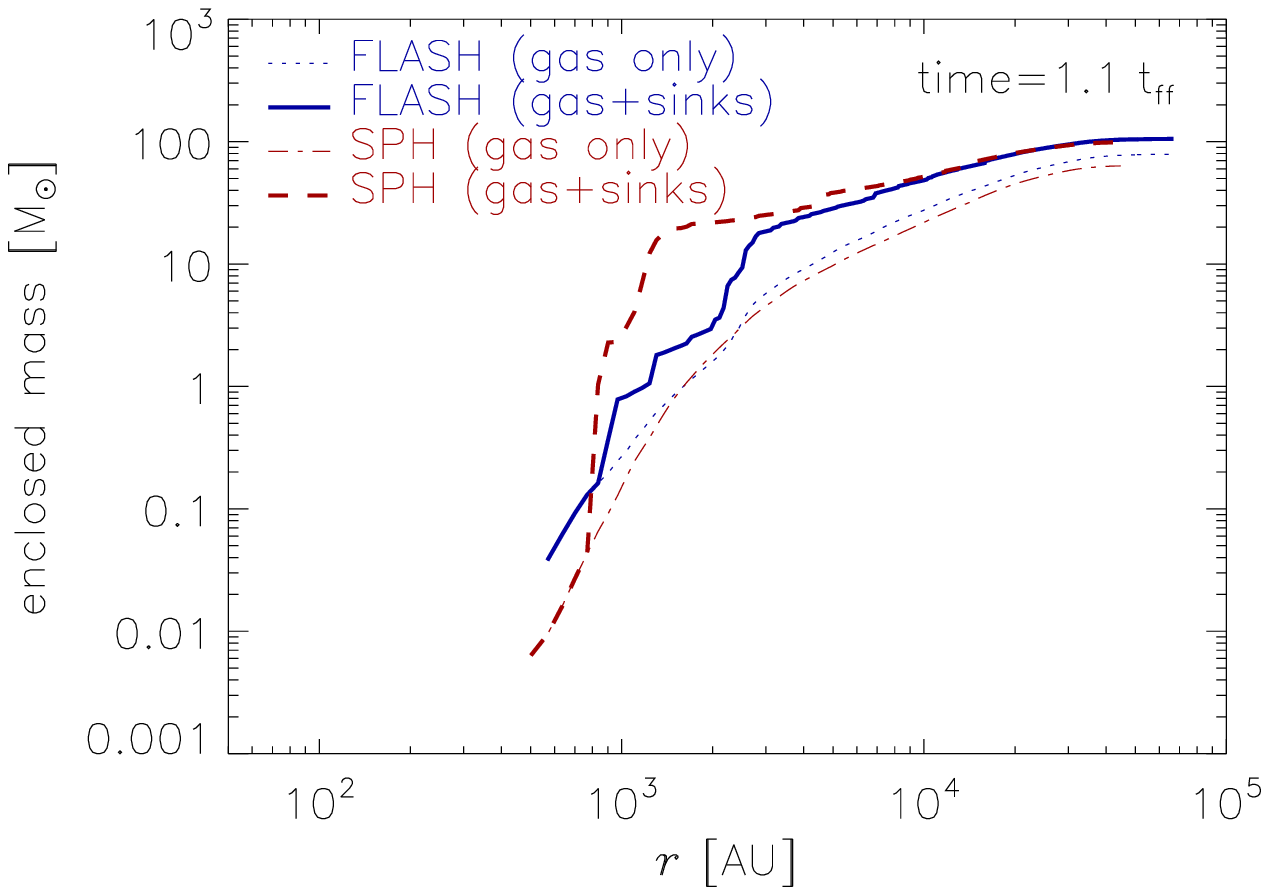} &
\includegraphics[width=0.32\linewidth]{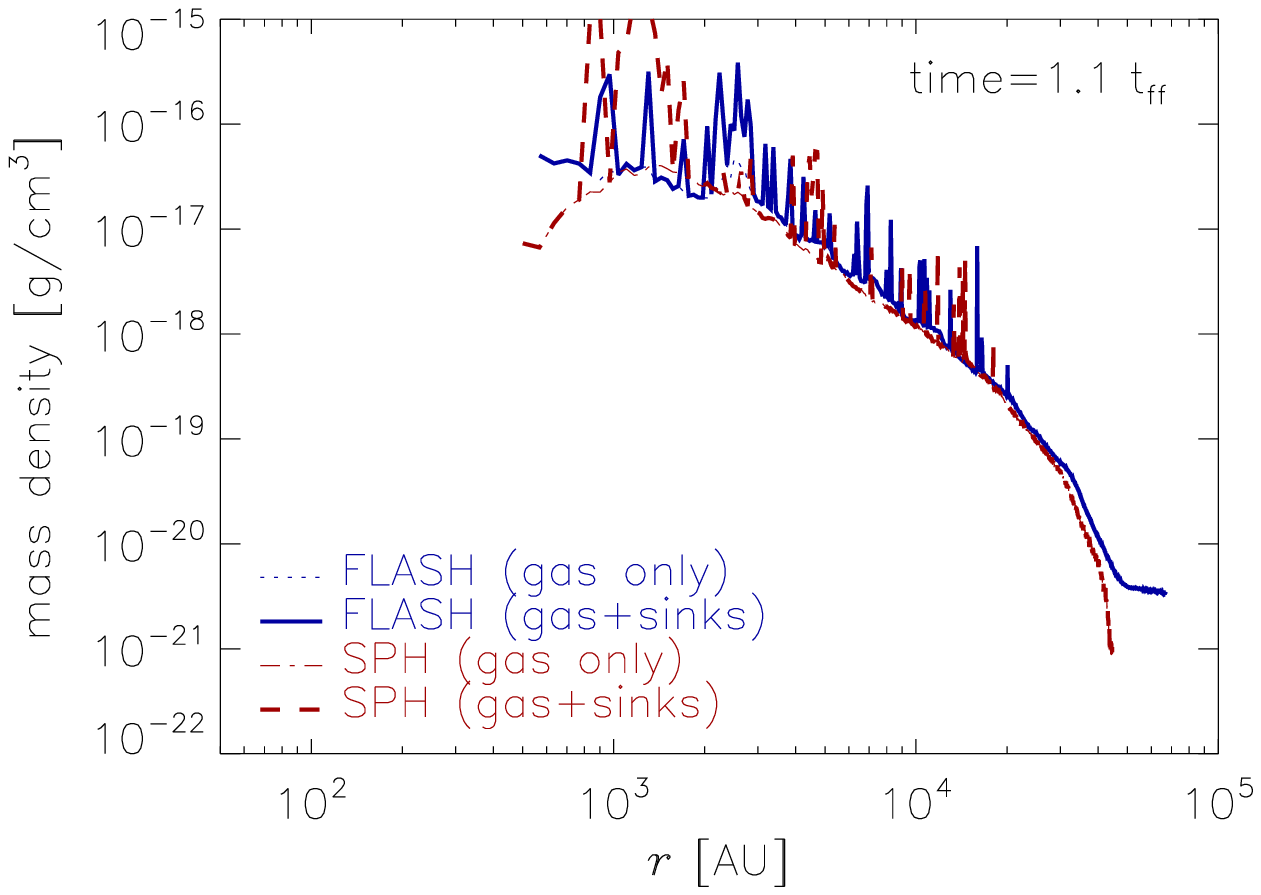} &
\includegraphics[width=0.32\linewidth]{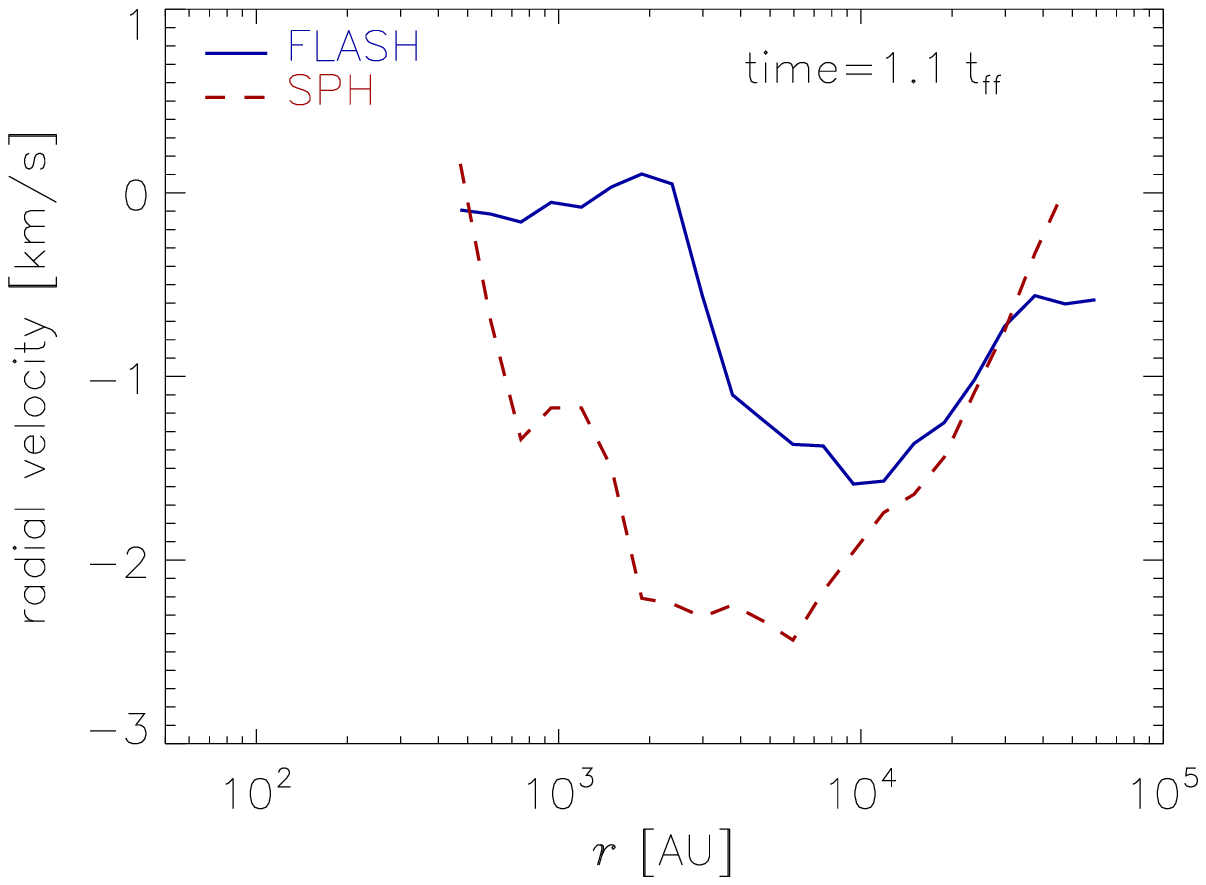}
\end{tabular}
\end{center}
\caption{Comparison of radial profiles (cumulative mass, density, and radial velocity from left to right), obtained in the \textsc{flash} and \textsc{sph} runs as a function of global freefall time ($t/\tff=0.0,\;0.5,\;0.8,\;0.9,\;1.0,\;1.1$, from top to bottom).}
\label{fig:rad_profiles}
\end{figure*}

\begin{figure*}[t]
\begin{center}
\def\arraystretch{0.2}
\begin{tabular}{cc}
\includegraphics[width=0.47\linewidth]{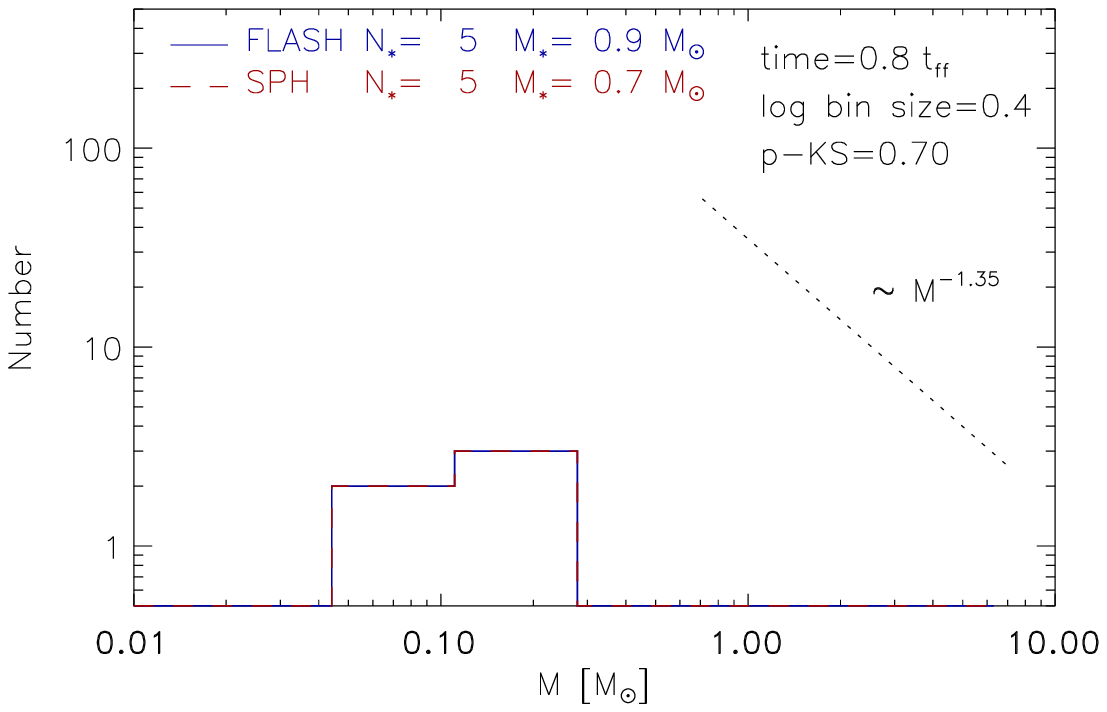} &
\includegraphics[width=0.47\linewidth]{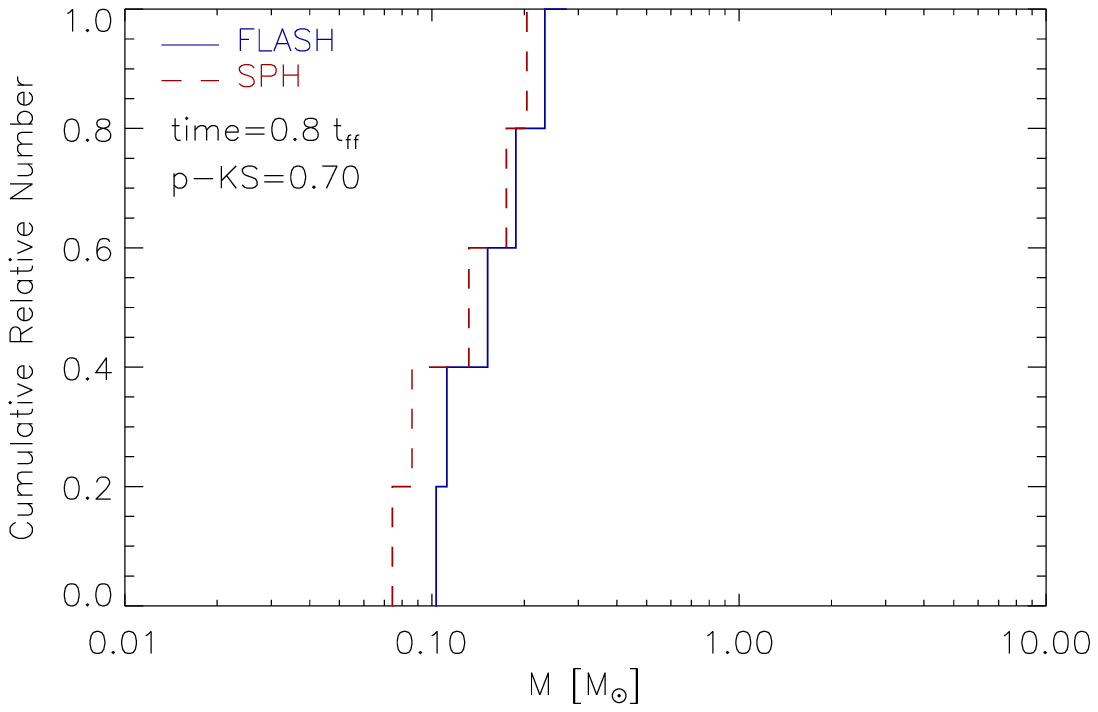} \\
\includegraphics[width=0.47\linewidth]{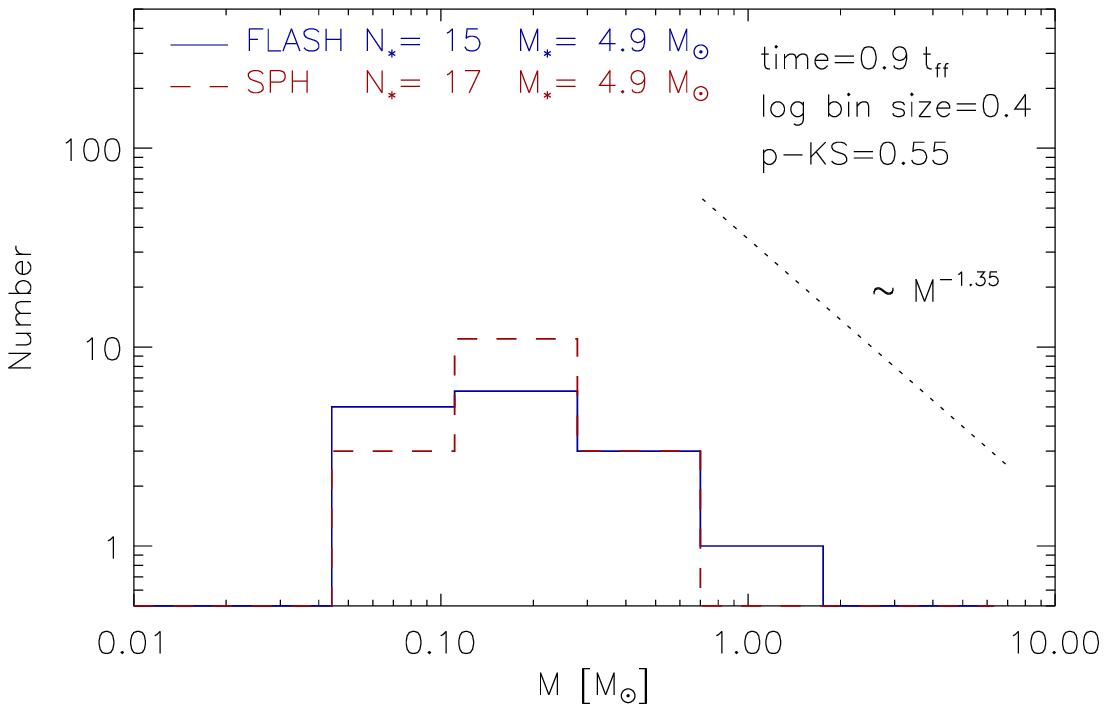} &
\includegraphics[width=0.47\linewidth]{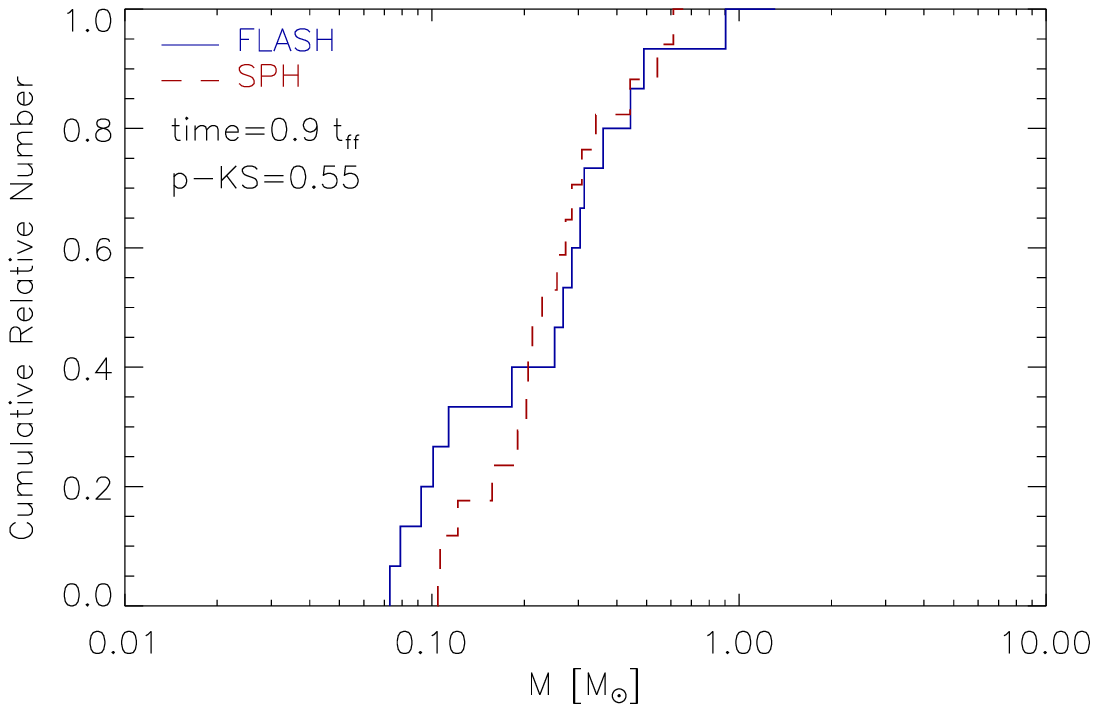} \\
\includegraphics[width=0.47\linewidth]{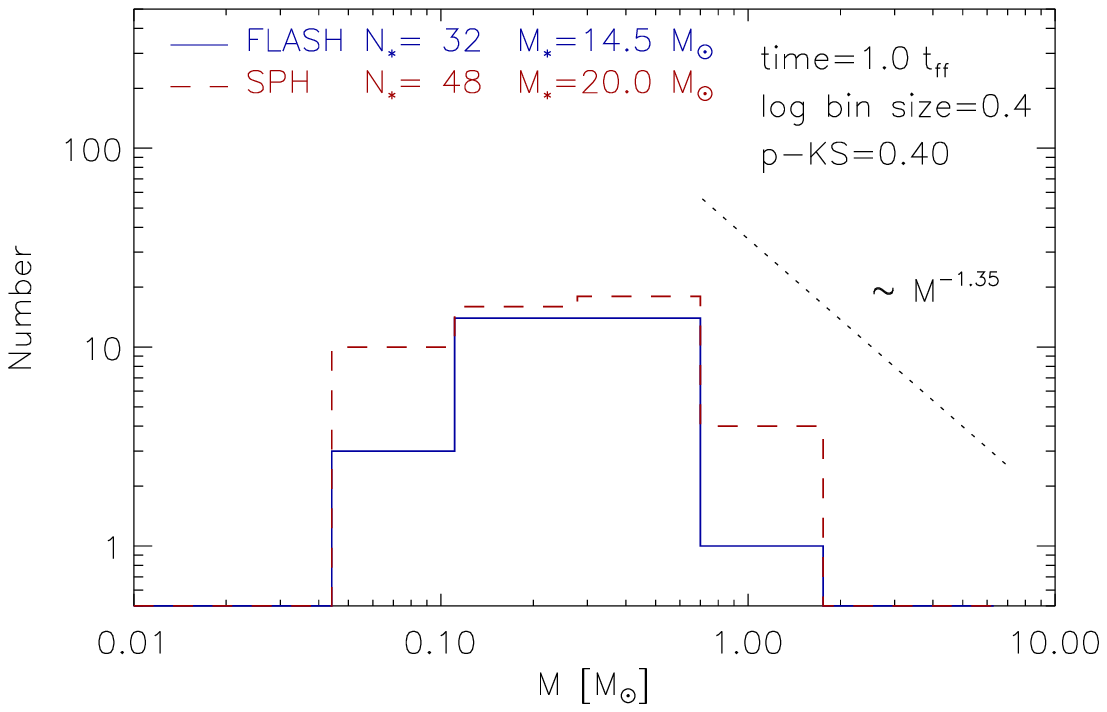} &
\includegraphics[width=0.47\linewidth]{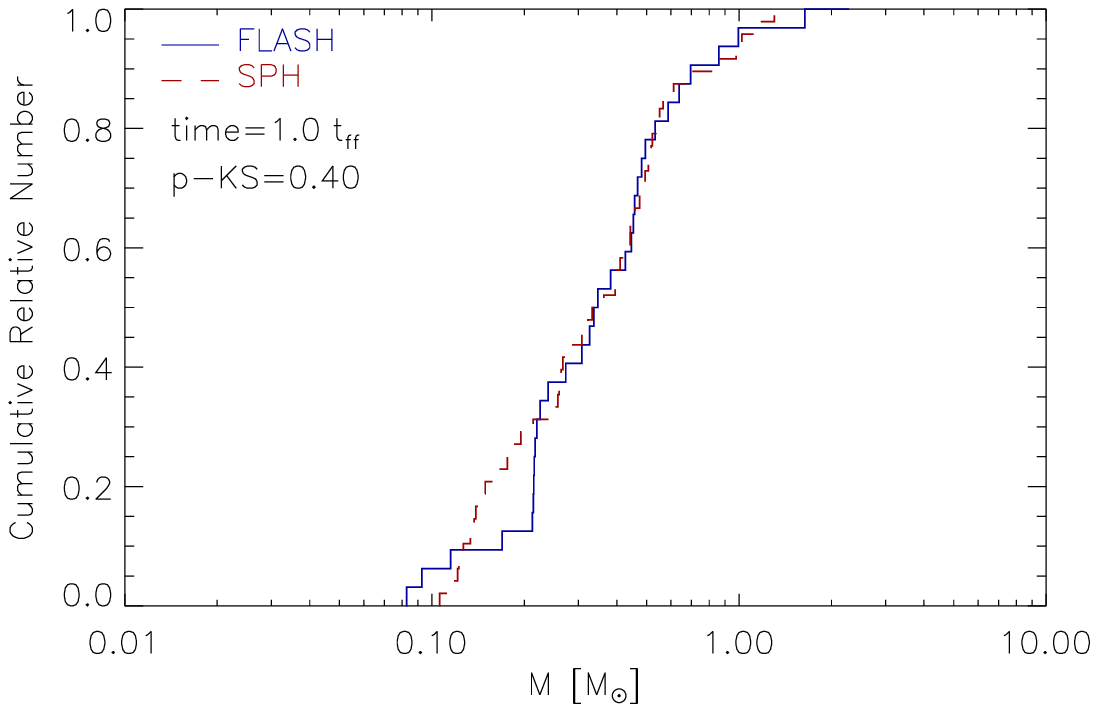} \\
\includegraphics[width=0.47\linewidth]{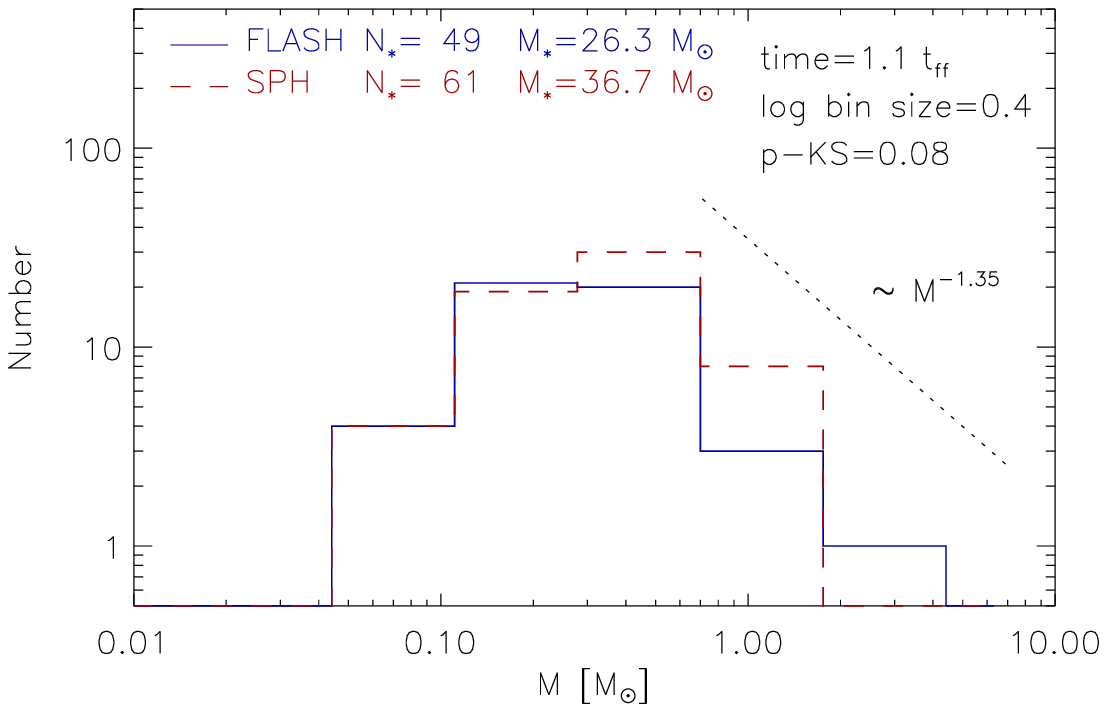} &
\includegraphics[width=0.47\linewidth]{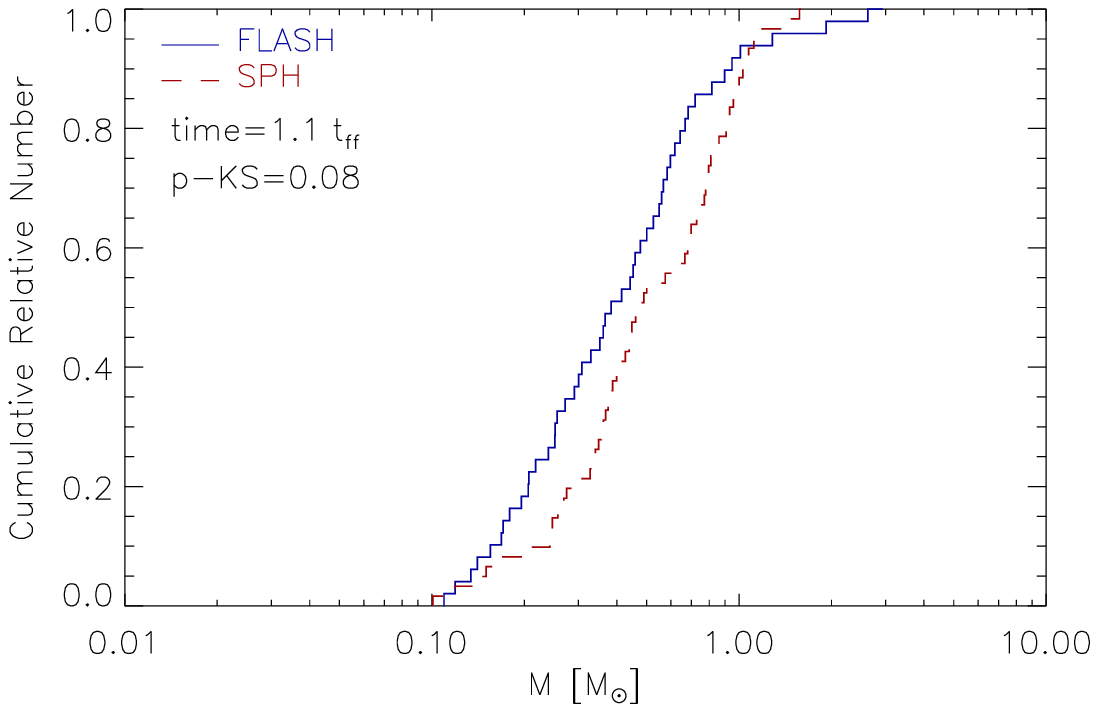}
\end{tabular}
\end{center}
\caption{Comparison of the sink particle mass distributions (left) and cumulative mass distributions (right), obtained with \textsc{flash} and \textsc{sph} as a function of global freefall time ($t/\tff=0.8,\;0.9,\;1.0,\;1.1$, from top to bottom).}
\label{fig:smfs}
\end{figure*}

\begin{figure*}[t]
\begin{center}
\def\arraystretch{0.2}
\begin{tabular}{cc}
\includegraphics[width=0.47\linewidth]{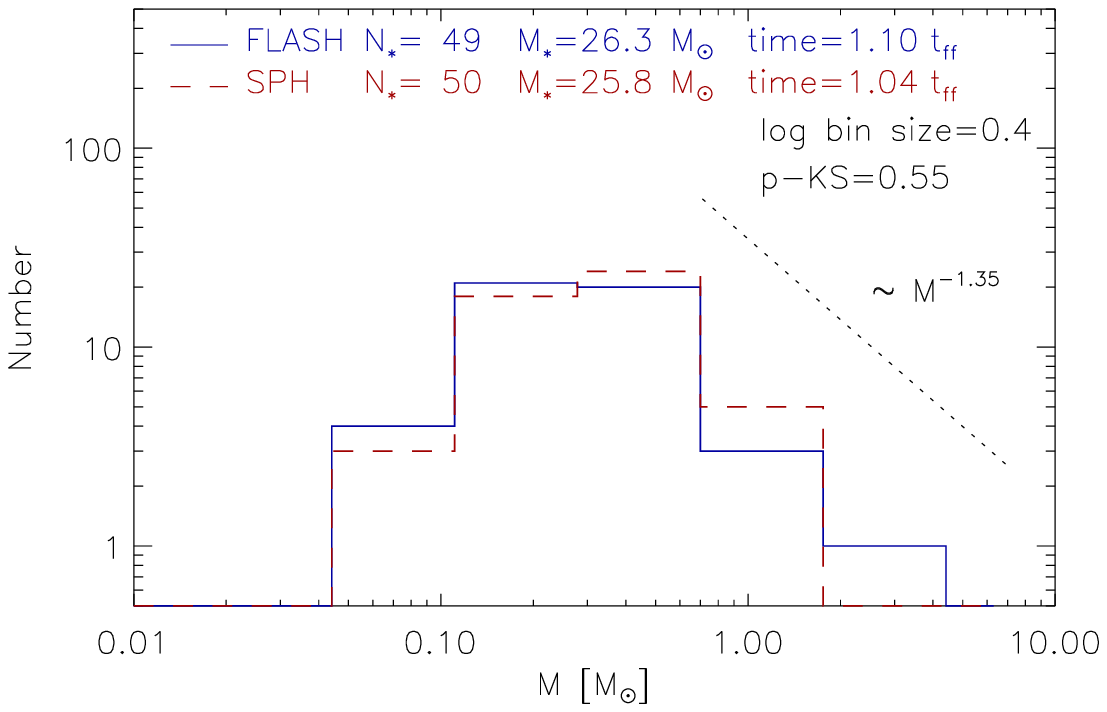} &
\includegraphics[width=0.47\linewidth]{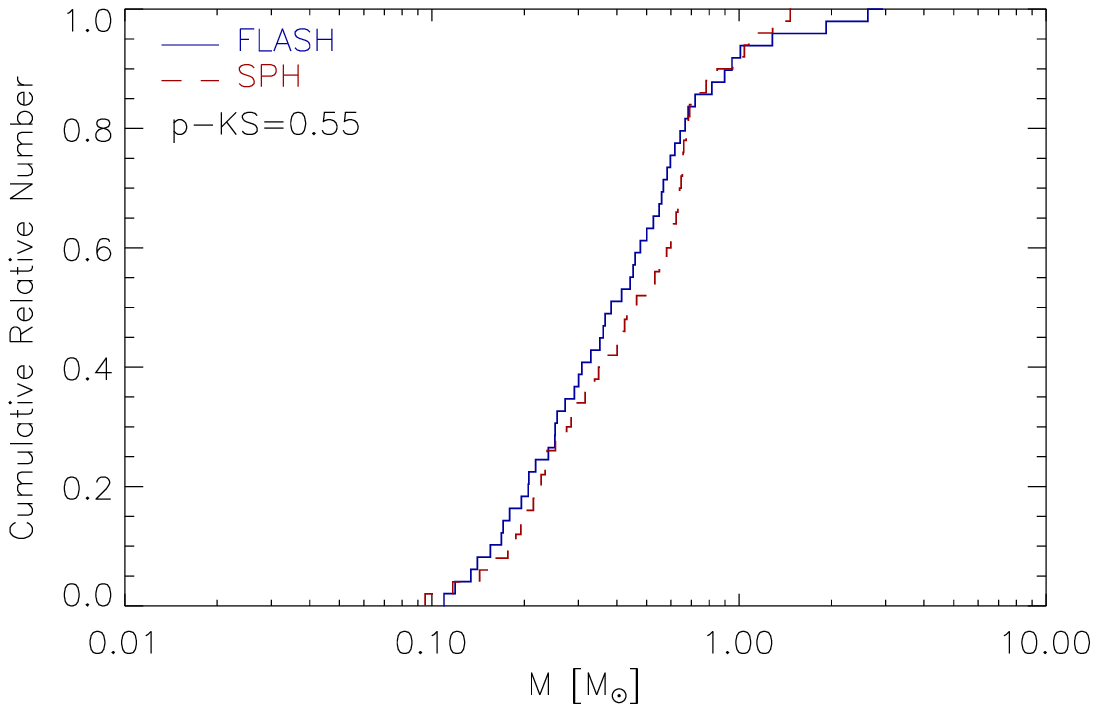}
\end{tabular}
\end{center}
\caption{Same as the bottom panels of Figure~\ref{fig:smfs}, but the mass function of \textsc{flash} at $t=1.10\,\tff$ is compared to the mass function of \textsc{sph} at $t=1.04\,\tff$, when the accreted gas mass is roughly the same for both runs ($M_\star\sim26\,\Msol$). The number of fragments (sink particles) is in excellent agreement between the two codes. The mass distributions are also very similar as shown by a Kolmogorov-Smirnov probability of 55\%, which indicates that the two samples are likely drawn from the same fundamental distribution.}
\label{fig:smfs_sfe}
\end{figure*}

\end{document}